\documentclass[useAMS,usenatbib,usegraphicx]{mn2e}

\usepackage{amsmath}
\usepackage{amsfonts}

\newcommand{\aap}{A\&A} 
 
\newcommand{\gca}{Geochim. Cosmochim. Acta} 
\newcommand{\mnras}{MNRAS} 
\newcommand{\apj}{ApJ} 
\newcommand{\araa}{ARA\&A} 
\newcommand{\apjl}{ApJ}
\newcommand{\apjs}{ApJS}
\newcommand{\pasa}{PASA}
\newcommand{\aj}{AJ}

\newcommand{\physrep}{Phys. Rep.}

\newcommand{\pasj}{PASJ}
\newcommand{\nat}{Nature} 
\newcommand{\cjaa}{Chinese J. Astron. Astrophys.} 
\newcommand{\pasp}{PASP} 
\newcommand{\prl}{Phys. Rev. Lett.} 
\newcommand{\rmp}{Rev. Mod. Phys.}

\title[PCA on Chemical Abundances Spaces]{Principal Component Analysis on Chemical Abundances Spaces}
\author[Y.~S.~Ting, K.~C.~Freeman, et al.]{Y.~S.~Ting,$^{1,2}$\thanks{E-mail: ting.yuansen@gmail.com} K.~C.~Freeman,$^{1}$ C.~Kobayashi,$^{1,3}$ G.~M.~De~Silva$^{4}$ and \newauthor J.~Bland-Hawthorn$^{5}$\\
$^{1}$Research School of Astronomy $\&$ Astrophysics, The Australian National University, Cotter Road, Weston Creek, ACT 2611, Australia\\
$^{2}$Department of Physics, National University of Singapore, Singapore 119260\\
$^{3}$Centre for Astrophysics Research, University of Hertfordshire, Hatfield AL10 9AB, United Kingdom\\
$^{4}$Australian Astronomical Observatory, PO Box 296, NSW 1710, Australia \\
$^{5}$Institute of Astronomy, School of Physics, University of Sydney, NSW2006, Australia}
\begin{document}



\maketitle

\label{firstpage}

\begin{abstract}
In preparation for the {\it HERMES} chemical tagging survey of about a million Galactic FGK stars, we estimate the number of independent dimensions of the space defined by the stellar chemical element abundances [X/Fe{]}. This leads to a way to study the origin of elements from observed chemical abundances using Principal Component Analysis. We explore abundances in several environments, including solar neighbourhood thin/thick disk stars, halo metal-poor stars, globular clusters, open clusters, the Large Magellanic Cloud and the Fornax dwarf spheroidal galaxy. By studying solar-neighbourhood stars, we confirm the universality of the $r$-process that tends to produce [$n$-capture elements/Fe{]} in a constant ratio. We find that, especially at low metallicity, the production of $r$-process elements is likely to be associated with the production of $\alpha$-elements. This may support the core-collapse supernovae as the $r$-process site. We also verify the over-abundances of light $s$-process elements at low metallicity, and find that the relative contribution decreases at higher metallicity, which suggests that this lighter elements primary process may be associated with massive stars. We also verify the contribution from the $s$-process in low-mass AGB stars at high metallicity. Our analysis reveals two types of core-collapse supernovae: one produces mainly $\alpha$-elements, the other produces both $\alpha$-elements and Fe-peak elements with a large enhancement of heavy Fe-peak elements which may be the contribution from hypernovae. Excluding light elements that may be subject to internal mixing, K and Cu, we find that the [X/Fe{]} chemical abundance space in the solar neighbourhood has about 6 independent dimensions both at low metallicity ($-3.5 \la \mbox{[Fe/H{]}} \la -2$) and high metallicity ($\mbox{[Fe/H{]}} \ga -1$). However the dimensions come from very different origins in these two cases. The extra contribution from low mass AGB stars at high metallicity compensates the dimension loss due to the homogenization of the core-collapse supernovae ejecta. Including the extra dimensions from [Fe/H{]}, K, Cu and the light elements, the number of independent dimensions of the [X/Fe]+[Fe/H] chemical space in the solar neighbourhood for {\it HERMES} is about 8 to 9. Comparing fainter galaxies and the solar neighbourhood, we find that the chemical space for fainter galaxies such as Fornax and the Large Magellanic Cloud has a higher dimensionality. This is consistent with the slower star formation history of fainter galaxies. We find that open clusters have more chemical space dimensions than the nearby metal-rich field stars. This suggests that a survey of stars in a larger Galactic volume than the solar neighbourhood may show about $1$ more dimension in its chemical abundance space.  
\end{abstract}


\section[]{Introduction}
Stars are believed to form in aggregates which are mostly short-lived.  They disrupt through the action of mass loss from stellar evolution, two-body effects and the tidal field of the Galaxy \citep[e.g.][]{che90,ode03}.  After the aggregates dissolve, their debris disperses and after several Galactic rotation periods becomes mixed throughout an annular region around the Galaxy.  In this way, the stellar disk was gradually built up.  The goal of chemical tagging is to use element abundances to reconstruct these ancient clusters in which the stars were born \citep{ken02}.  Individual clusters are observed to be chemically homogeneous, at least in the elements heavier than Na \citep[e.g.][]{siv06,mik10,mik11}.  Stars that were born in the same cluster and have now dispersed will have similar element abundance patterns reflecting the chemical evolution of the gas from which they formed.  This gas had its own history of  pollution by ejecta  from core-collapse supernovae,  type Ia supernovae (SNe Ia) and Asymptotic Giant Branch (AGB) stars.

Chemical space ($\mathcal{C}$-space)  is a space defined by the abundances of the chemical elements that we are able to measure.  Finding the debris from long-dispersed clusters using chemical tagging is an exercise in group finding in this multidimensional $\mathcal{C}$-space.  For stars which formed in clusters within the Milky Way (MW), chemical tagging appears relatively straightforward. The stars were born in chemically homogeneous clusters and the detailed abundance pattern changes from cluster to cluster.  The debris from a single star cluster will be tightly clustered in chemical space.   Stars can also come into the Galaxy in small accreted galaxies; their stars are again disrupted by the Galactic tidal field and dispersed.   If these small galaxies are like the dwarf spheroidal galaxies (dSph) and ultra-faint galaxies now around the Milky Way, they would not have been be chemically homogeneous \citep{tol09}. Their broad and different element abundance patterns are defined by their individual star formation histories (SFH) and are unlike those of the Galactic disk.  The debris of such accreted galaxies would not lie in a tight clump but in a streak through $\mathcal{C}$-space which reflects the element abundance pattern in the parent dwarf.  

To make a significant recovery of dispersed aggregates, simulations show that a large sample of stars, of order one million, is needed \citep{bla04}. The GALAH\footnote{GALactic Archaeology with {\it HERMES}} survey using the {\it HERMES}\footnote{High Efficiency and Resolution Multi-Element Spectrograph} instrument on the Anglo Australian Telescope is designed to do such a chemical tagging study.  It will measure abundances of about 25 elements using multi-object  high resolution (R = 28,000) spectroscopy of about a million stars.  Its  $\mathcal{C}$-space will have about 25 dimensions but the abundances of these 25 elements do not all vary independently.   The abundances of some elements are correlated because of the underlying nucleosynthetic processes, and the effective dimensionality of the $\mathcal{C}$-space will be less than 25.  Based on the existence of the various element groups (light, light odd-Z, $\alpha$, Fe-peak, light and heavy $s$-process and mostly $r$-process) represented in the {\it HERMES} spectra, and from various abundance patterns observed in field and cluster stars,  we think that the dimensionality of the $\mathcal{C}$-space will be about 7 to 9. The higher the dimensionality of $\mathcal{C}$-space, the more power chemical tagging will have to identify the debris of disrupted systems. We would like to determine the dimensionality of  $\mathcal{C}$-space more rigorously and that is one of the main purposes of this paper.  

It seems likely that the dimensionality of $\mathcal{C}$-space will depend on the metallicity\footnote{We assume that the iron abundance [Fe/H] is a pertinent tracer of the metallicity in this paper.} of the stars defining the $\mathcal{C}$-space.  For example,  stars of low metallicity may have formed within a brief period of time and the gas from which they formed may have been enriched by only a few supernovae events \citep{aud95}.  Among the neutron-capture ($n$-capture) elements, there would not have been time for AGB stars to provide enrichment of $s$-process elements, so the $r$-process may dominate the chemical evolution. On the other hand, the more metal-rich stars may have had a long history of chemical evolution with more than one process contributing to the $n$-capture element abundances.

We will use Principal Component Analysis (PCA) to determine the dimensionality of $\mathcal{C}$-space for various samples of stars, including field stars in different metallicity intervals, star clusters and stars in the Fornax dSph galaxy and the Large Magellanic Cloud (LMC). This is because the components will be different in the different situations.  We will attempt to interpret the principal components in terms related to nucleosynthesis processes and hopefully gain some insight into the dominant processes in each situation.

This paper is organized as follows: Section~\ref{sec:chemev} gives a brief discussion of some related aspects of chemical evolution  processes and sites.  In Section \ref{sec:data}, we give details on the samples that we chose to study. In Section~\ref{sec:math}, we will describe the PCA method and a way to estimate intrinsic correlation using Monte-Carlo simulations. In Section~\ref{sec:results}, we will present the results of PCA analysis and then in Section~\ref{sec:discussion}, we will interpret the results and discuss their origins. We summarize our main conclusions in Section~\ref{sec:conclusion}.

\section[]{Chemical evolution processes}\label{sec:chemev}
Stellar element abundance studies are usually presented in terms of  [X/Fe]--[Fe/H]\footnote{By definition, $\mbox{[X/Y]} \equiv \log_{10} (\mbox{N}_{X}/\mbox{N}_{Y})_{\mbox{star}} - \log_{10} (\mbox{N}_{X}/\mbox{N}_{Y})_{\odot}$, where $\mbox{N}_{X}$ and $\mbox{N}_{Y}$ are the abundances of element $X$ and element $Y$ respectively. } diagrams. These illustrate many basic element behaviour patterns, such as decreasing trend of [$\alpha$/Fe] toward higher [Fe/H] which is usually attributed to the increasing contribution of SNe Ia to the chemical evolution at later times.  These two-dimensional plots do not readily  reveal the interplay between different families of elements during the different stages of chemical evolution.   This interplay is of particular interest to us in trying to determine the dimensionality of  $\mathcal{C}$-space.  As chemical evolution proceeds, different groups of elements evolve together and define individual dimensions of the $\mathcal{C}$-space.  We seek an approach that is able to display these groups of elements, using PCA that we will present here.  This will hopefully benefit the analysis of elemental abundances that will be obtained soon for large samples of stars with high resolution multi-object surveys such as {\it APOGEE}\footnote{The Apache Point Observatory Galactic Evolution Experiment}\citep{all08}, {\it HERMES} and the {\it ESO-Gaia VLT} survey. First we briefly discuss major production sites for different families of elements.

$\alpha$-elements such as O, Ne, Mg, Si, S, and Ca are mainly produced by core-collapse supernovae (i.e. Type II, Ib and Ic Supernovae),  while Fe-peak elements such as Cr, Mn, Ni, and Fe are mainly produced by SNe Ia. Although the progenitor models are still debated, the onset delay for these SNe is in a range of $0.1$ -- $2$ Gyr after star formation begins \citep[see model from][]{kob09} .  Observations show complications such as over-abundances of heavy Fe-peak elements such as Zn and Co at low metallicity \citep*{mcw95,rya96,nor01,cay04,che04}.  These cannot be understood from core-collapse supernovae models with explosion energy of about $10^{51}$ erg. \citet{ume02,ume05} showed that more energetic supernovae (hypernovae) are responsible for the over-abundances of Zn and Co. This is supported by \citet{chi02}. \citet{heg10} argued that over-abundances of Zn can be explained without involving hypernovae. However, \citet{izu10} showed that it is necessary to include hypernovae in order to explain both Co and Zn observations.

Neutron-capture elements ($A \ga 65$) can provide independent probes for Galactic chemical evolution. There is not yet a consistent scenario to explain all observations of $n$-capture elements. Physically, the $n$-capture processes are divided into slow ($s-$) and rapid ($r-$) processes. In the $r$-process  the neutron flux is intense and the time-scale of $n$-capture is much shorter than that of the $\beta$-decay. This causes the seed nucleus to grab many neutrons before it $\beta$-decays to the valley of stability. The observed abundance patterns of $r$-process elements show very small star-to-star variation, and are in excellent agreement with the scaled solar $r$-process curve, at least for elements $56 < Z < 72$ \citep[e.g.][]{sne96,wes00,hil02,cow02,cow05}. This suggests the universality of the $r$-process nucleosynthesis. However, the site(s) of the $r$-process remain uncertain. The requirements on the physical conditions are neutron rich (i.e. low electron fraction $Y_e$), high entropy and short dynamic time-scales \citep{wan06}; but see caveat from \citet{fre99a}. A possible site is the neutrino-driven neutron-wind during the formation of neutron stars in core-collapse supernovae \citep*[e.g.][]{woo94}. Various scenarios of this possibility have been studied, including association with low-mass supernovae \citep[e.g.][]{wan09} or massive supernovae \citep[][]{tru02,wan06}, such as the $\nu$-driven He-shell mechanism \citep*[][]{ban11}. Another alternative site is neutron star mergers \citep*[][]{fre99b,ros99}. It has been argued that the stringent mass range might be responsible for the large scatter that has been observed for $n$-capture elements \citep{wan06}. The homogenizing effect of $n$-capture element evolution is less apparent than for the $\alpha$-elements that are produced by the whole mass range of core-collapse supernovae. 

In contrast, for the $s$-process, the neutron flux is not so intense and the $n$-capture rate is comparable to the $\beta$-decay rate. The major sites for the $s$-process are believed  to be low-mass ($1.5$ -- $3~\mbox{M}_\odot$) AGB stars \citep*[for a review, see][]{bus99,her05,kap11} with the input of the $^{13}$C pocket \citep{ibe82,hol88}. However, the progenitor mass range, metallicity dependency and the impact of rotation \citep{lan99,mey02} are still uncertain. In the $s$-process,  the major proposed neutron sources are from $^{13}\mbox{C}(\alpha,\mbox{n})^{16}\mbox{O}$ \citep[e.g.][]{kap90,sta07,cri09}.  More massive stars ($> 4~\mbox{M}_\odot$) achieve higher temperatures and another neutron source $^{22}\mbox{Ne} (\alpha, \mbox{n}) ^{25}\mbox{Mg}$ could be dominant \citep{ibe75,tru77}. Stars in the mass range of $8$ -- $12~\mbox{M}_\odot$ might evolve into super AGB phase \citep{sie06,sie07,sie10}. However, the $s$-process in super AGB phase is still yet to be understood. $S$-process also operates in massive stars that do not go through AGB/super AGB phase. This {\it weak} $s$-process can produce elements significantly up to Sr \citep*[e.g.][]{pra90,pig10}. Due to the long time-scales of low-mass AGB stars, it is believed that AGB stars will not contribute significantly below metallicity [Fe/H] $= -2$ \citep[e.g.][]{roe10,kob11c}. However, the contribution of AGB stars can appear even at low metallicity in the form of some classes of carbon enhanced metal-poor (CEMP) stars through binary mass transfer \citep{mcc84,joh02b,joh04,luc03,siv04,gos06,aok07,aok08,mas10}. The AGB star ejecta is transfered to the companion star and causes them to be enriched in carbon and $s$-process elements \citep[see models from][]{sta08,sta09,bis09}. In addition to $n$-capture element production from the $s$-process, low-mass stars also produce a small amount of Mg and O in the dredge-up \citep{mar01,kar03}, and might pollute the interstellar medium (ISM) with lighter elements due to the mass loss from the outer envelope \citep[e.g.][]{rei77,vas93}. However, these contributions to the ISM are marginal, especially at high metallicity \citep[e.g.][]{kob06,kar10}.

For the elements at  $A \simeq 90$ including Sr, Y and Zr, the origin is even more problematic \citep[e.g.][]{hon04b,aok05,cow05}. The observed abundance ratios of $n$-capture elements suggest the overproduction of these elements. \citet{tra04} showed that this overproduction cannot be explained by the weak $s$-process alone and named it as the Lighter Element Primary Process (LEPP). It has been shown that these elements could be produced by collapse of rotating massive stars, i.e. collapsars \citep*{pru04} or the {weak} $r$-process with slightly low $Y_e$ matter that is naturally expected in core-collapse supernovae \citep{izu09}. \citet{qia07,qia08} pointed out that these elements might be formed through charged particle reaction (CPR). More recently, \citet{boy11} suggested that this anomaly could be explained by truncated $r$-process for massive stars, in which the produced heavier $r$-process elements are consumed by the collapse of neutron star to a black hole.

Light elements (Li, C, N, O, Na) will be depleted or internally-mixed during their Red Giant Branch (RGB)/AGB phase either from the CNO cycle, the NeNa cycle \citep[e.g.][]{and07} or the hot bottom burning process \citep[e.g.][]{kar03}.  For instance, Li in giants will be depleted \citep{nor97,spi05,bon07,sbo10}, whereas N will be enhanced at the expense of C and O.  To avoid complication due to internal mixing and to better study the evolutionary state of the ISM when the stars were formed, we will leave the discussion of Li, C, N, O, and Na to a later paper.  

We will also study the $\mathcal{C}$-space properties for some dwarf galaxies, because dwarf galaxies are often proposed as building blocks for larger galaxies like the Milky Way \citep[for a general review on local group dwarf galaxies, see][]{mat98}. The chemical properties of the dwarf galaxies that are orbiting the Milky Way,  such as Sculptor, Fornax, Sagittarius, Sextans and the LMC\footnote{Some exclude LMC from the dwarf galaxies category \citep[e.g.][]{mat98}. However, for simplicity, we do not distinguish fainter satellite galaxies and dwarf galaxies in this paper} \citep*{she01,she03,gei05,sbo07} reveal that the present-day dwarfs are chemically too dissimilar to the Milky Way to be realistic building blocks \citep[but see][]{fre10b}.  However the ultra-faint dSph galaxies such as Coma Berenices, Hercules, Leo IV, Leo T, Ursa Major I and II have abundance patterns more like the pattern of the Galactic metal-poor halo and appear to be plausible building blocks for the halo \citep{kir08,koc08,fre10a}, although their baryonic masses are very small.  

We briefly summarize the difference in abundance patterns between the brighter dwarf galaxies and the Milky Way; for a more complete discussion, see  \citet*{ven04,ven08,tol09}.  Some dwarf galaxies show lower $[\alpha/\mbox{Fe}]$ at $-2 \la$ [Fe/H] $\la -1$ than the metal-poor stars in the solar neighbourhood, but reach the same metal-poor plateau ($[\alpha/\mbox{Fe}] = 0.3$ -- $0.5$) at [Fe/H] $\la -2$. This is often explained by the slower star formation histories (SFH) of the dwarf galaxies, so SNe Ia and AGB stars that have longer time-scales can contribute more to the chemical evolution at lower metallicity than they do for the nearby metal-poor stars. This produces a `knee' in the $[\alpha/\mbox{Fe}]$--[Fe/H] diagram but at lower [Fe/H]. However, \citet{kob09} argued that if this is due to the SNe Ia contribution, [Mn/Fe] should also show an increasing trend at the same [Fe/H], while [Mn/Fe] ratios in dwarf galaxies are as low as in the Galactic halo stars \citep{mcw03}. They claimed that the low $[\alpha/\mbox{Fe}]$ and low [Mn/Fe] abundances patterns in the dwarfs are more consistent with the lack of massive star contribution. Or perhaps both the slower SFH and the lack of massive stars play a role in explaining the chemical profile of the dwarf galaxies.

The slow SFH also coincides with the observed over-abundances of heavy $s$-process elements ({\it hs}) such as Ba and La relative to their Galactic counterparts at [Fe/H] $\ga-1.5$, since there is more time for the {\it hs} elements to be produced by the $s$-process. On the other hand, at low metallicity [Fe/H] $\la-2.5$, \citet{aok09} and \citet{taf10} found that dwarf galaxies do not show over-abundances of [{\it hs}/Fe] and tend to follow the Galactic halo trend. They argued that, at this metallicity, all of the $n$-capture elements are produced by the $r$-process. The light $s$-process elements ({\it ls}) such as Sr, Y and Zr do not however show over-abundances for dwarf galaxies at [Fe/H] $\ga-1.5$, and are sometimes under-abundant, similar to $[\alpha/\mbox{Fe}]$.  This is not unexpected. Although the $s$-process in Galactic low-mass AGB stars produces both {\it ls} and {\it hs} elements \citep{her05,kap11}, it has been proposed that the relative under-abundances of the {\it ls} elements could be explained by the metallicity-dependence of the $s$-process \citep{tol09,kap11}. Due to the {\it primary} nature (i.e. independent of metallicity) of the $^{13}$C pocket as the major neutron source, theoretical studies suggest that metal-poor AGB stars will preferentially produce heavier $s$-process elements due to the high neutron-to-seed ratio at low metallicity \citep{gal98,bus01,cri09,bis10}.

\section[]{Data selection}\label{sec:data}
Different mechanisms contribute to different metallicity intervals. For example, mechanisms that are associated with more massive progenitors will preferentially contribute at lower [Fe/H] due to the short evolution time, whereas the  $s$-process in low-mass AGB stars will only contribute at higher [Fe/H] due to their longer evolution time. The $\mathcal{C}$-space dimensionality and its interpretation may then be different in different metallicity ranges.  Therefore we separate our discussion according to two major metallicity intervals: $-3.5 \la$ [Fe/H] $\la -2$ and $-1 \la$ [Fe/H] $\la 0$.  These correspond roughly to the low metallicity halo and the high metallicity (thick + thin) disks of the Galaxy.  In some cases we will also study the intermediate interval of $-2.5\la$ [Fe/H] $\la -1$.  These metallicity ranges are just a rough guide and depend on the samples that we adopt from the literature. 

There are many observational studies of elemental abundances in the literature \citep*{edv93,mcw94,mcw98,nis97,nis10,han98,pro00,isr01,car02,joh02a,ste02,che02,che03,nis02,nis04,nis07a,nis07b,gra03a,gra03b,ben03,ben05,ake04,arn05,jon05,caf05b,asp05a,asp06,gar06,all06,pre06,ful07,lai07,lai08,ruc10,fuh11}, but the analyzing models, solar abundances and the sources of stellar parameters vary from author to author. This may cause systematic differences in elemental abundances, and produce spurious dimensions in the PCA analysis. Therefore, where possible we do not use compilations of abundances from multiple sources. Many surveys are relatively restricted in the number of elements measured or number of stars  ($<50$) observed.  Some surveys are strongly biased to specific classes of metal-poor stars, such as CEMP stars \citep[e.g.][]{coh06}, r-enhanced stars or related to planet hunting studies \citep[e.g.][]{sou08,nev09} and therefore not suitable for our study, because our goal is to measure the overall dimensionality of the $\mathcal{C}$-space. For each metallicity range, we focus on a few relatively unbiased, large samples of homogeneous data, and adopted their abundances directly from the original papers.  For the open and globular clusters,  we had no alternative but to use compilations.
\begin{figure*}
\begin{center}
\begin{minipage}{170mm}
\hspace{-0.9cm}
$\begin{array}{ccc}
\includegraphics[width=2.55in]{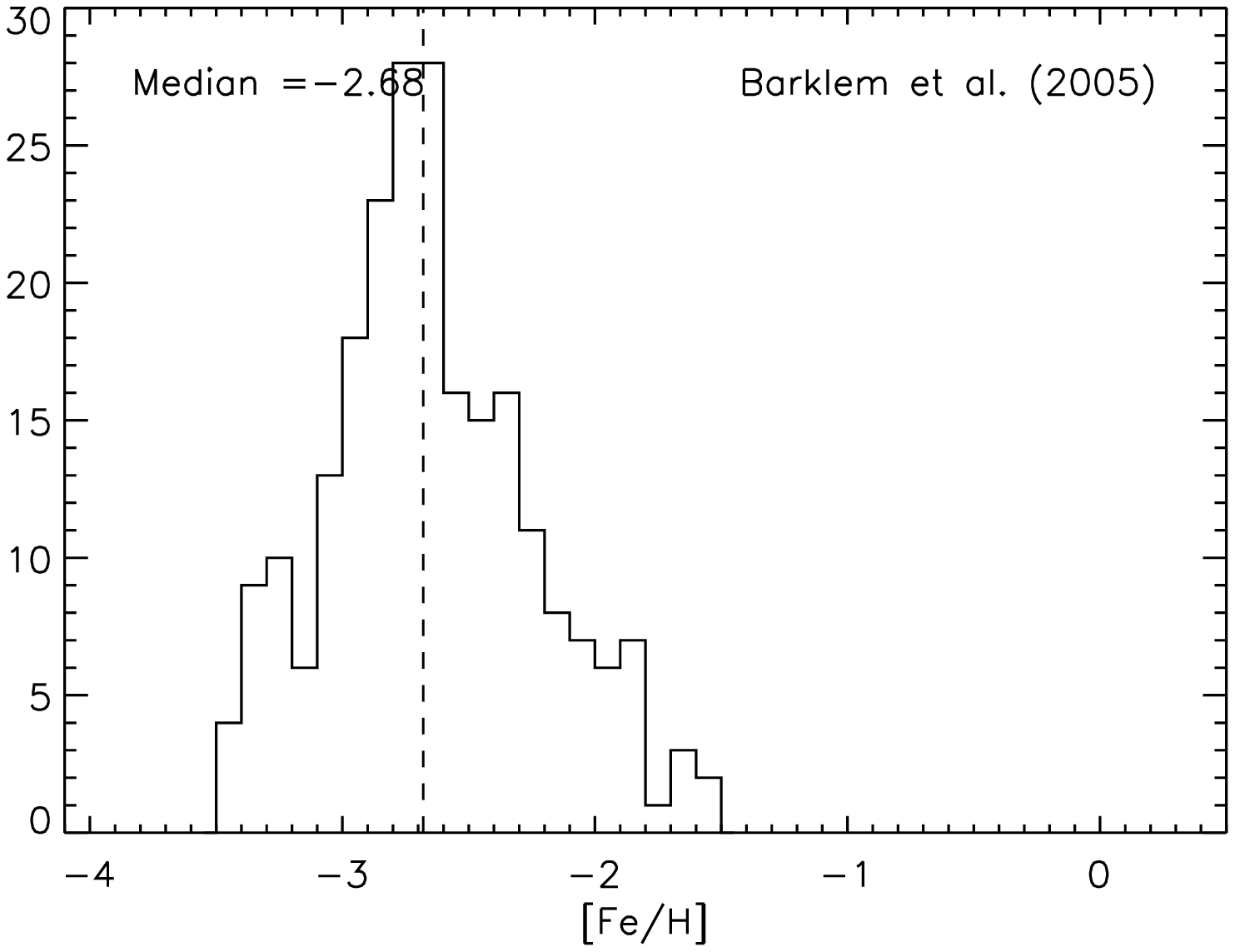} \hspace{-1.3cm} & 
\includegraphics[width=2.55in]{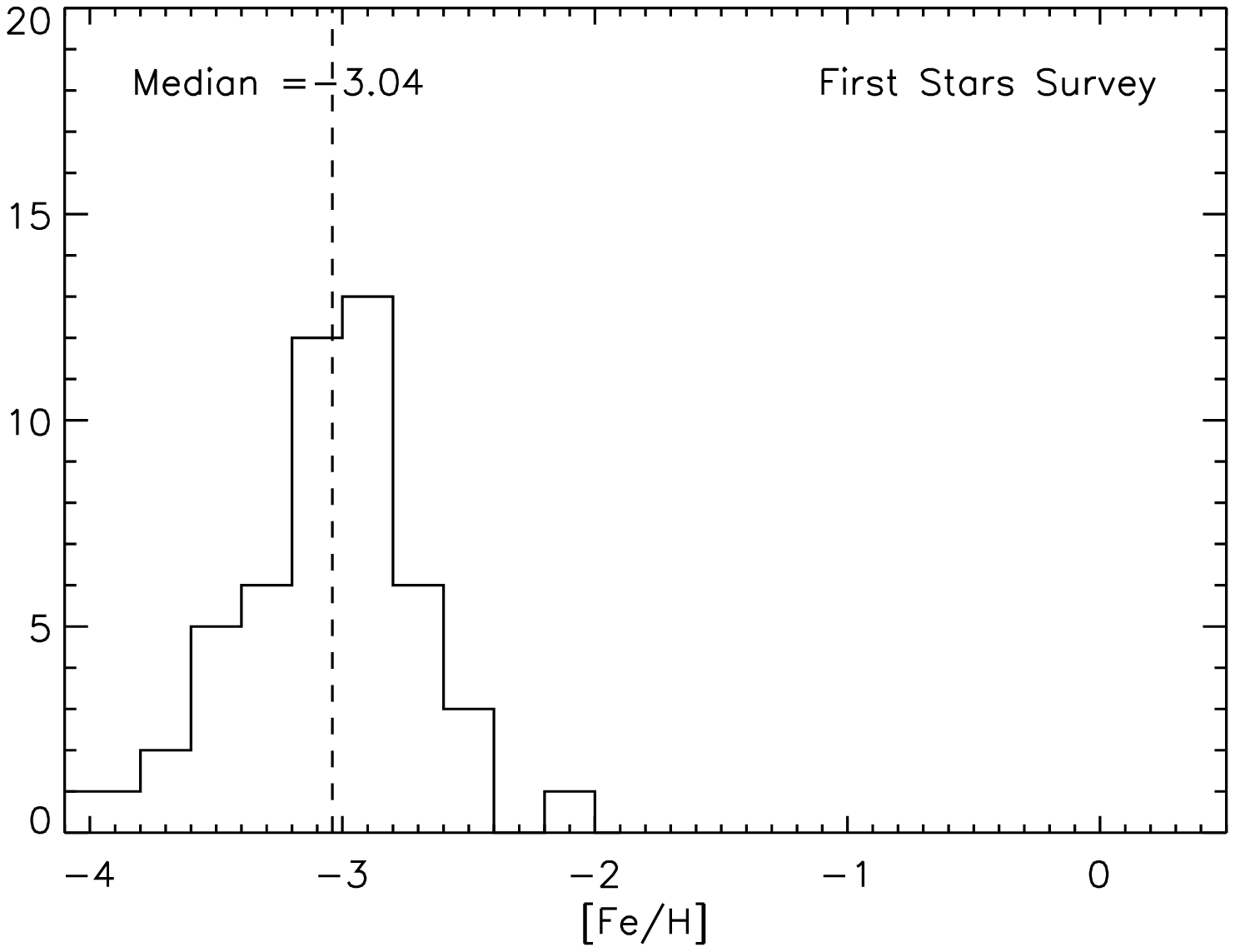} \hspace{-1.3cm} & 
\includegraphics[width=2.55in]{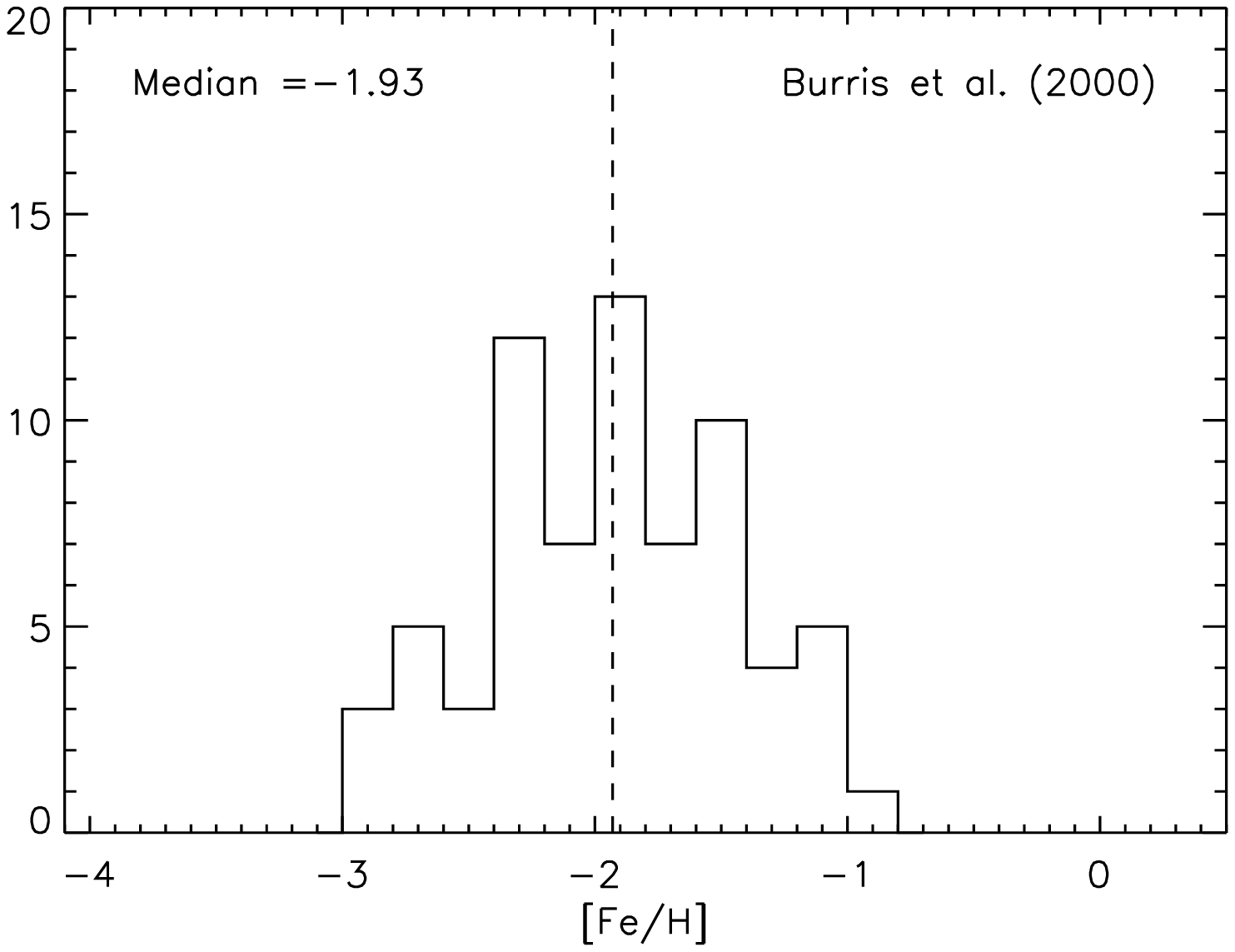} \vspace{-0.4cm}\\ 
\includegraphics[width=2.55in]{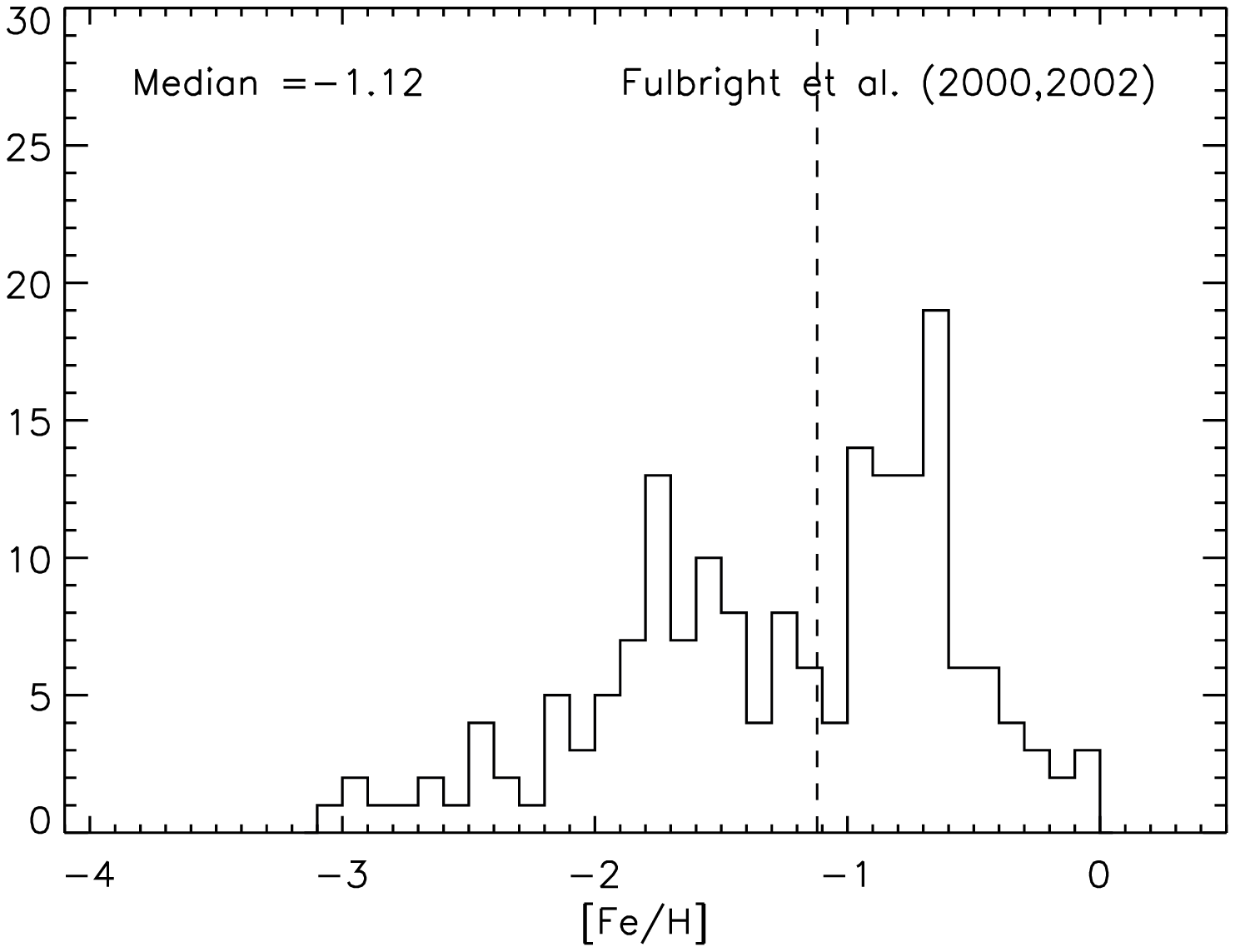} \hspace{-1.3cm} & 
\includegraphics[width=2.55in]{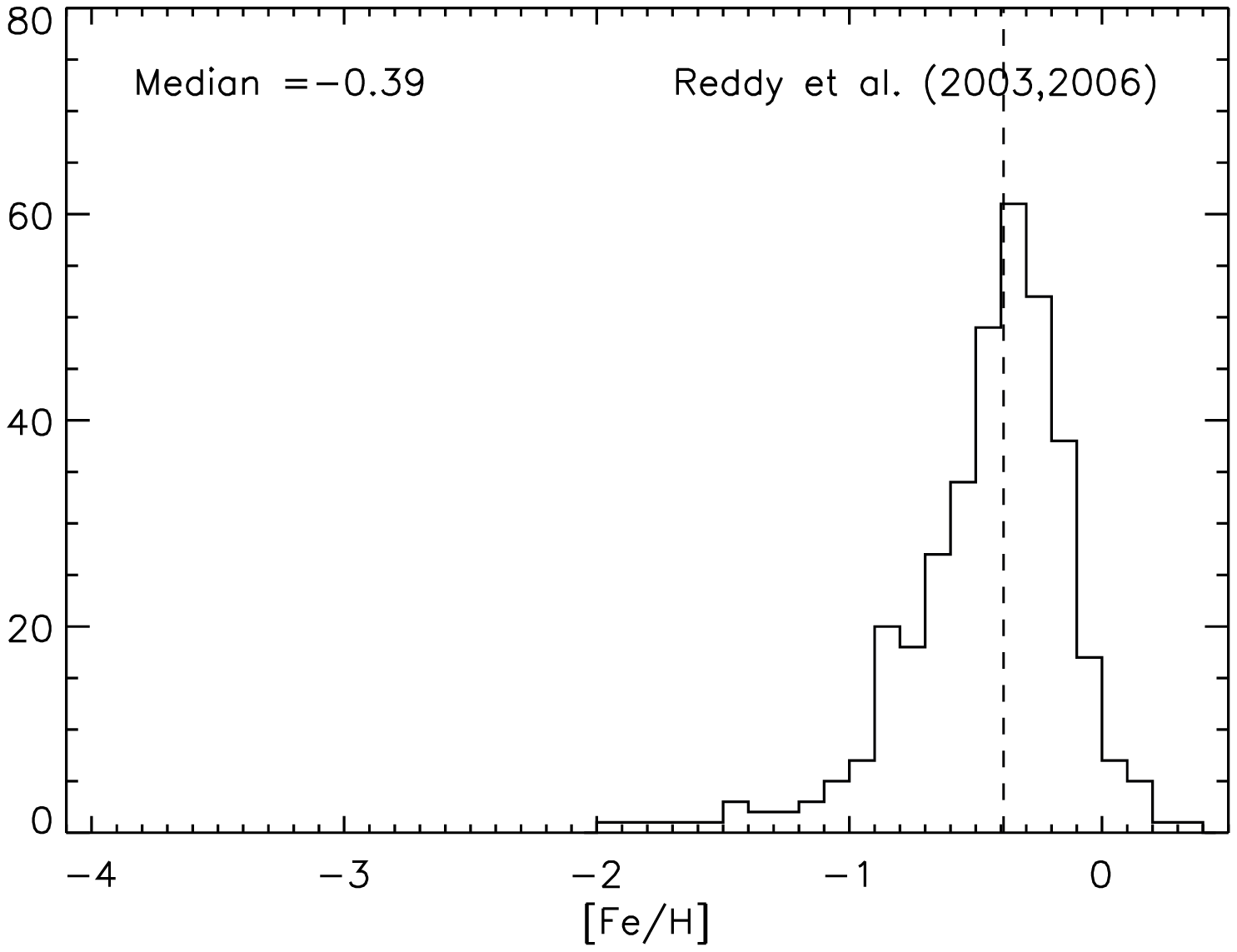} \hspace{-1.3cm} & 
\includegraphics[width=2.55in]{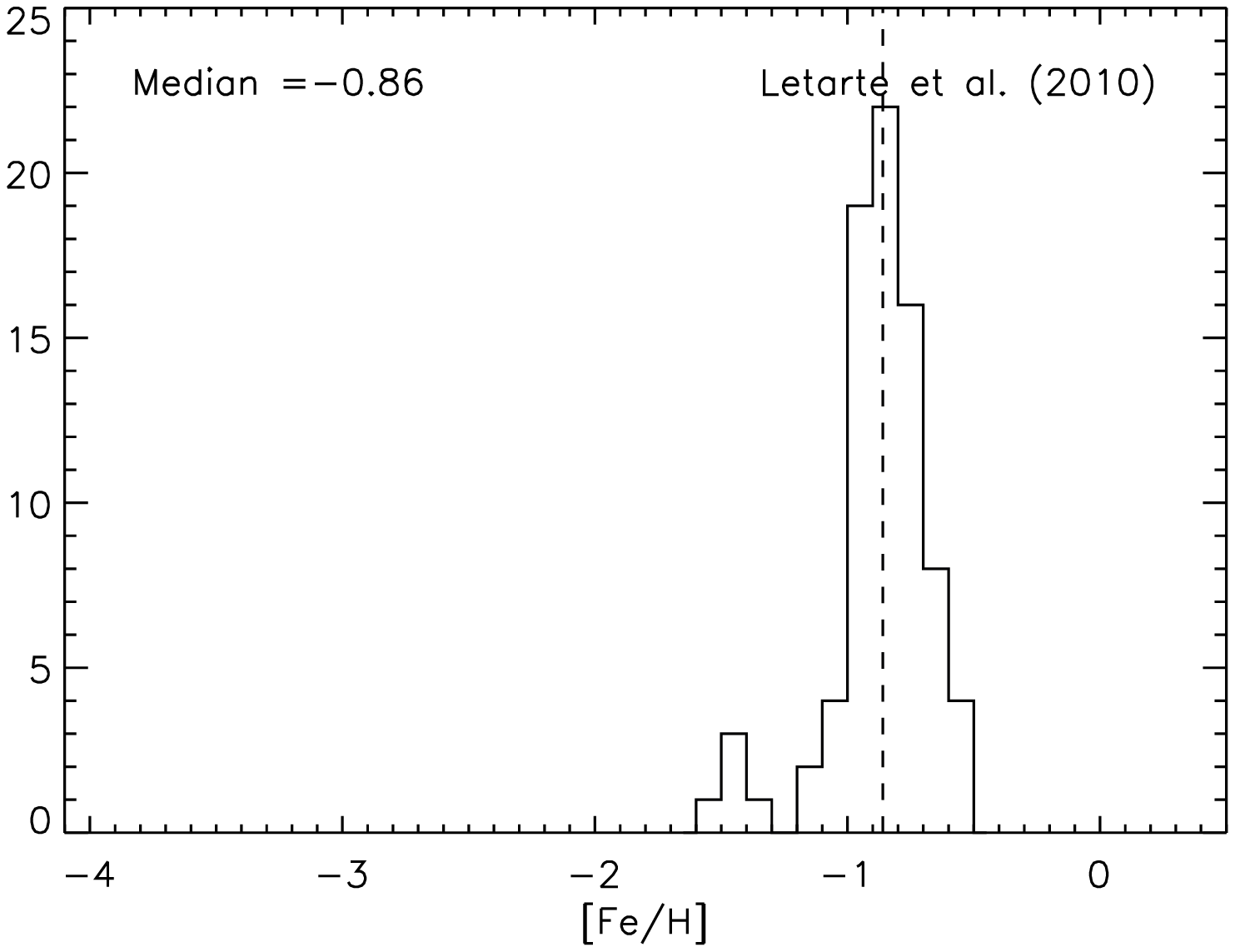} \vspace{-0.4cm}\\
\includegraphics[width=2.55in]{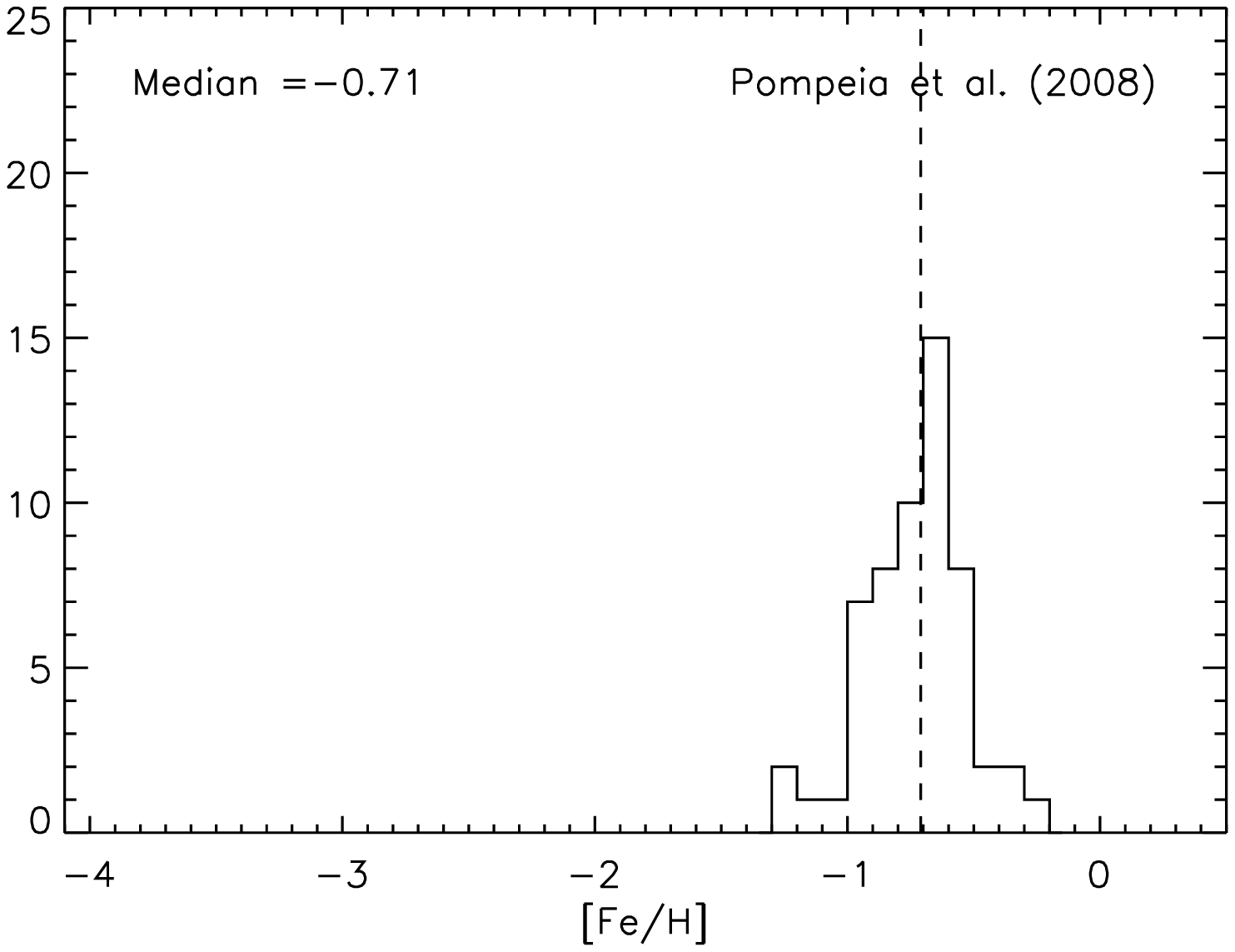} \hspace{-1.3cm} & 
\includegraphics[width=2.55in]{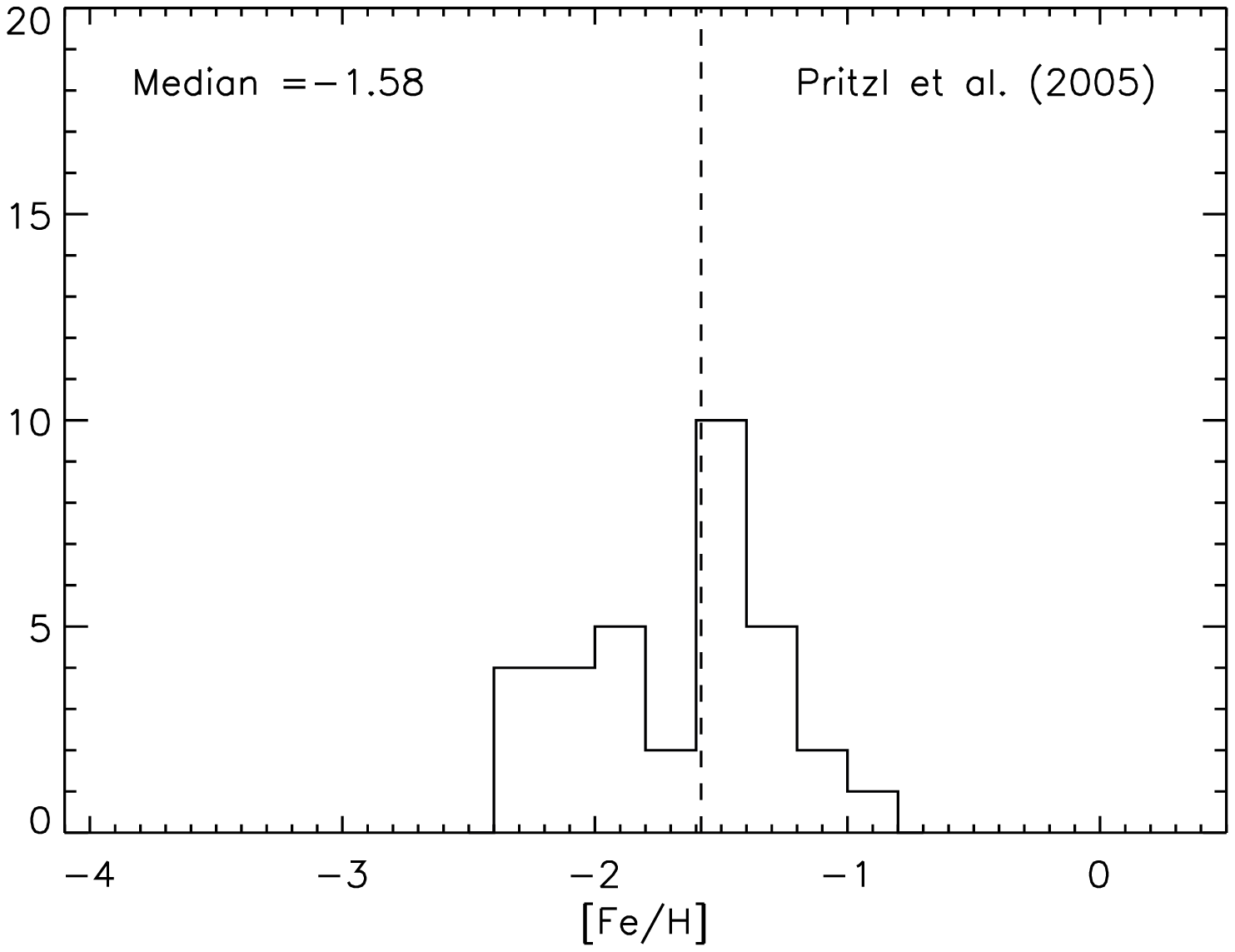} \hspace{-1.3cm} &
\includegraphics[width=2.55in]{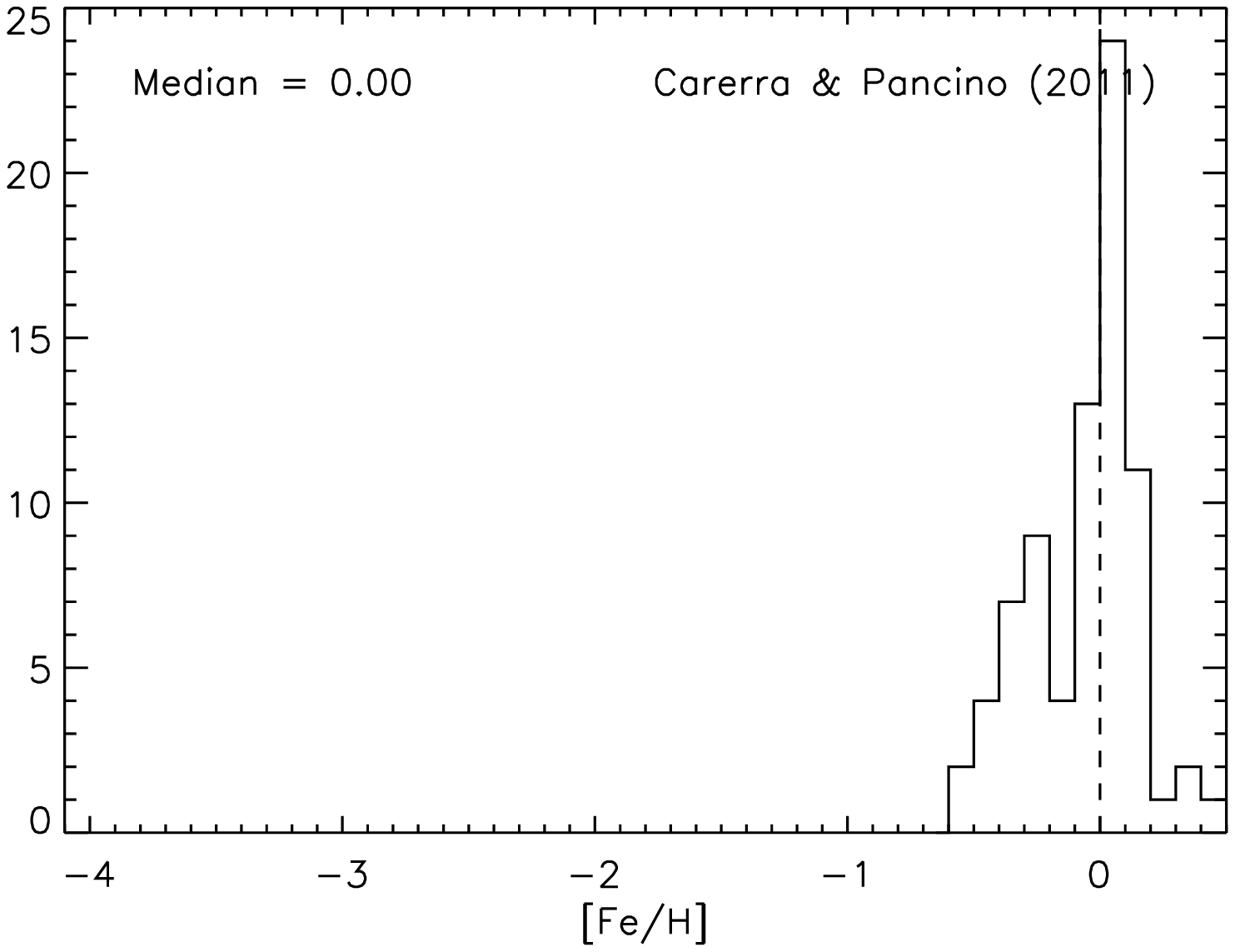} 
\end{array}$
\caption{This figure shows the metallicity distribution of each sample that we have adopted. For solar neighbourhood stars, we have {\it low metallicity} stars from \citet{bar05} and First Stars Survey; {\it intermediate metallicity} stars from \citet{bur00}, \citet{ful00,ful02}; and {\it high metallicity} stars from \citet{red03,red06}. We also have the {\it Fornax dSph galaxy} sample from \citet{let10}, the {\it LMC} sample from \citet{pom08}, {\it MW globular cluster} compilation from \citet{pri05} and {\it MW open cluster} compilation from \citet{car11}. The dashed lines show the median abundance of each sample. }\label{fig:metallicity}
\end{minipage}
\end{center}
\end{figure*}

\subsection[]{Low metallicity}
 For solar neighbourhood low metallicity halo stars in this study, we use the observational data from \citet{bar05} and the First Stars Survey \citep{cay04,fra07,bon09}.  From \citet{bar05}, we consider only stars in the range $-3.5 <$ [Fe/H] $< -1.5$. We exclude CEMP stars and blue stragglers as discussed below. In total, after culling, we have 231 stars from \citet{bar05} and 50 stars from First Stars Survey. 

Blue stragglers \citep{pre00} and {\it some classes} of CEMP stars are believed to have suffered binary mass transfer (cf. Section~\ref{sec:chemev}) and therefore might not reflect the abundances profile of the ISM from which they formed. We exclude known blue stragglers and CEMP stars with [C/Fe] $>1$: this criterion is from \citet{bee05}.  Having made this exclusion, the dimensionality of $\mathcal{C}$-space derived by our PCA method should be regarded as a lower limit, in case the carbon enhancement of some CEMP stars is not due to binary mass transfer \citep[e.g.][]{aok02,dep02}, but to some distinctive astrophysical origins such as {\it faint} supernovae \citep*{tsu03,ume03,karl06,kob11b} or massive rotating stars \citep{mey06}.

As the metallicity decreases, departures from Local Thermodynamic Equilibrium (LTE) are expected to become more pronounced \citep[for a review, see][]{asp05b}. Therefore Non-LTE (NLTE) calculation are needed \citep*[e.g.][]{bau98,gra99,mas01,kor03,tak03,and09,and10,and11,ber10a}. However, note that \citet{bar11} cautioned that estimating the inelastic hydrogen atom collisions with Drawin's formula in these studies might not be appropriate.  As a large homogeneous NLTE-corrected sample is not yet available, and combining heterogeneous samples will give spurious dimensions in the PCA analysis due to the systematic differences, we decided to use available homogeneous 1D-LTE abundances. All the chemical abundances from 1D-LTE models are taken directly from the original papers, with the exception of the element Al.
 
We include Al in our analysis because, unlike Na and O, Al abundances do not suffer from significant internal mixing in Galactic stars \citep{and08,bon09}. However, Al abundances are affected by NLTE effects at low metallicity. We adopted a $+0.6$ NLTE correction for [Al/Fe] as suggested by \citet{bau97} and \citet{coh04}. Our results would not alter if we were to exclude Al. Al is the only NLTE-corrected abundance in this study.  Our results should be reviewed once a large homogenized NLTE-corrected sample is available.

\subsection[]{Intermediate, high metallicity}
For solar neighbourhood intermediate metallicity stars, we chose to use the \citet{bur00} sample (70 stars) and \citet{ful00,ful02} sample (178 stars). For high metallicity disk stars, we use the \citet*{red03} and \citet*{red06} samples (357 stars). These samples are adopted without further modification beyond culling obvious outliers (e.g. [Al/Fe] $>1$, [V/Fe] $>0.8$, [Co/Fe] $>0.7$, [Nd/Fe] $>5$,  [Eu/Fe] $>5$).

\subsection[]{Dwarf galaxies}
To compare the state of chemical evolution in the Milky Way and fainter satellite galaxies, we study the Fornax dSph galaxy using the \cite{let10} sample (80 stars). We exclude the star B058 because it is a metallicity outlier ([Fe/H] $=-2.58$) for this sample ($-1.2 < \mbox{[Fe/H]} < -0.6$). We also studied the LMC, using the homogeneous sample (57 stars) from \citet{pom08}. For dwarf galaxies, to our knowledge, these are the only public available datasets with large ($>50$ stars) homogeneous samples and elemental abundances measured at high resolution.

\subsection[]{Globular and open clusters}
The halo/disk star samples are mostly in the solar neighbourhood with $7.5 \la r_G \la 8.5$ kpc, where $r_G$ is the distance from the Galactic centre. We would like to know whether stars over a larger Galactic volume would give us more inhomogeneity and therefore more independent dimensions in the $\mathcal{C}$-space. To probe wider regions, globular clusters \citep{sea78}, moving groups \citep[e.g.][]{siv07,bub10} and open clusters \citep*[e.g.][]{yon05,fri10,pan10,andr11,jac11} are suitable objects. 

No large homogeneous survey of globular clusters including a wide variety of elements is available to date \citep[but see][for a homogeneous survey with a restricted number of elements]{car09}.  We have to use the compilation of \citet*{pri05} from different authors. In this compilation, mean abundances for each globular cluster were derived. Since we do not study light-elements like C, N, O, Na that have been shown to have star-to-star dispersion within a cluster, other elements should have small star-to-star dispersion within a cluster and therefore it is justified to take mean abundances for most of the elements in this study. However, we are aware that recently \citet{roe11} have shown star-to-star dispersion in heavy $n$-capture elements (like La, Eu) and therefore taking mean abundances of each globular cluster will only give a lower limit of the $\mathcal{C}$-space dimensionality. We exclude objects identified with a dwarf spheroidal galaxy debris stream, such as Rup106, Pal12, Ter7 \citep[e.g.][]{caf05a}, and M68. Metallicity outliers are excluded by restricting [Fe/H] to the range $-2.5$ to $-1$. This leaves us with a total of 33 clusters.

We also study a recent open clusters compilation from \citet{car11} (private communication). We exclude those open clusters that have only [Fe/H] measurements. For clusters with multiple measurements, we take the mean abundance for each element.  The sample then contains a total of 78 clusters, with Galactocentric radii $6.4 \leq r_G \leq 20.8$ kpc and $-0.57 \leq \mbox{[Fe/H]} \leq 0.41$. For this compilation, we found that different model parameters lead to systematic differences of up to about $0.1$ dex.  Other sources of systematics such as differences in model atmospheres and methodology (e.g. equivalent width vs. spectrum synthesis), are difficult to quantify but are likely to be smaller. Thus we estimate that the systematic differences can be up to $0.2$ dex depending on the elements.

Since we will not perform PCA combining multiple samples other than globular clusters and open clusters, and the solar abundance differences for elements heavier than Na are usually small \citep*[e.g.][]{ande89,gre98,asp05c,asp09,lod09}, we do not homogenize the solar abundances adopted by different authors. The metallicity distribution of each sample is shown in Fig.~\ref{fig:metallicity} and we summarize our adopted samples in Table~\ref{tab:samples}.
\begin{table}
\begin{center}
\caption{Summary of adopted samples in this study.\label{tab:samples}}
\begin{tabular}{lll}
\hline
References & Categories & Count$^*$ \\
\hline
{\underline {\it MW Solar Neighbourhood}} & & \\
\cite{bar05} & Metal-poor halo stars& 231 \\
First Stars Survey$^\dagger$ & Metal-poor halo stars & 50 \\
\cite{red03,red06} & Metal-rich disk stars & 357\\
\cite{ful00,ful02} & In between & 178 \\
\cite{bur00} & In between & 70 \\
{\underline{\it Others}} & & \\
\cite{pri05} & MW globular clusters & 33$^{**}$\\
\cite{car11} & MW open clusters & 78$^{**}$ \\
\cite{let10} & Fornax dSph galaxy & 80 \\
\cite{pom08} & The LMC & 57 \\
\hline
\end{tabular}
\end{center}
$^*$ After restricting metallicity range; culling outliers, CEMP and blue stranglers\\
$^\dagger$ \citet{cay04,bon07,fra07}\\
$^{**}$ Number of clusters
\end{table}

\section[]{Analysis method}\label{sec:math}
\subsection[]{PCA}
We start with the $\mathcal{C}$-space defined by the set of element ratios [X$_i$/Fe]. We form the matrix of the correlation coefficients between all pairs of element ratios [X$_i$/Fe], [X$_j$/Fe] and diagonalize this matrix.  One can show (Appendix~\ref{appendix:PCA}) that the eigenvector corresponding to the largest eigenvalue is the direction where we have the largest variance in the {\it mean-shifted, normalized} data set and so on for the successively smaller eigenvalues. Furthermore, the variances along those directions are given by the corresponding eigenvalues. Note that the matrix is symmetric, therefore the eigenvectors are orthogonal.  These eigenvectors are the {\it principal components} and we can, {\it to some extent}, interpret these eigenvectors in terms of nucleosynthetic processes.  We illustrate the procedure with toy models.

\subsubsection[]{Toy models}
\begin{figure*}
\centering
$\begin{array}{c}\hspace{-0.5cm}
\includegraphics[width=0.35\textwidth]{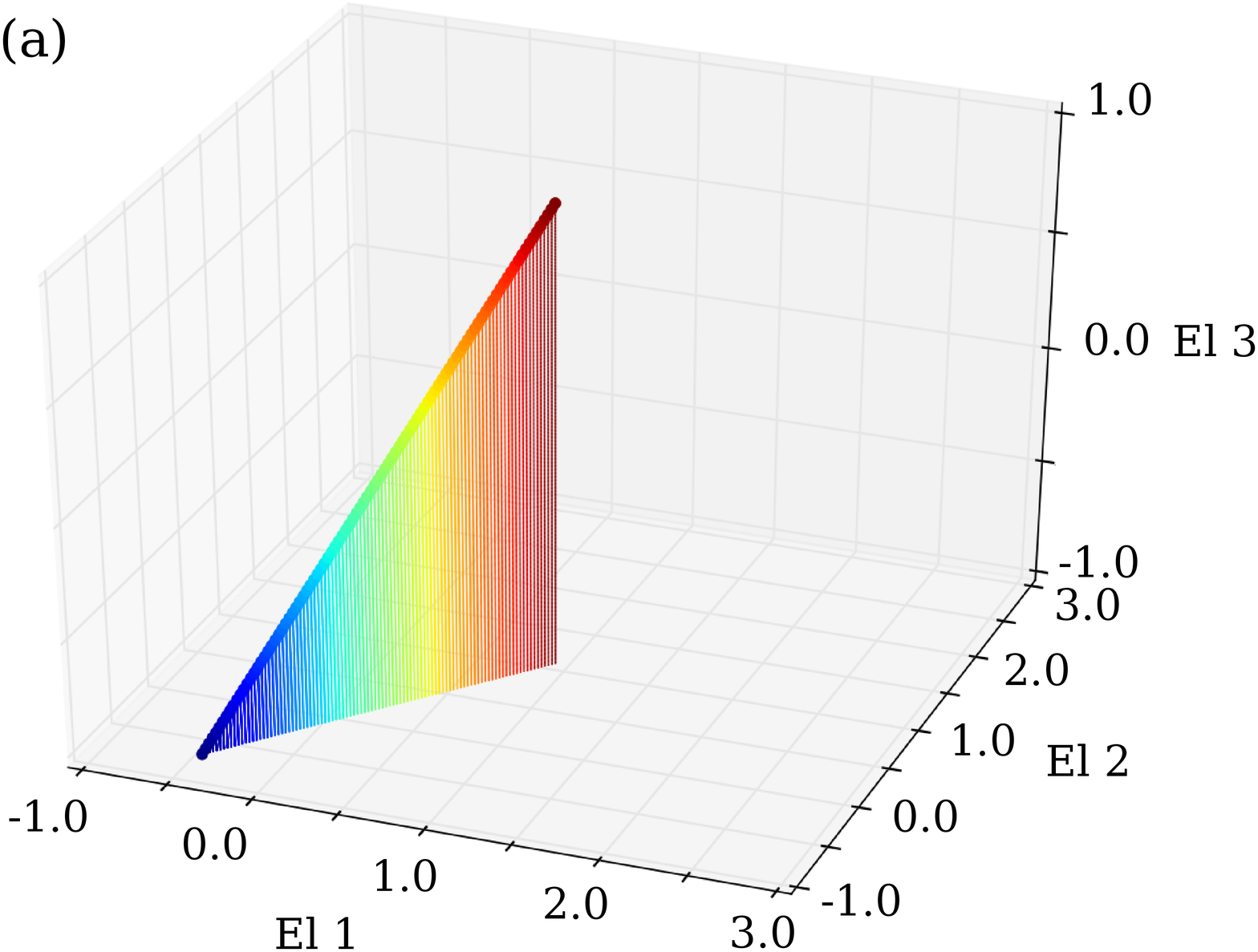} 
\includegraphics[width=0.35\textwidth]{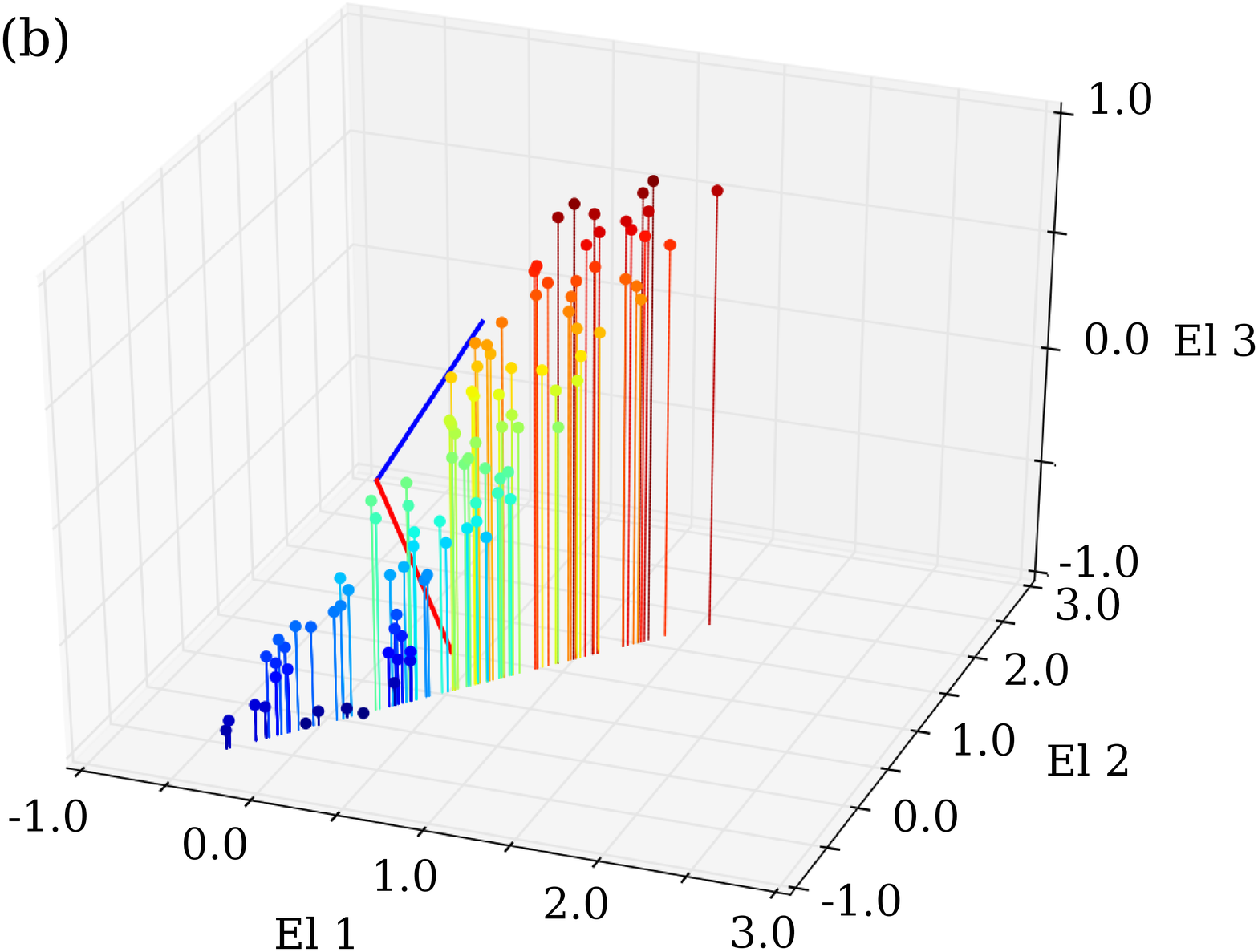} 
\includegraphics[width=0.35\textwidth]{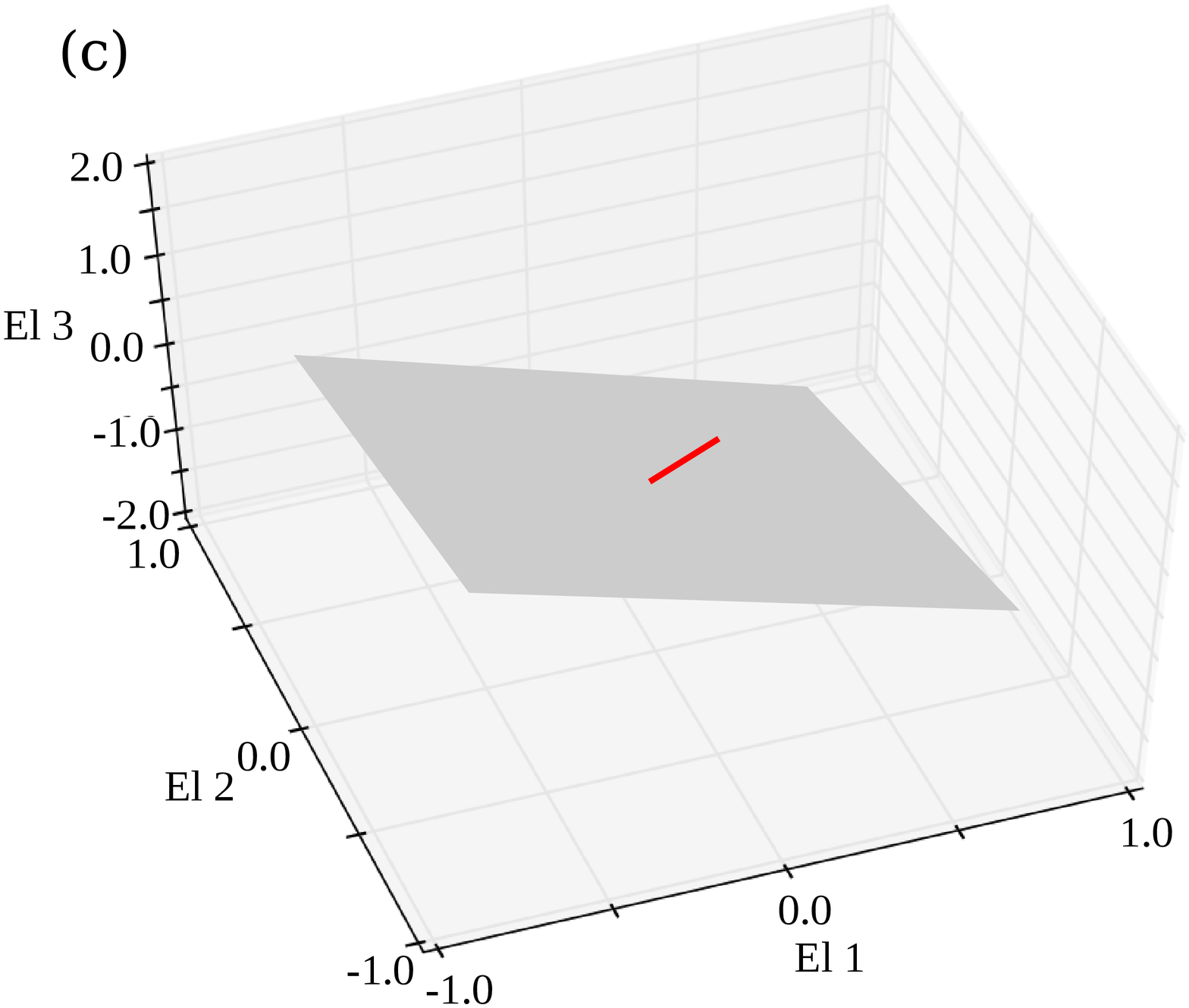}  \\
\includegraphics[width=0.35\textwidth]{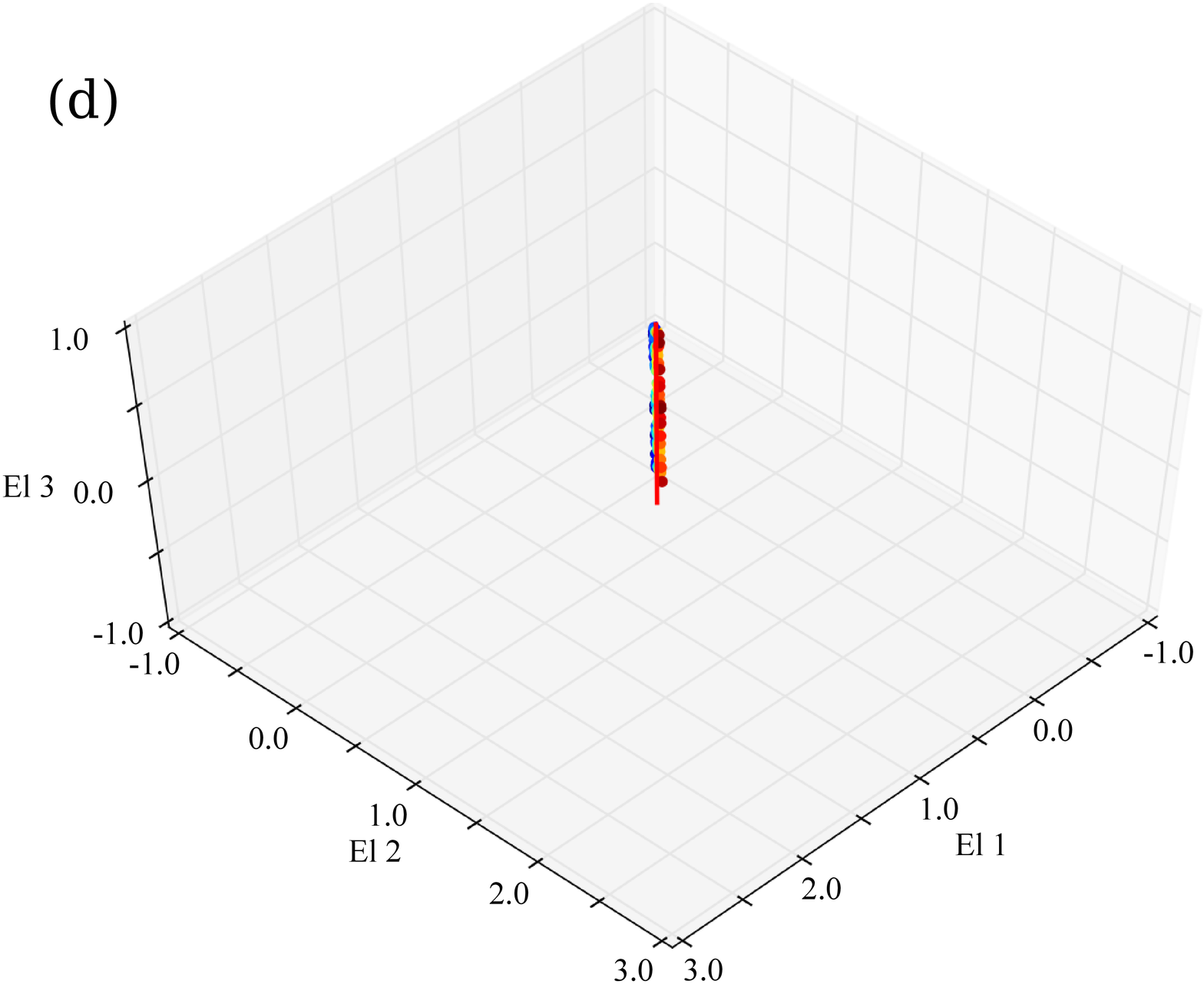} 
 \includegraphics[width=0.35\textwidth]{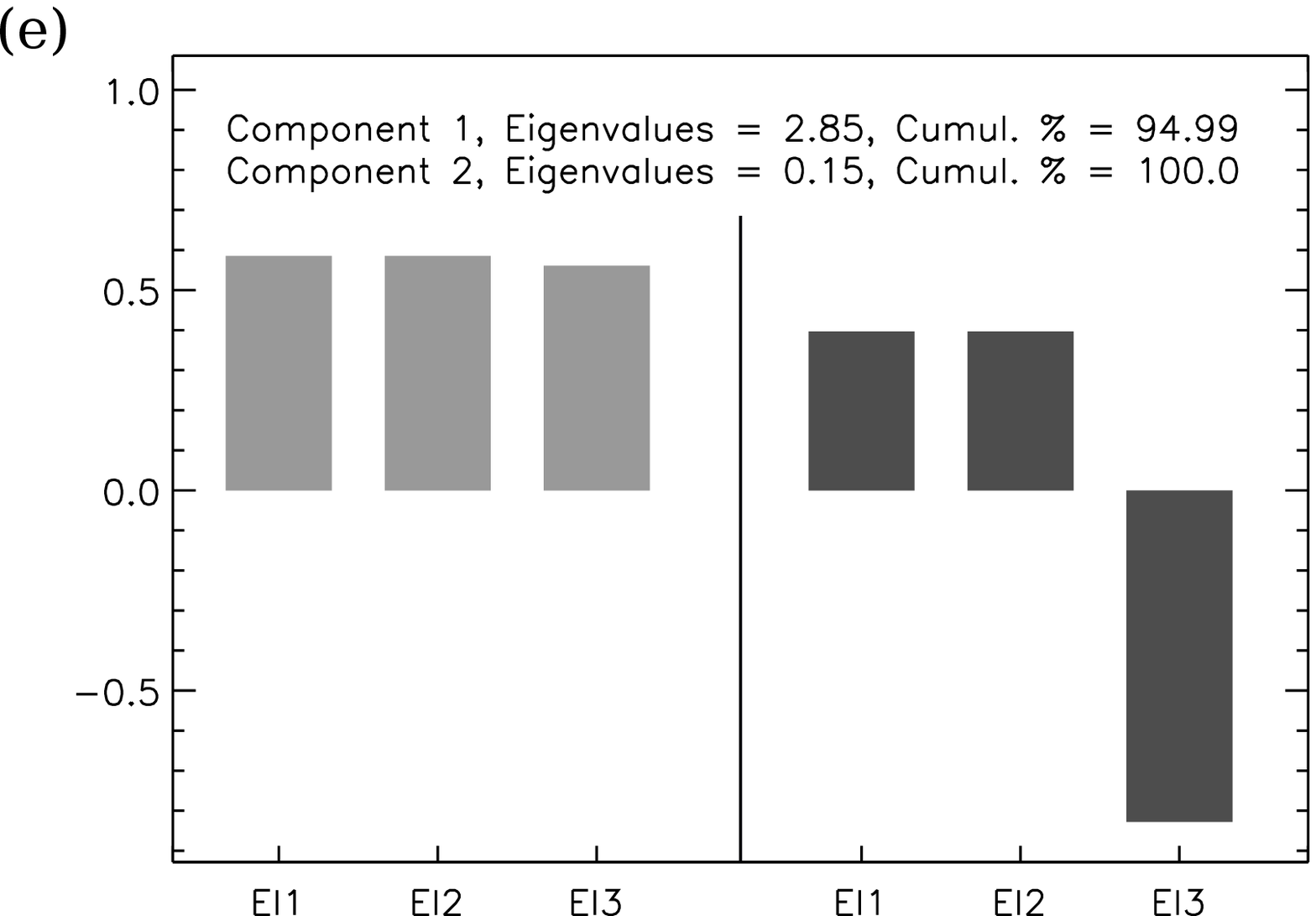}  
\end{array}$
\caption{This is a model ($\gamma =1$) to illustrate how PCA works. The vertical lines in panel (a) and (b) are visual guides. Panel (a) shows the scatter plot of the chemical abundances where only mechanism 1 is working (see text). Panel (b) is the same as (a) but both mechanisms 1 \& 2 are contributing to the chemical abundances. Panel (b) also shows the first principal component in {\it blue solid line} and the second principal component in {\it red solid line}. Panel (c) shows the the hyperplane of the first principal component in {\it grey} and the second principal component in {\it red solid line}. Panel (d) shows the data points projected onto the hyperplane of the first principal component and the {\it red solid line} is the second principal component. Panel (e) shows the composition of the first two normalized eigenvectors. Colour version of all figures in this paper are available in the online version of the journal.} 
\label{fig:illustrate-pca1} 
\end{figure*}

\begin{figure*}
\centering
$\begin{array}{c}\hspace{-0.5cm}
\includegraphics[width=0.35\textwidth]{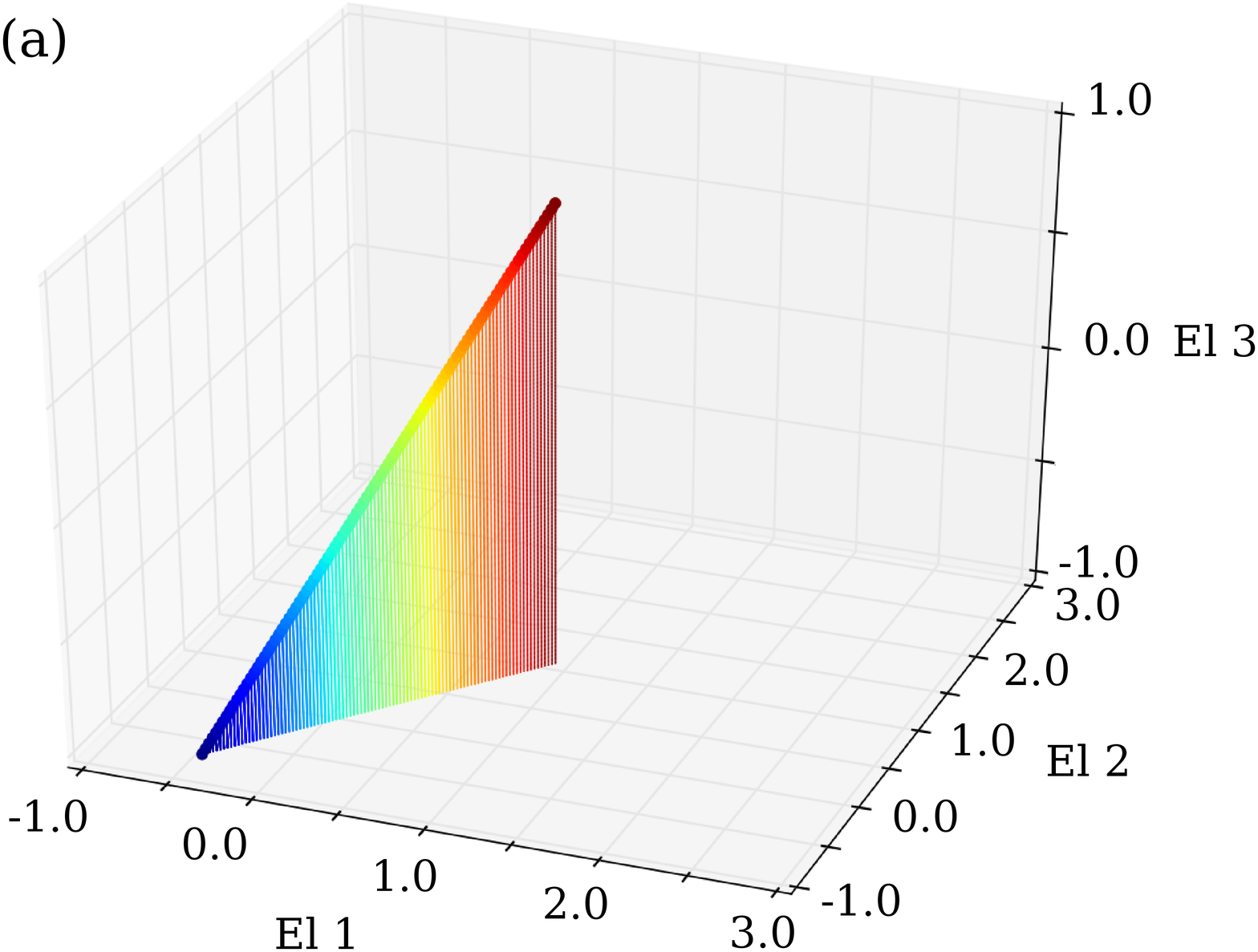} 
\includegraphics[width=0.35\textwidth]{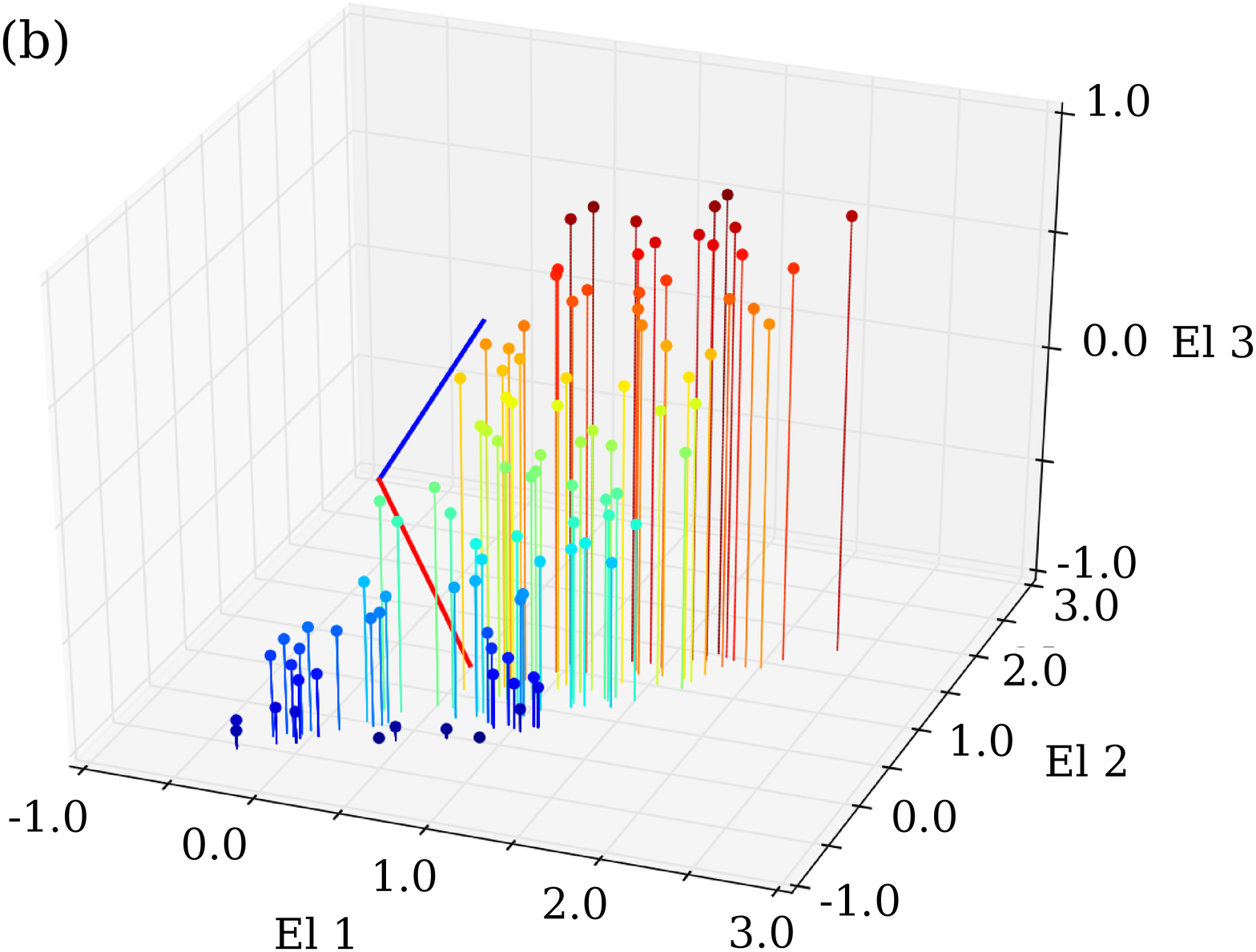} 
\includegraphics[width=0.35\textwidth]{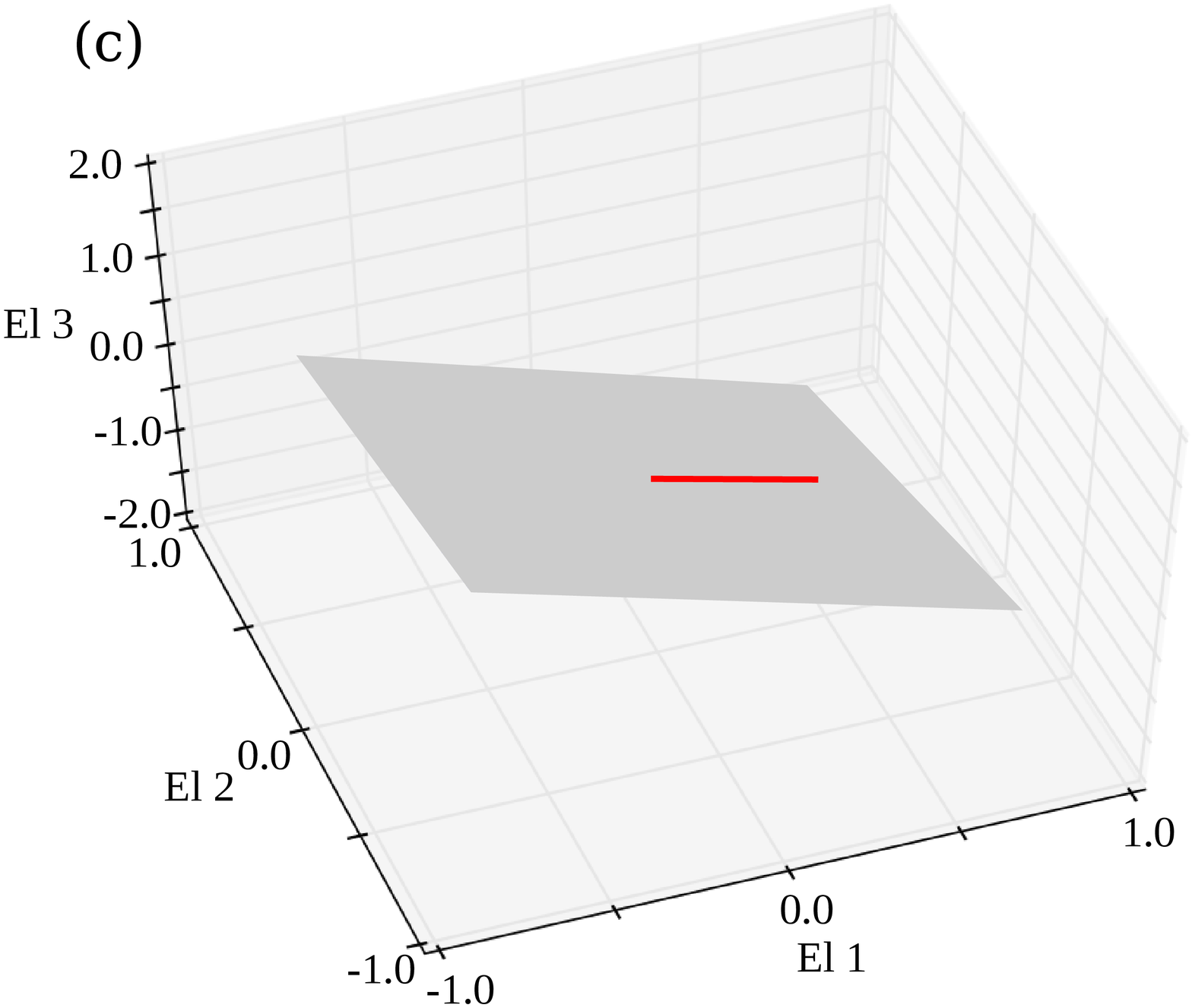}  \\
\includegraphics[width=0.35\textwidth]{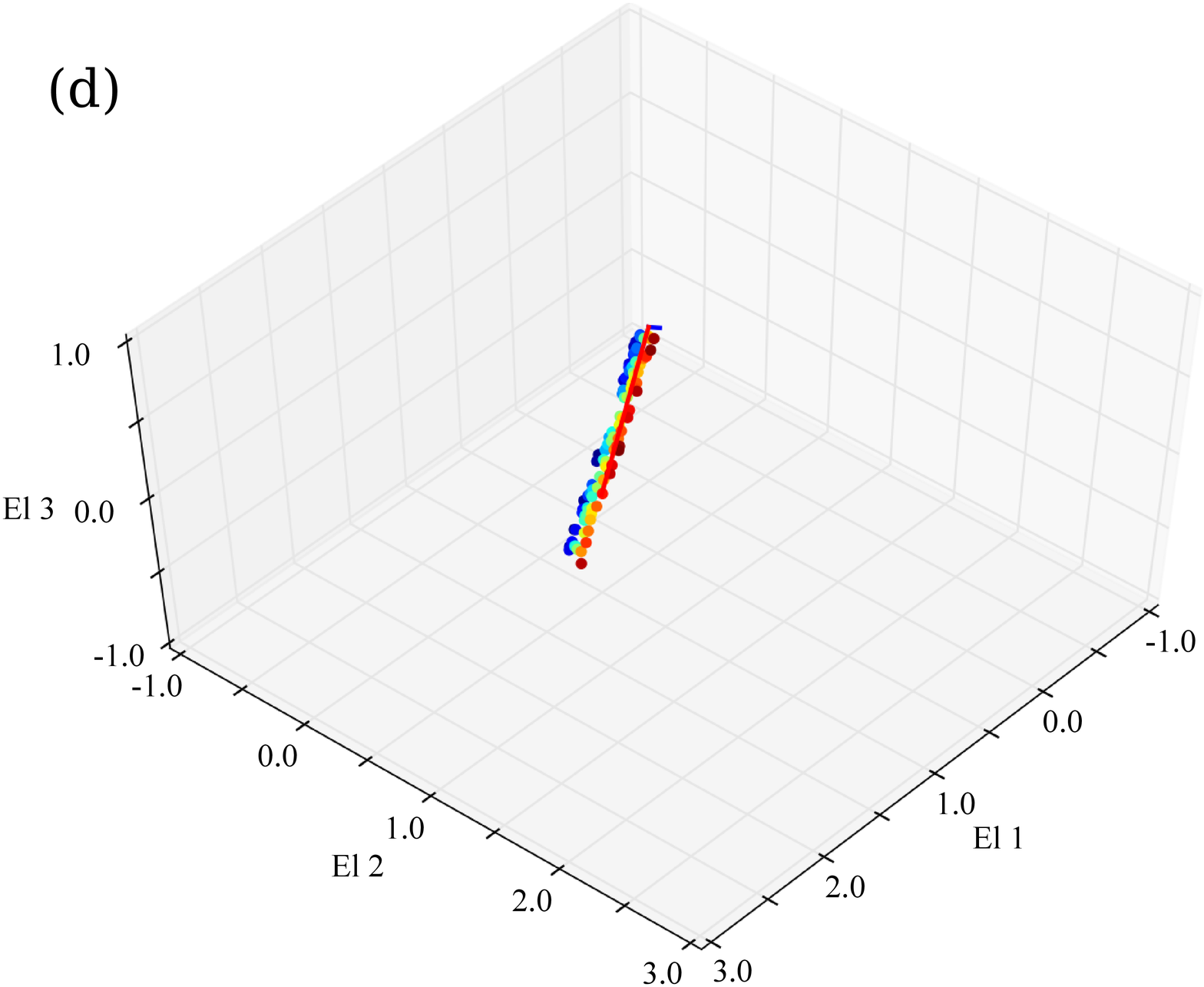} 
 \includegraphics[width=0.35\textwidth]{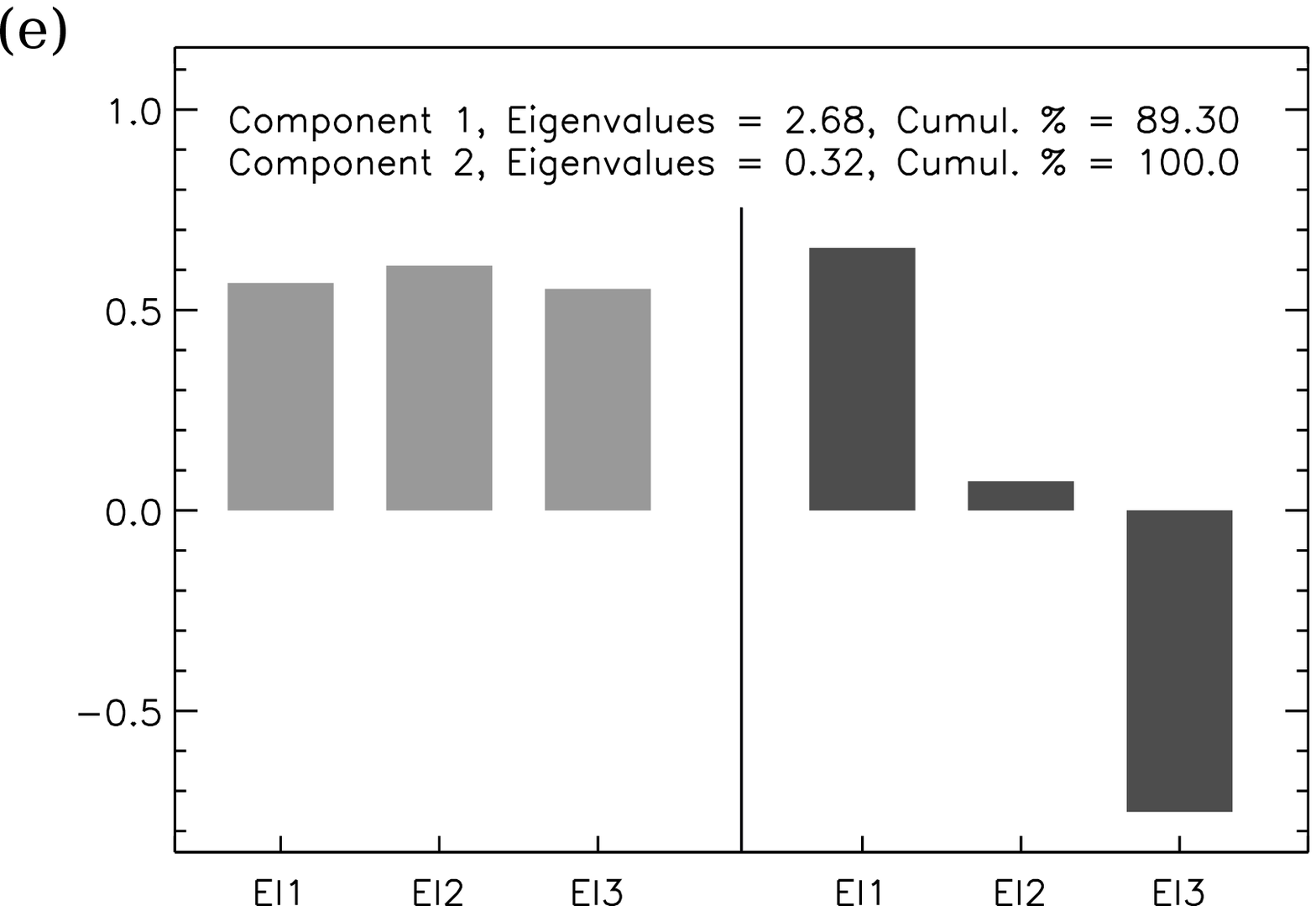}  
\end{array}$
\caption{Same as Fig.~\ref{fig:illustrate-pca1}, but with $\gamma =2$.}
\label{fig:illustrate-pca2}
\end{figure*}

Take a 3-dimensional space defined by three elements which we denote as El1, El2 and El3, and assume that there are two element-producing mechanisms. The first produces exactly the same amount of El1, El2 and El3 and the second mechanism only produces El1 and El2 with production ratio El1/El2 $=\gamma$.  Figs.~\ref{fig:illustrate-pca1} and \ref{fig:illustrate-pca2} illustrate the cases $\gamma = 1$ and $\gamma =2$. For each figure, panel (a) shows the scenario where only mechanism 1 is working: all elements are produced in exactly the same ratio. Panel (b) shows the scatter plot with both mechanism 1 and mechanism 2 contributing to the element abundances: e.g. in panel (b), the points have now been randomly translated in the direction (1,1,0) by the action of the second mechanism.

The first principal component of the PCA analysis, shown in panel (b) as the  {\it blue solid line}, represents the direction that has largest variance in the {\it normalized, mean shifted} $\mathcal{C}$-space. After determining the first component (with the largest eigenvalue), the PCA machinery then projects orthogonally all data points on to the hyperplane normal to the first principal component, as shown in panel (c) and panel (d). 

The second principal component is then the direction in the hyperplane that shows the largest variance among the projected points.  Panel (d) shows the data points projected on the hyperplane of the first principal component. The second principal component is shown as the {\it red solid line}. The composition of the normalized eigenvectors in the 3-dimensional $\mathcal{C}$-space are shown in the bar chart (e), with the first eigenvector as the three bars on the left and the second eigenvector as the three bars to the right.  The first eigenvector has approximately equal components in each element and represents the contribution of mechanism 1.  Mechanism 1 has equal components, but the eigenvector shows small departures from equality which come from projections of random amounts of mechanism 2 on to this eigenvector.

The first eigenvector has a clear interpretation: in the bar chart (panel e), the contributions from mechanism 1 have similar components for all elements, as expected.    The second principal component represents the contribution of mechanism 2.  In the bar chart,  El1 and El2 have the same sign but El3 has the opposite sign. It is important to note that, although the second mechanism does not produce El3, the contribution of El3 to the second principal component is not zero. This is intuitively clear, because the second component must be orthogonal to the first and must lie in the hyperplane shown in panel (c). The direction where we have both positive El1 and El2 in the tilted hyperplane is pointing in the negative direction of El3. This illustrates an issue with interpreting the eigenvectors in terms of nucleosynthetic processes in more realistic systems.  Although both principal components are well defined, the first eigenvector has a clear interpretation, but the interpretation of the second and later eigenvectors is less straightforward.

Comparing the second eigenvector for  $\gamma =1$ and  $\gamma =2$,  El1 and El2 show similar contributions for $\gamma =1$, but for  $\gamma =2$ the contribution from El1 is larger.   This is not surprising; as we have seen in panel (d), the second eigenvector on the hyperplane shows a ratio of El1 to El2 $>1$ when $\gamma =2$.  As before, although both principal components are well defined by the PCA,  the first eigenvector has a clear interpretation but it is less obvious for the second eigenvector.  When one component is in the positive diagonal direction (i.e. all elements contribute with the same sign) and the second eigenvector components show opposite signs of different families (here El1 \& El2 vs El3), this suggests that there is contribution from a first mechanism that produces both families, and contribution from a second mechanism that preferentially produces El1 \& El2 or preferentially produces El3 but we cannot tell which without further diagnostics.  Section~\ref{subsec:results-low-metallicity-stars} gives an example of a situation in which further diagnostics can illuminate the interpretation of second and later eigenvectors.

\subsubsection[]{Dealing with incomplete data sets} 
Given an incomplete data set (i.e. some of the element abundances are not available for all stars), in principle we can still calculate Pearson's correlation for any two chemical abundances by using only the data points that have both chemical abundances, and therefore we can construct the correlation matrix entry by entry. However in this case the correlation matrix is clearly not a Gram matrix\footnote{A $n \times n$ matrix is Gramian if and only if for every $i, j \in \{ 1,2, \ldots, n\}$, the ($i$,$j$)-th entry is given by the inner product $\langle \bmath{v_i}, \bmath{v_j} \rangle$ of the same set of vectors $\bmath{v_1}, \bmath{v_2}, \ldots, \bmath{v_n}$} and therefore might not necessarily be semi-positive-definite and we might have unphysical negative eigenvalues.

This is a well-known problem in finance analysis. We use the algorithm suggested by \citet{reb99}. The idea is to search for a closest semi-positive-definite matrix $\mathbfss{C'}$ that resembles the correlation matrix (cf. Appendix~\ref{appendix:incomplete-data-set})\footnote{Another method to deal with missing data is by guessing the missing data from the best fit model, where the best fit model is deduced from the rest of the data. However in this case, unavoidably, we will strengthen our correlation unnecessarily. This illustrates a problem with most of the manifold learning methods such as Locally Linear Embedding (LLE), ISOMAP, Diffusion Map etc. These methods require a complete data set. Therefore there is no alternative to guessing the missing data. We argue that the method that we introduce here is a more democratic way to deal with missing data without imposing any prior.}. We can measure the differences between these two matrices using the canonical matrix norm or the quadratic sum of the differences of the eigenvalues (cf. Appendix~\ref{appendix:incomplete-data-set}). Throughout our study, the differences of either norm over the number of entries gave value $\ll 1$ and therefore the results are robust.

After {\it orthogonally} diagonalizing the matrix $\mathbfss{C'}$, we rank the eigenvalues in decreasing order and calculate their cumulative percentages. The cumulative percentages of ranked-eigenvalues represents the accounted total variances of the data cloud. 

\subsubsection[]{Best cut-off for ranked-eigenvalues cumulative percentages}\label{sec:best-cut-off-4-PCA}
Given that the measurements are not perfect, what percentage of the variance of the data cloud is due to measurement uncertainty? For example, Section~\ref{subsec:results-low-metallicity-stars} shows the cumulative percentages for the ranked-eigenvalues for the $n$-capture elements at low metallicity abundance. The first component provides about 80\% of the variance. The next two provide almost all of the rest of the variance. Our goal is to find the dimensionality of the $\mathcal{C}$-space. How many of the principal components should we accept as real? We attempted to answer this question by performing Monte Carlo simulations, in which we created mock data sets lying in an $n$-dimensional spaces and tracing a non-tilted $m$-dimensional flat manifold\footnote{A $m$-dimensional flat manifold in $n$-dimensional space is the generalization of a $2$-dimensional plane in $3$-dimensional space. In this study, we made two assumptions: (1) If correlated, [$X_i$/Fe] are correlated linearly; (2) The errors of [$X_i$/Fe] are Gaussian-distributed. The first assumption is reasonable because [$X_i$/Fe] are abundances in log scale-- the assumption holds if $N_{X_i} \propto N^m_{X_j}$, for all $m \in \mathbb{R}$, where $N_{X}$ is the abundance of element $X$. Furthermore, our study samples are always restricted to a small metallicity abundance range. To justify the second assumption, we compared the [Fe/H] of stars in common of RAVE Survey \citep{ste06} and Geneva-Copenhagen Survey \citep{nor04}. We found that the differences can be approximated by a Gaussian distribution. This suggests that the abundance errors in log scale could be Gaussian-distributed.}. The mock data set has a spread of $\Delta = 0.3$ to $0.7$ dex mimicking the real situation with element abundance distributions. We performed the simulation by rotating the manifold with a random orthogonal matrix. For each simulation, we added $0.1$ dex of measurement uncertainty and then performed PCA on the noisy data. We recorded the cumulative percentages corresponding to the $m$-th principal component. In this way, we tried to estimate the cumulative percentage at which we should stop accepting principal components as real, in order to deduce the correct $m$-dimensions of the manifold. 

For each set of values of $n$ and $m$, we performed 10000 simulations. We varied $n$, $m$ and $\Delta$ in the range $n \in [7,15]$, $m \in [2,5]$, $\Delta \in [0.3,0.7]$. The simulation showed that a reasonable cutoff for identifying real principal components in the PCA analysis is about $85\%$. 

In summary, assuming a cosmic spread of $\Delta = 0.3$--$0.7$, measurement uncertainty of $0.1$ dex and a flat manifold (which can be tilted), the rank of the eigenvalue corresponding to the cumulative percentage of about $85\%$ gives a robust estimate of the independent dimensions of the chemical space. The remaining variance of the data cloud comes from measurement uncertainty. If the measurement uncertainty is less than $0.1$ dex, we should take the cut off larger than $85\%$ and vice versa. In this study we will perform PCA analysis on the random variables [X/Fe]\footnote{We chose to work in [X/Fe] space because all elements are highly correlated in [X/H], i.e. if we were to perform PCA on [X/H], the dominant dimension will consume more than 80\% of the variance for all cases.}, where X can include Al, Sc (light odd-Z elements); Mg, Si, Ca, Ti ($\alpha$-elements), V, Mn, Cr, Co, Ni, Zn (Fe-peak elements) and Y, Zr, Ba, La, Nd, Eu ($n$-capture elements). Among $n$-capture elements, we consider Y and Zr as light $s$-process ({\it ls}) elements, Ba and La as heavy $s$-process ({\it hs}) elements and Nd, Eu as mostly $r$-process elements \citep[refer to][]{arl99, bur00}. This choice of elements is determined by those in the available abundance surveys  (see Table 1). The {\it HERMES} survey will include some other elements, and their contribution is discussed later.

It is important to note that the assumption $\Delta = 0.3$--$0.7$ might not hold for some elements, such as Ni. Our simulations show that if the intrinsic cosmic scatter for {\it all} [X/Fe] is 0.3 dex or less, the simulated noise will dominate over the intrinsic cosmic scatter and the the estimated cut-off will start to drop significantly (about 70\%). However, this will not qualitatively alter our conclusion: for n-capture elements subspace, the cosmic scatters of n-capture elements are believed to be larger than 0.3 dex. For the all elements space, we only interpreted the first four eigenvectors, for which the cumulative eigenvalues percentage is about 75\%, cf. Section~\ref{sec:results}. 

\subsection[]{Estimate of intrinsic correlation}\label{sec:estimate-intrinsic-scatter}
For typical samples of stellar abundances with typical measuring errors, we showed in Section~\ref{sec:best-cut-off-4-PCA} that we can take as real the eigenvectors or principal components contributing to the first 85\% or so of the cumulative percentages for the ranked eigenvalues.  Now we address a different issue. We may need to compare two samples which have different measurement uncertainties. For example, when comparing the principal components for abundances in dwarf galaxies and the Milky Way, it is important to find a way to correct for the different measuring uncertainties.

We can do this by estimating via simulations the intrinsic correlations for each sample, i.e. the values of the correlations if the measurement were perfect, without uncertainties. Then we can directly compare the principal components for each sample. Having reduced the contribution of noise to the correlations, the level of the cumulative percentages for the real ranked eigenvalues is now larger than 85\%, and close to 100\%, though not exactly 100\% because of the residual uncertainties in the noise reduction process.

In the simulations, given two correlated elements $\mathcal{X}_1$ and $\mathcal{X}_2$, first we need to find the best fit line to $[\mathcal{X}_1$/Fe] and $[\mathcal{X}_2$/Fe] to create the mock data set. Instead of deriving the best fit line using weighted or bi-weighted  {\it linear} least squares, we searched for the best fit line by weighted {\it total} least squares. For more details, refer to \citet{kry07}, and Appendix~\ref{appendix:weight-TLS}. 

This is crucial to make sure that the best fit line is symmetric in fitting y to x and fitting x to y and therefore minimize the differences between estimating in the forward and reverse directions. Let $(x_k ,y_k)$ be the k-th data point, and $u_{x,k}$ and $u_{y,k}$ be the corresponding measurement uncertainties. In the special case $u_{x,k} = u_{y,k} = \sigma$, minimizing
\begin{eqnarray}
\label{eq:minimize-WTLS}
\chi^2 = \frac{1}{n-2} \sum^n_{k=1} \Bigg[ \frac{(x_k - X_k)^2}{u^2_{x,k}} +  \frac{(y_k - Y_k)^2}{u^2_{y,k}} \Bigg] 
\end{eqnarray}

\noindent
is equivalent to minimizing the weighted orthogonal distance of the measured points $(x_k,y_k)$ to the fitting line $y =ax+b$ where $(X_k,Y_k)$ is the orthogonal projection of the data point $(x_k,y_k)$ to the line.

\begin{figure}
\begin{center}
\hspace{-0.53cm} \includegraphics[scale=0.5, angle=0]{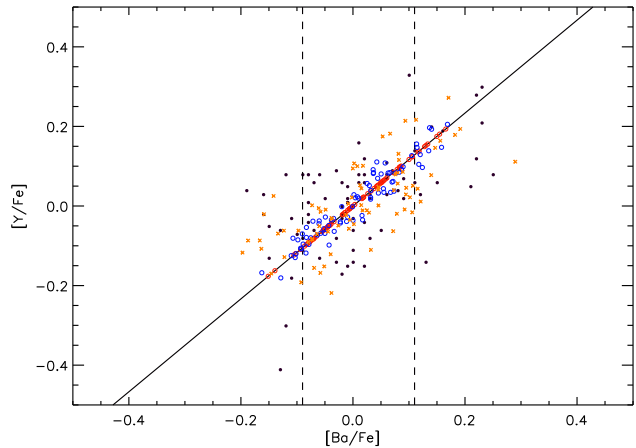}
\caption{A example of simulated mock data to calculate the intrinsic correlation between [Y/Fe] and [Ba/Fe] for the high metallicity sample, in metallicity range $-1.2<$ [Fe/H] $<-0.6$. Here we assume a measurement uncertainty of 0.05 dex for both [Y/Fe] and [Ba/Fe]. The symbols are explained in the text. The {\it black solid line} is the best fit line calculated using weighted total least square. The {\it black dashed vertical lines} are the $1 \sigma$ of the observed data points distribution, assuming Gaussian distribution.}\label{fig:mock-intrinsic-scatter}
\end{center}
\end{figure}

In the simulation, we start with perfectly correlated mock data points lying on the best fit line, as shown in Fig.~\ref{fig:mock-intrinsic-scatter}, the {\it red open circles}.  The {\it black filled circles} are the observed data. We then add simulated cosmic scatter on the {\it red open circles}. The updated mock data points are now denoted as {\it blue open circles}. We then add simulated measurement uncertainty (known amplitude) to the {\it blue open circles} and the updated mock data points are denoted with {\it orange crosses}. The mock data are adjusted until the mock data with simulated noise (cosmic scatter and measurement uncertainty) have the same scatter as the observed data, i.e. the program is iterated until the {\it orange crosses} have the same correlation as the {\it black filled circles}. The average of Pearson's correlation of the {\it blue open circles} from forward and reverse fitting is used as the estimated intrinsic correlation. For every element pair $\mathcal{X}_1$ and $\mathcal{X}_2$, we performed 100 Monte Carlo simulations and adopted the median as the best estimate. This procedure works well for element pairs with measured correlations $> 0.1$ for which the correlations are probably real.  Some element pairs have smaller correlations, and the procedure is not so obvious.  

The goal of much of our analysis is to compare the eigenvectors and dimensionality for various domains of [Fe/H] and subsamples like the dwarf galaxies and star clusters.  When the correlation between an element pair is small, its sign may not be correct because of noise.  The Monte Carlo procedure described above will not change the sign of the correlation so cannot account for an error in sign of a small correlation. We therefore also explored the possibility that the true sign of the correlation is different from the one observed and then estimated what effect this has on the dimensionality;  it is always very small.  For the results reported below, we were guided by the larger stellar samples in adopting the sign of a small correlation for the Monte Carlo modeling.

\section[]{Analysis results}\label{sec:results}
In this section, we show the PCA analysis results of $n$-capture elements alone and then all elements, including light odd-Z, $\alpha$-, Fe-peak and $n$-capture elements.  We discuss stars of low metallicity and high metallicity separately, and then consider open clusters, dwarf galaxies and globular clusters. 

\subsection[]{Low metallicity stars}\label{subsec:results-low-metallicity-stars}
\begin{figure}
\begin{center}
\hspace{-0.7cm} \includegraphics[scale=0.51, angle=0]{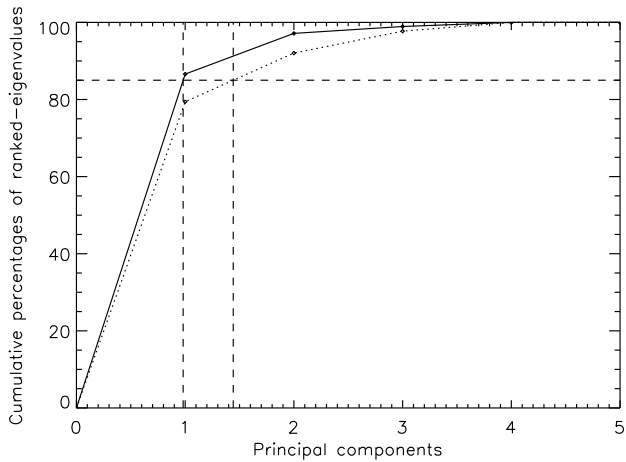} \vspace{-0.5cm}
\caption{This ranked-eigenvalues cumulative percentages of $n$-capture elements (Y, Zr, Ba, Nd, and Eu) for the low metallicity samples-- \citet{bar05} sample (solid line) and First Stars Survey sample (dotted line).}\label{fig:ubiquitous-r}
\end{center}
\end{figure}

We first study the dimensionality of the low metallicity stars using the ranked-eigenvalues cumulative percentages. Fig.~\ref{fig:ubiquitous-r} shows the results of the Barklem's sample (solid line) and First Stars Survey sample (dotted line) considering firstly the $n$-capture elements (Y, Zr, Ba, Nd, and Eu) alone. The vertical dashed line shows the principal component dimensions that corresponds to $85\%$ of the cumulative percentage. Both samples showed that the $n$-capture elements at low metallicity have a strong component that makes up almost all of the variances. 
\begin{figure}
\begin{minipage}{80mm}
$\begin{array}{c}
\vspace{-0.8cm} \hspace{-1cm}\includegraphics[width=3.7in]{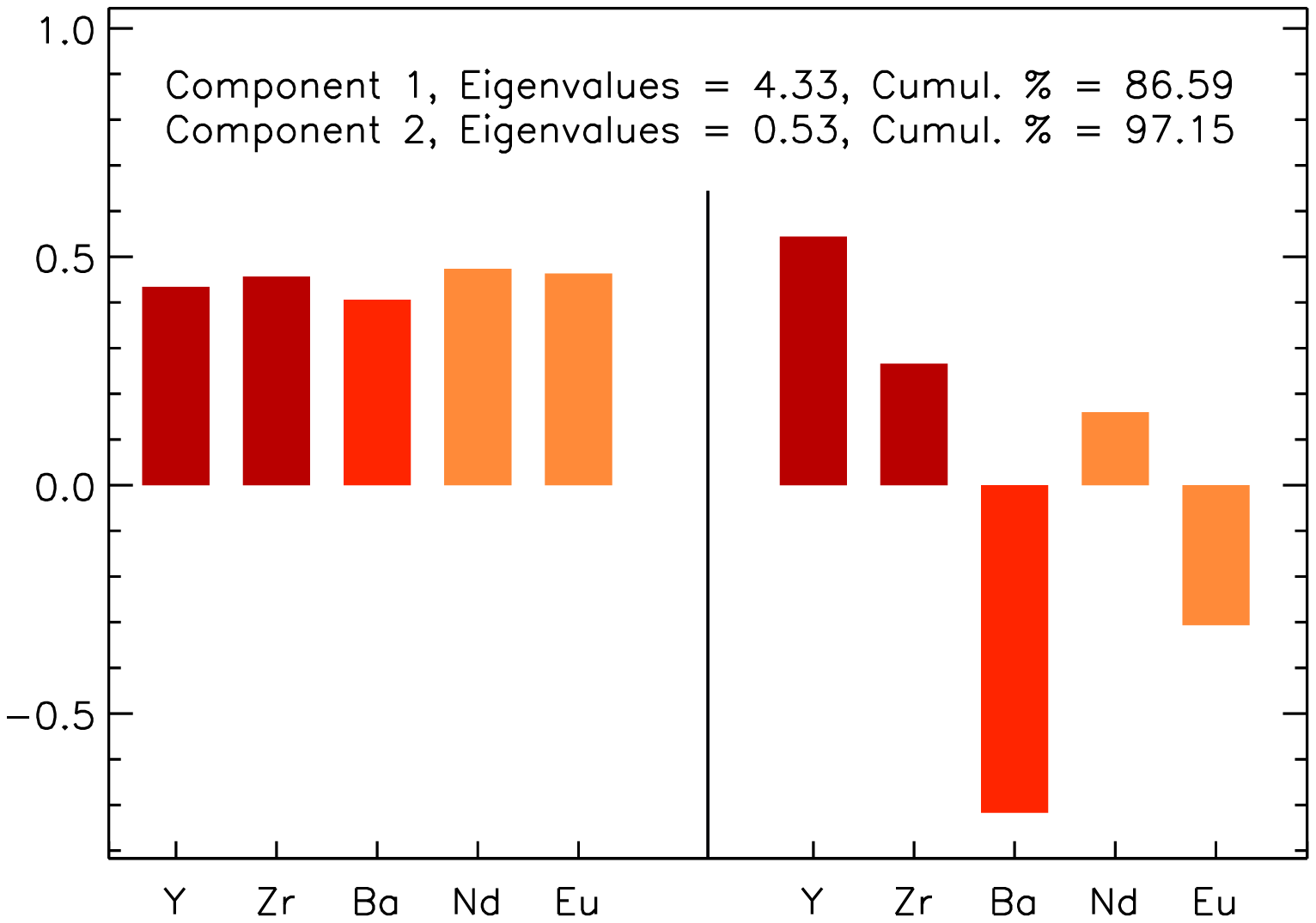}\\
\vspace{-0.5cm} \hspace{-1cm}\includegraphics[width=3.7in]{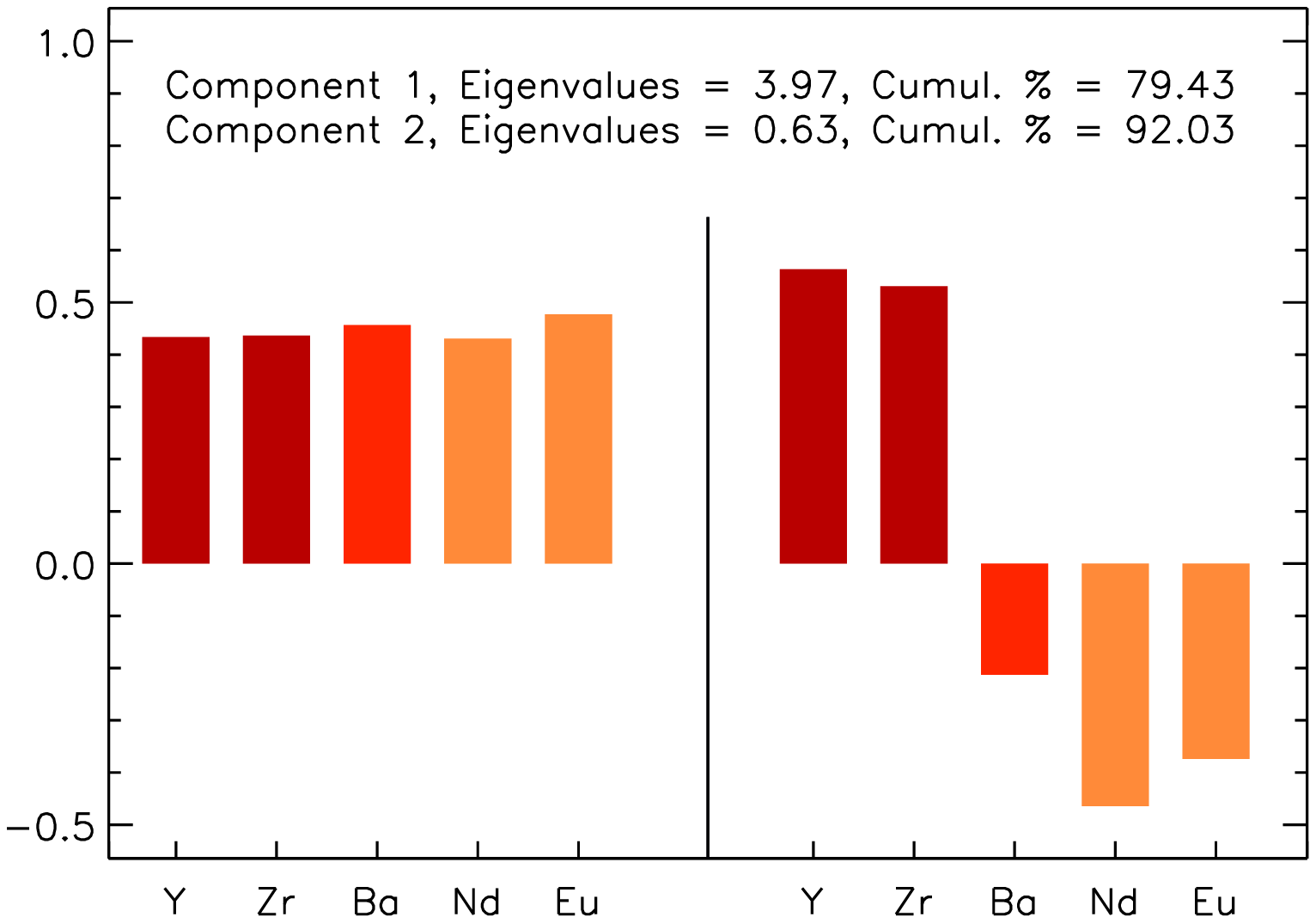}
\end{array}$
\caption{The composition of the first two $n$-capture elements normalized principal components, i.e. the eigenvector components of the $n$-capture elements correlation matrix, using the low metallicity samples: Barklem's sample (upper plot) and First Stars Survey sample (lower plot). The horizontal axis shows the 5 n-capture elements and the vertical axis shows the component of the eigenvector in the direction of each of the 5 elements.. See the text for the colour code.}\label{fig:Barklem-neutron}
\end{minipage}
\end{figure}
\begin{figure}
\begin{center}
\hspace{-0.7cm}\includegraphics[scale=0.51, angle=0]{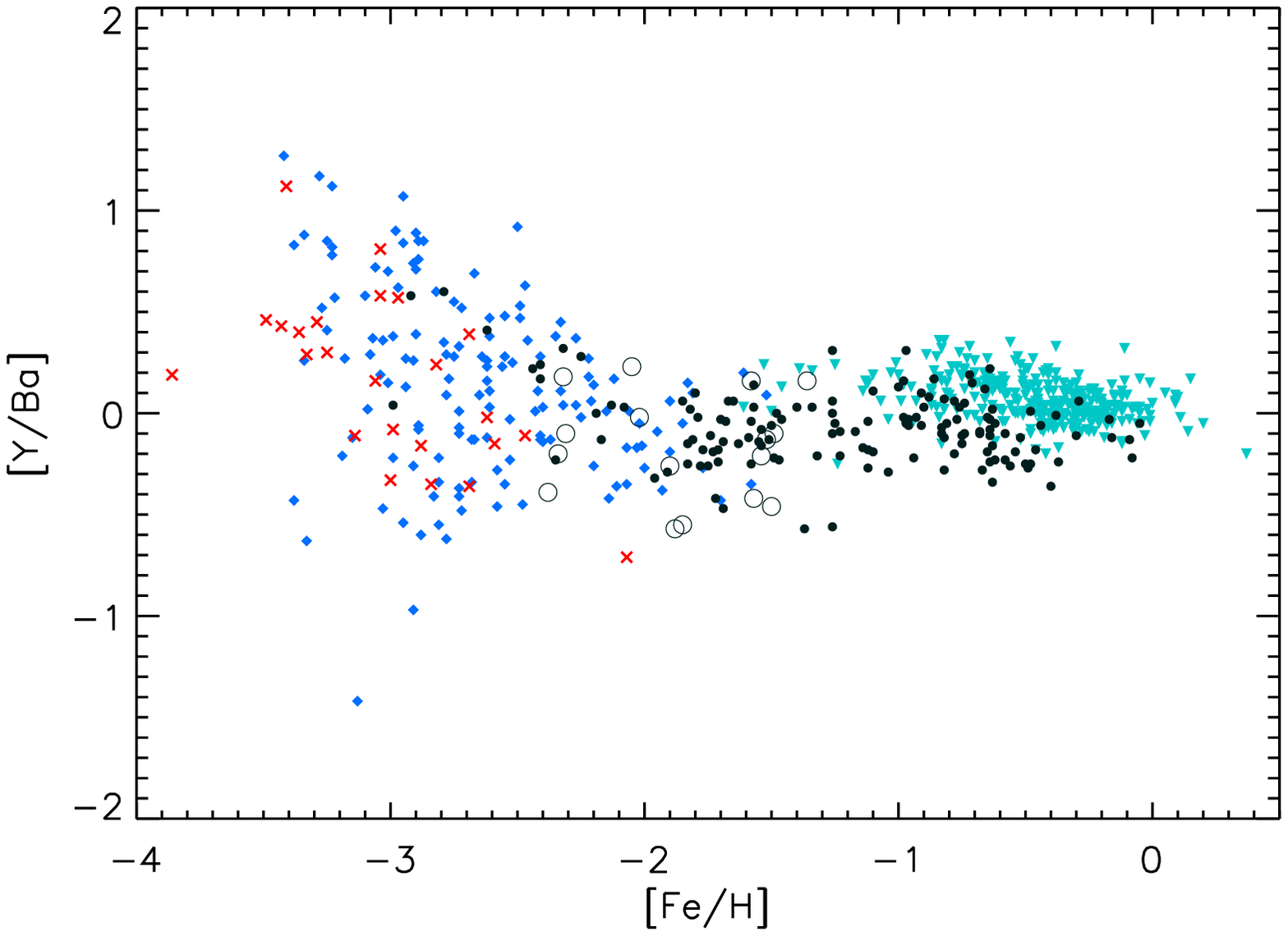}
\caption{Y ({\it ls}) to Ba ({\it hs}) ratios as a function of [Fe/H] for the Milky Way halo/disk stars from First Stars Survey ({\it red crosses}), \citet{bar05} ({\it blue diamonds}), \citet{ful00,ful02} ({\it black filled circles}), \citet{red03,red06} ({\it cyan triangles}) and globular clusters from \citet{pri05} ({\it black open circles}) All samples assume 1D-LTE stellar model.}\label{fig:light-s-over-abundances}
\end{center}
\end{figure}

We then examine the individual components. Fig.~\ref{fig:Barklem-neutron} shows the composition of the first two eigenvectors (first two principal components), i.e. the first two directions that are orthogonal and account for most of the variance in the normalized mean shifted $\mathcal{C}$-space. The upper plot and lower plot show the results of the Barklem sample and First Star Survey sample respectively.  The colour is coded as a rough grouping to differentiate families of elements from different major production sources: {\it dark red} represents light $s$-process elements, {\it red} stands for heavy $s$-process elements, and mostly $r$-process elements are coded as {\it orange}. For simplicity, on the x-axis we denote [X/Fe] as X in all such figures in this paper. Since the eigenvectors are normalized, the quadratic sum of the five [X/Fe] abundances contribution for each principal component equals 1. 

In the first component, we see that all elements have the same sign. This illustrates that this dominant principal component is pointing in the positive diagonal direction in this 5-dimensional $\mathcal{C}$-subspace. We can infer that the first component represents a mechanism that produce all 5 elements simultaneously. This suggests that all of these $n$-capture elements are mostly produced by a dominant primary process at low redshift, which could be the $r$-process \citep{tru02,wan06}. 

The second eigenvector points in the anti-diagonal direction of the light $s$-process elements (Y, Zr) and the heavier $n$-capture elements (Ba, Nd, Eu), i.e. this component represents a mechanism that produces mainly light $s$-process elements or mainly heavy $n$-capture elements. Fig.~\ref{fig:light-s-over-abundances} helps to resolve the nature of this mechanism:  we see that the light $s$-process elements are over-abundant at low metallicity, and therefore the former interpretation is preferred. This second principal component might correspond to the contribution of LEPP \citep{tra04,qia07,izu09}. However, as shown in the ranked-eigenvalues cumulative percentages plot (cf. Fig.~\ref{fig:ubiquitous-r}), the contribution of this component is small relative to the first principal component. We discuss this further in Section~\ref{sec:discussion}. 

To show the relative importance of the two components as a function of metallicity, we performed the PCA analysis on stars in [Fe/H] bins of $0.7$ dex. We calculated the ratio of the second component eigenvalue to  the sum of the eigenvalues for the first two components to estimate the contribution of the second component to the overall $n$-capture element yield. The results is shown in the upper panel of Table~\ref{tab:LEPP}.  The same calculation was repeated for the First Stars Survey sample and is shown in lower panel of Table~\ref{tab:LEPP}.  The fractional contribution of the second component appears to decrease with increasing metallicity.  We have performed Kendall's $\tau$ and Spearman's $\rho$ tests: both tests showed that this trend is significant ($<  5\%$ probability of a false positive correlation).
\begin{table}
\begin{center}
\caption{Fraction of the second component contribution to the $n$-capture elements yield: Barklem's sample (upper panel), the First Stars Survey sample (lower panel).\label{tab:LEPP}}
\begin{tabular}{cc}
\hline
[Fe/H] range & Fraction($\%$) \\
\hline
$(-3.4 , -2.7)$ & $10.15$ \\
$(-3.3 , -2.6)$ & $10.09$ \\
$(-3.2 , -2.5)$ & $8.41$ \\
$(-3.1 , -2.4)$ & $7.88$ \\
$(-3.0 , -2.3)$ & $7.09$ \\
$(-2.9 , -2.2)$ & $5.62$ \\
\hline
\end{tabular}
\\
\begin{tabular}{cc}
\hline
[Fe/H] range & Fraction($\%$) \\
\hline
$(-3.5 , -2.8)$ & $18.10$ \\
$(-3.4 , -2.7)$ & $16.78$ \\
$(-3.3 , -2.6)$ & $13.67$ \\ 
$(-3.2 , -2.5)$ & $10.96$ \\
\hline
\end{tabular}
\end{center}
\end{table}
\begin{figure}
\begin{minipage}{80mm}
$\begin{array}{c}
\hspace{-0.3cm}\includegraphics[width=3.1in]{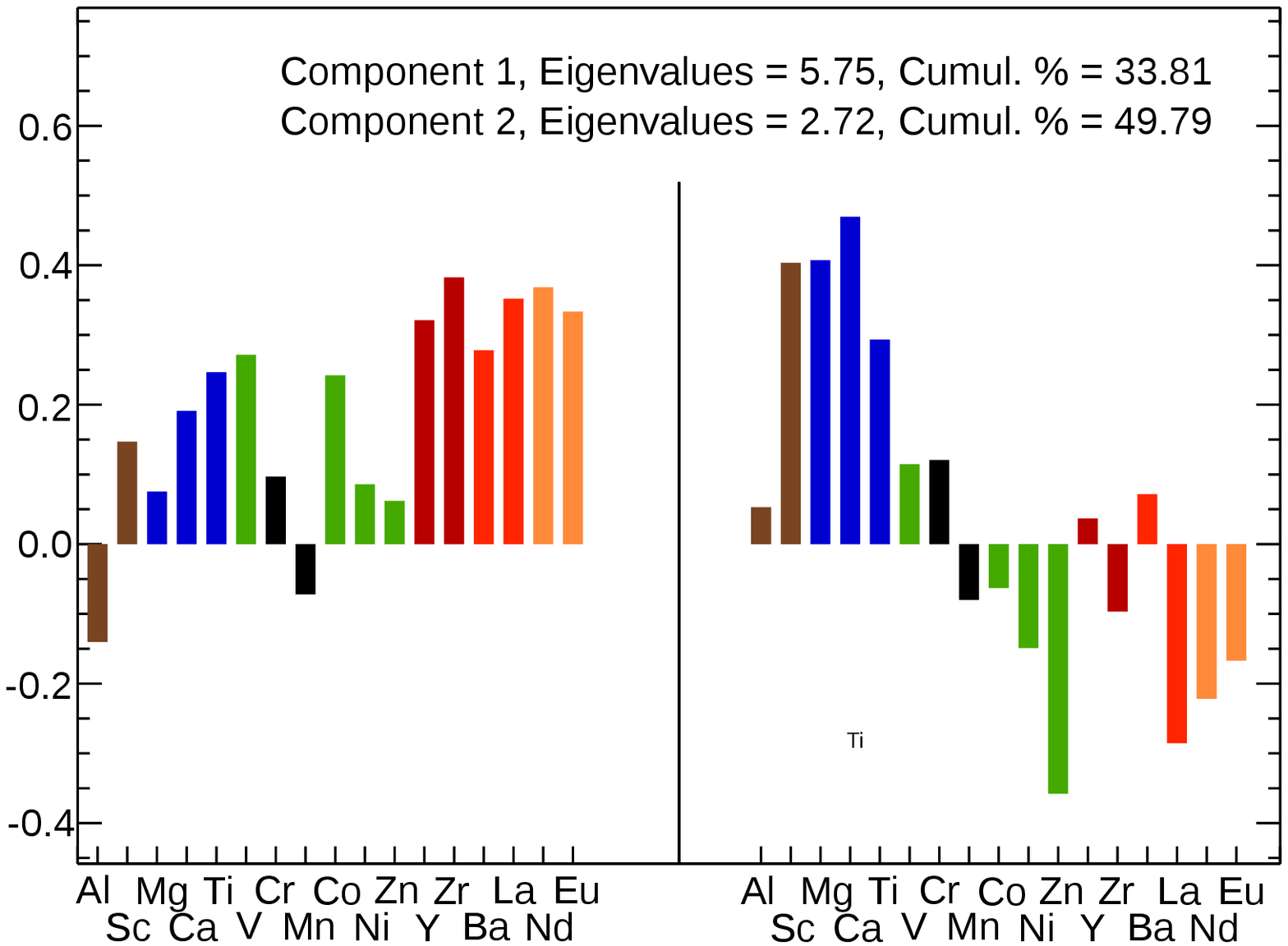} \vspace{0.5cm}\\
\hspace{-0.3cm}\includegraphics[width=3.1in]{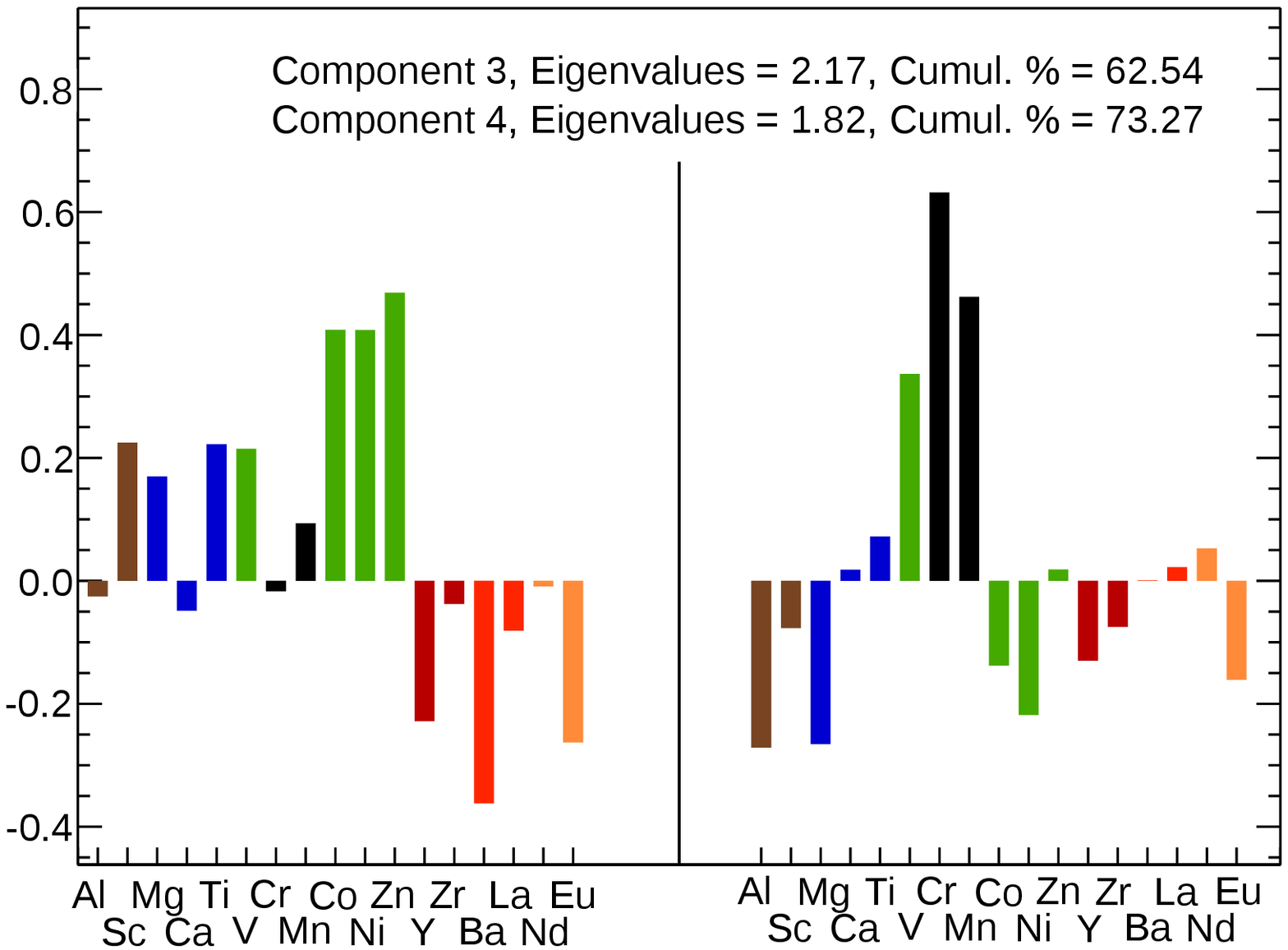}
\end{array}$
\vspace{0.3cm}
\caption{The normalized principal components of 17 elements for the low metallicity sample \citep{bar05}. [Al/Fe] is corrected for NLTE effect with $+0.6$ dex. The upper plot and lower plot show the first two principal components and the third and fourth principal components, respectively.}\label{fig:all-Barklem}
\end{minipage}
\end{figure}

Note that La is not included in the analysis of $n$-capture elements, in order to compare the low metallicity results with those for high metallicity. La is not measured in our high metallicity Reddy's sample. However, we have tested our results by adding La or restricting ourselves to a smaller subspace. In either case, it does not alter the results qualitatively. The rest of the principal components are probably due to measurement uncertainty since the first two components already account for $>90\%$ of the data cloud variance.
\begin{figure}
\hspace{-0.5cm}\includegraphics[scale=0.5, angle=0]{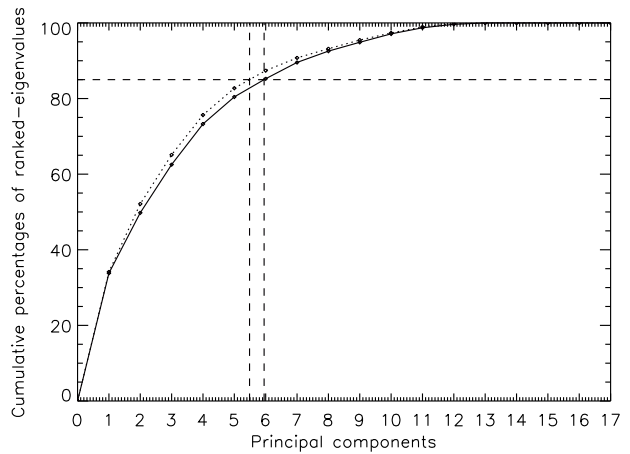}
\caption{The ranked-eigenvalues cumulative percentages of 17 elements for the low metallicity sample \citep{bar05}. The {\it solid line} and {\it dotted line} show the results with and without the NLTE correction of [Al/Fe], respectively.}\label{fig:cumul-Barklem-all}
\end{figure}
\begin{figure}
\begin{minipage}{80mm}
$\begin{array}{c}
\hspace{-0.3cm}\includegraphics[width=3.1in]{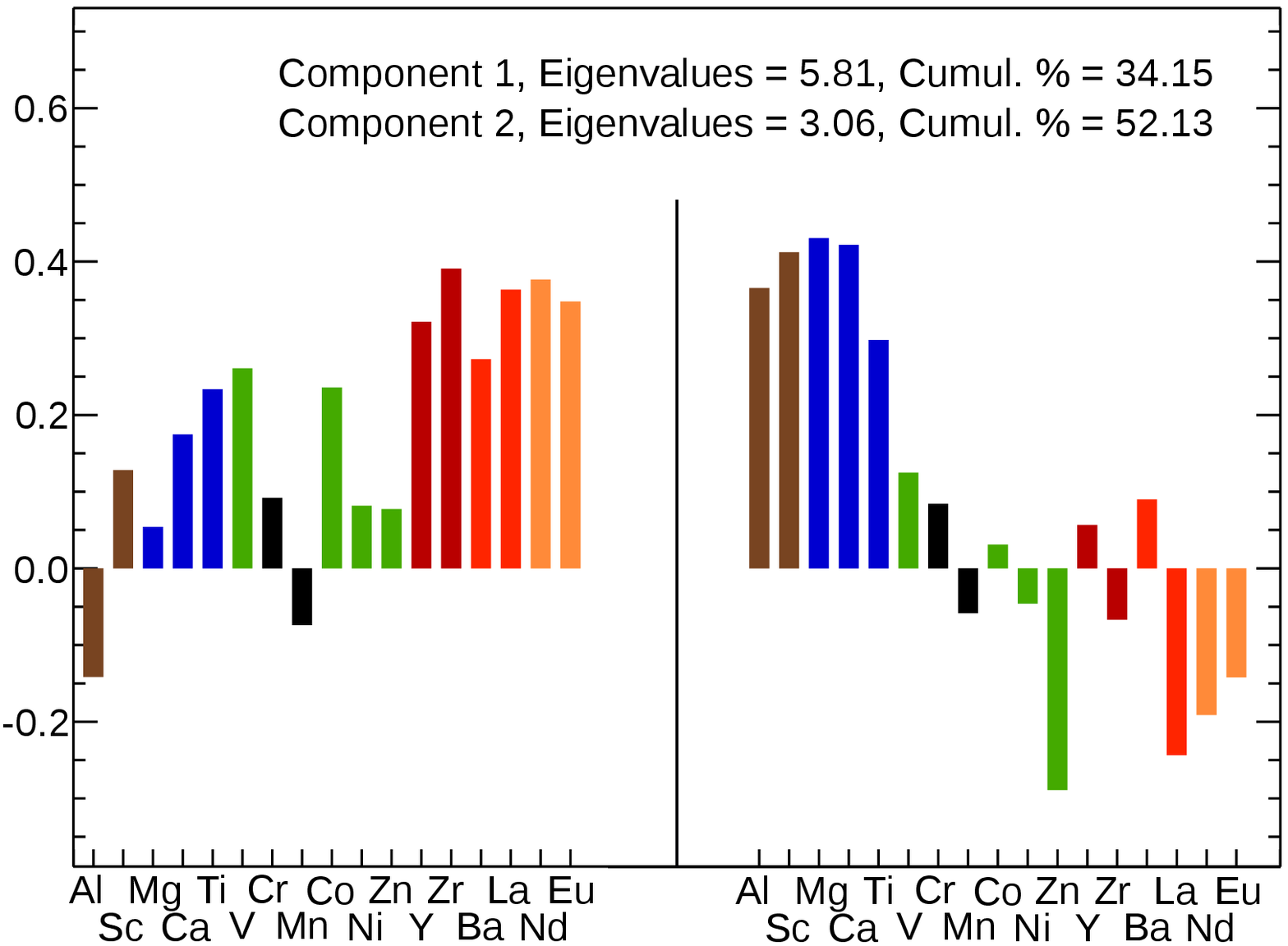} \vspace{0.5cm}\\
\hspace{-0.3cm}\includegraphics[width=3.1in]{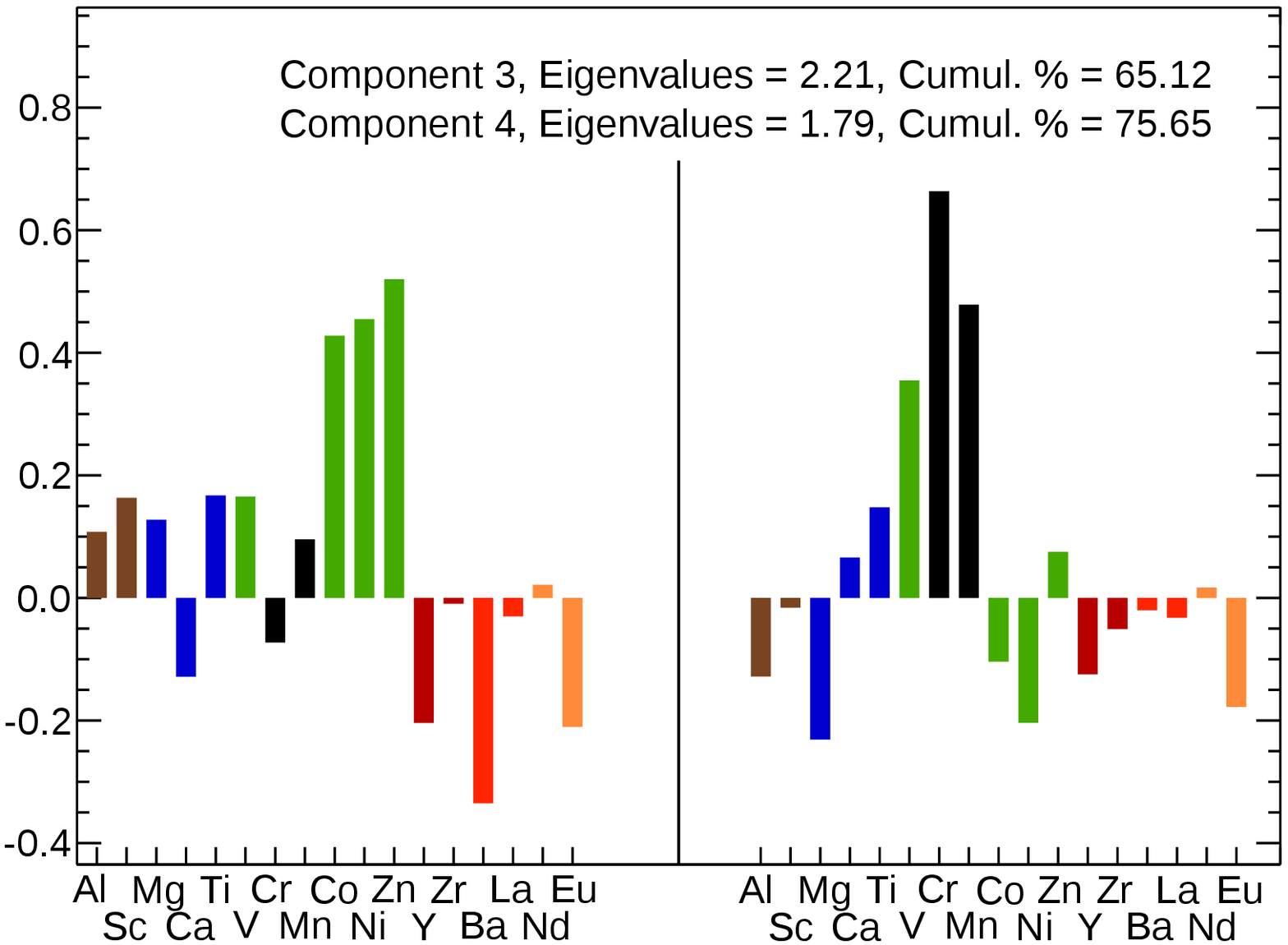}
\end{array}$
\vspace{0.3cm}
\caption{The same as Fig.~\ref{fig:all-Barklem} but without the NLTE correction for [Al/Fe].}\label{fig:LTE-al}
\end{minipage}
\end{figure}

Now we calculate the principal components for the whole set of elements available for the low metallicity stars. The principal components of PCA analysis on 17 elements (Al, Sc, Mg, Ca, Ti, V, Cr, Mn, Co, Ni, Zn, Y, Zr, Ba, La, Nd, and Eu) for Barklem's sample are shown in Fig.~\ref{fig:all-Barklem}. The colour is coded to differentiate categories of elements: the $n$-capture elements are coded as before, light odd-Z elements are colour-coded with {\it dark brown}, $\alpha$-elements with {\it blue}, Fe-peak elements (beside Cr \& Mn) with {\it green}, and Cr \& Mn with {\it black}.   If we adopt a $85 \%$ cutoff for the ranked-eigenvalues cumulative percentages, as shown in Fig.~\ref{fig:cumul-Barklem-all}, we find about 6 independent dimensions for the 17-dimensional $\mathcal{C}$-space. The number of dimensions is not altered if we apply a $+0.6$ dex NLTE correction for [Al/Fe], as shown in Fig.~\ref{fig:cumul-Barklem-all}.

The first four principal components of this 17-dimensional $\mathcal{C}$-space can be summarized as follows: 
\begin{itemize}
\item The first component shows contributions primarily from {\it all} $n$-capture elements and also contribution from $\alpha$-elements. This suggests that the production sites of the $r$-process also produce significant amounts of $\alpha$-elements, which may be consistent with core-collapse supernovae as the $r$-process site rather than neutron star mergers. The first component could be the contribution from core-collapse supernovae that involve the $r$-process. 
\item The second component shows an anticorrelation of $\alpha$-elements with Fe-peak elements and $n$-capture elements. This suggests that there is a mechanism that produces $\alpha$-elements but does not produce Fe-peak and $n$-capture elements. This component may correspond to the contribution from `normal' core-collapse supernovae that do not involve the $r$-process
\item The third component shows an anticorrelation of $\alpha$-elements and Fe-peak elements with the $n$-capture elements. This suggests the production of {\it both} $\alpha$-elements and Fe-peak elements. This component may correspond to the contribution of hypernovae \citep[e.g.][]{kob06} where significant amounts of Co, Ni, Zn are produced with high explosion energy. 
\item The fourth component shows a strong contribution from Cr and Mn. Different from heavier Fe peak elements, Cr and Mn are synthesized in the incomplete Si-burning region, which is the outer region of the complete Si-burning region. We discuss this further in Section~\ref{sec:discussion}. 
\end{itemize}

Note that with the NLTE correction, Al does not contribute to the first four components. If we do not apply NLTE correction for Al, Al contributes to the second component as shown in Fig.~\ref{fig:LTE-al}, but the results for other elements remain unaltered. We will discuss this further in Section~\ref{sec:discussion}.

\subsection[]{High metallicity stars}
\begin{figure}
\hspace{-1cm}
\includegraphics[scale=0.55, angle=0]{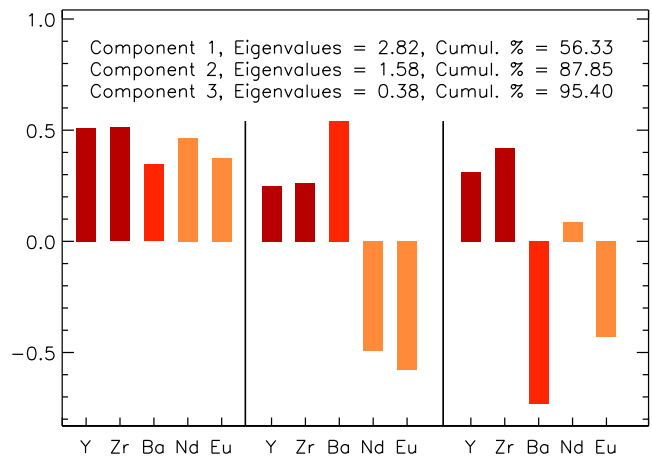}
\vspace{-0.8cm}
\caption{The composition of the first three normalized principal components for the $n$-capture elements for the high metallicity sample.}\label{fig:Reddy-neutron}
\end{figure}

\begin{figure}
\hspace{-0.5cm}\includegraphics[scale=0.51, angle=0]{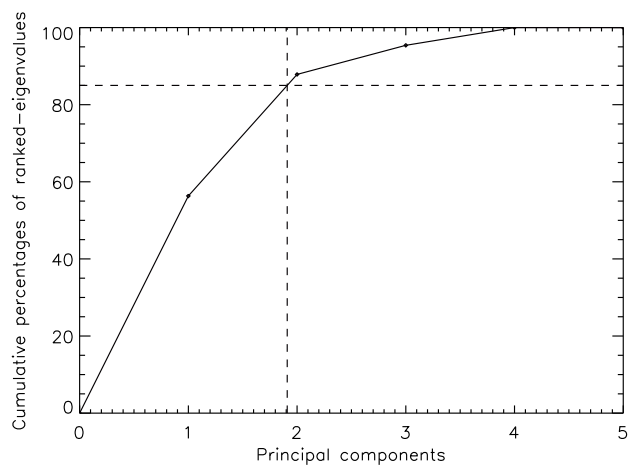}
\caption{The ranked-eigenvalues cumulative percentages for the $n$-capture elements for the high-metallicity sample.}\label{fig:cumul-Reddy-neutron}
\end{figure}

Fig.~\ref{fig:Reddy-neutron} and Fig.~\ref{fig:cumul-Reddy-neutron} show the PCA analysis results for the $n$-capture elements (Y, Zr, Ba, Nd, Eu) in the Reddy high metallicity sample. This results show two marked differences from the low metallicity sample (cf. Fig.~\ref{fig:ubiquitous-r}, Fig.~\ref{fig:Barklem-neutron}): (1) The first principal component contributes far less ($56\%$) to the overall variance compared with low metallicity sample ($79$ -- $87\%$). It is clear that at high metallicity, there is more than one mechanism contributing to the $n$-capture elements; (2) At high metallicity, the second component is made up of both light $s$-process elements (Y and Zr) and heavy $s$-process elements (Ba), instead of light $s$-process elements alone as in the low metallicity sample anticorrelating with mostly $r$-process elements (Nd and Eu). This component may correspond to the $s$-process in low-mass AGB stars. The $s$-process in low-mass AGB stars, operating in metal-rich environment, is expected to produce both light $s$-process and heavy $s$-process elements. The third component might be the remnant of LEPP contribution at low metallicity, or it might just be noise, since the first two components already account for $88 \%$ of the total variance. 

Fig.~\ref{fig:all-Reddy} shows the first four principal components from the PCA analysis of all 17 elements (Al, Sc, Mg, Si, Ca, Ti, V, Cr, Mn, Co, Ni, Zn, Y, Zr, Ba, Nd, and Eu) which are available for the high metallicity sample. The colour coding is the same as in Fig.~\ref{fig:all-Barklem}. The results are summarized as follows: 
\begin{figure}
\begin{minipage}{80mm}
\vspace{0.3cm}
$\begin{array}{c}
\hspace{-0.3cm}\includegraphics[width=3.1in]{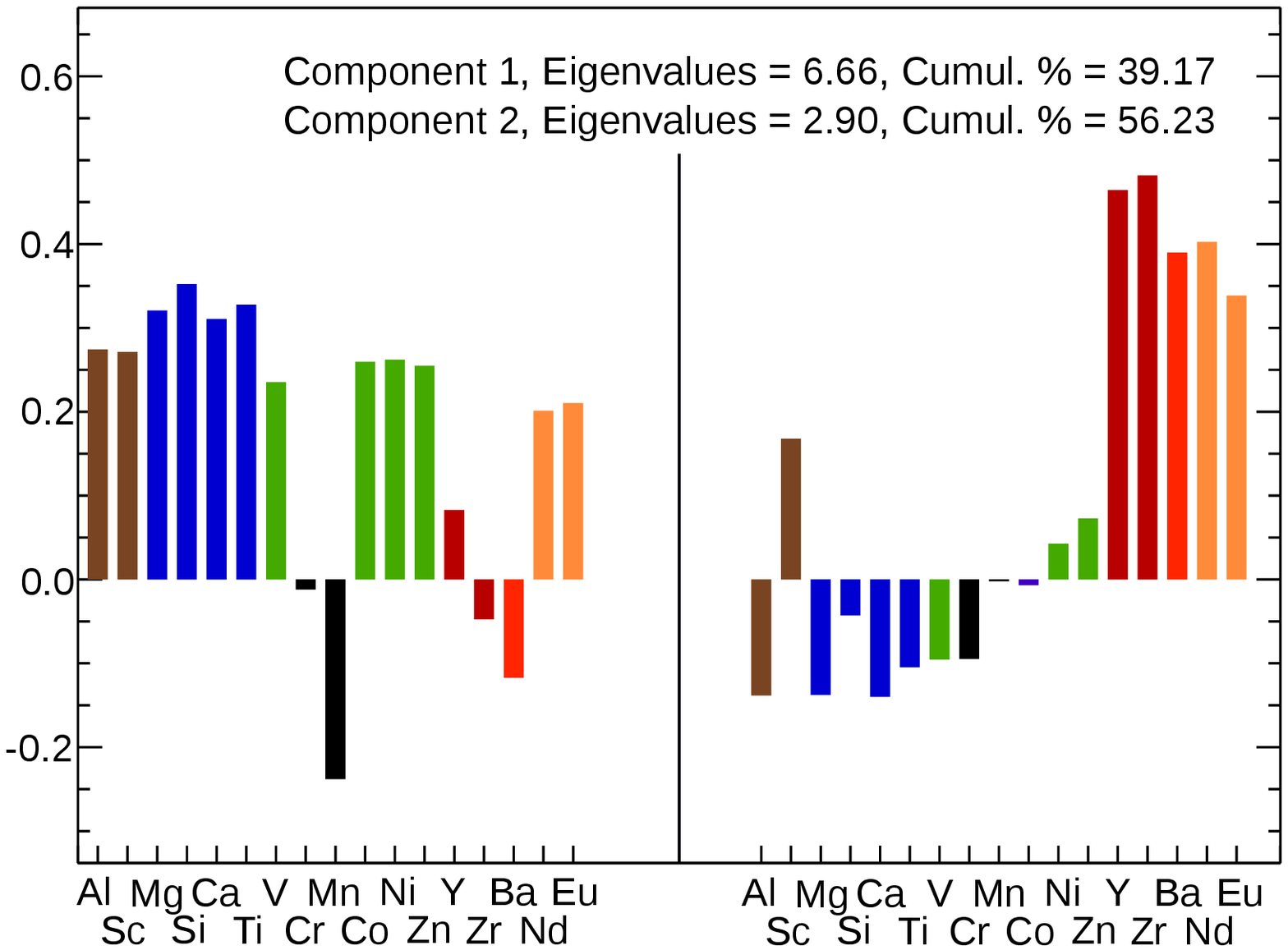} \vspace{0.6cm}\\
\hspace{-0.3cm}\includegraphics[width=3.1in]{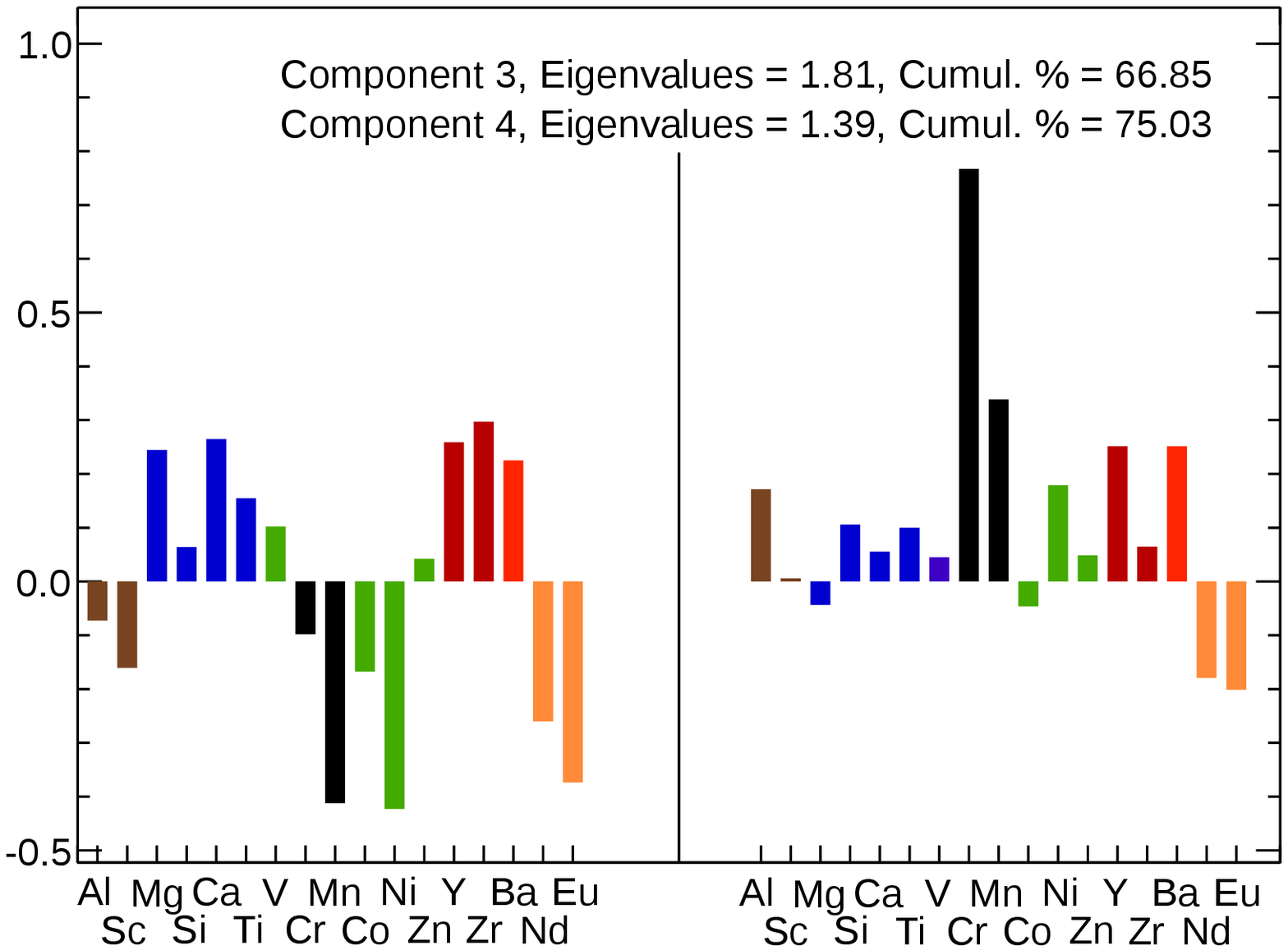}
\end{array}$
\vspace{0.4cm}
\caption{The normalized principal components of 17 elements for the high metallicity sample. The upper and lower plot show the first two principal components and the third and fourth principal components, respectively.}\label{fig:all-Reddy}
\end{minipage}
\end{figure}
\begin{itemize}
\item The first component is a strong component, accounts for about $40 \%$ of the total variance, and is made up of $\alpha$-elements and Fe-peak elements (except for Mn and Cr) as well as light odd-Z element Al (1D-LTE-based). Note that NLTE correction is supposed to be negligible at high metallicity. We recall that in the low metallicity sample, Al was absent from the first few principal components when NLTE-corrected. This component may correspond to the well-mixed ISM that is affected by many generations of core-collapse supernovae. The deficit of Mn suggests that this component does not include the contribution from SNe Ia.
\item The second component is made up mainly of all of the $n$-capture elements, which could be the contribution from $r$-process. However, unlike the first principal component of the low metallicity sample which also includes all of the $n$-capture elements,  the $\alpha$-elements contribution are not visible in this second component for the high metallicity sample. We discuss this in more detail in Section~\ref{sec:discussion}.
\item The third component shows anticorrelation between light $s$-process, heavy $s$-process, $\alpha$-elements and the other elements. This could correspond to the contribution of AGB stars ejecta.
\item The fourth component is again dominated by Cr and Mn. We will discuss this further in Section~\ref{sec:discussion}.
\end{itemize}

\begin{figure}
\hspace{-0.5cm}\includegraphics[scale=0.51, angle=0]{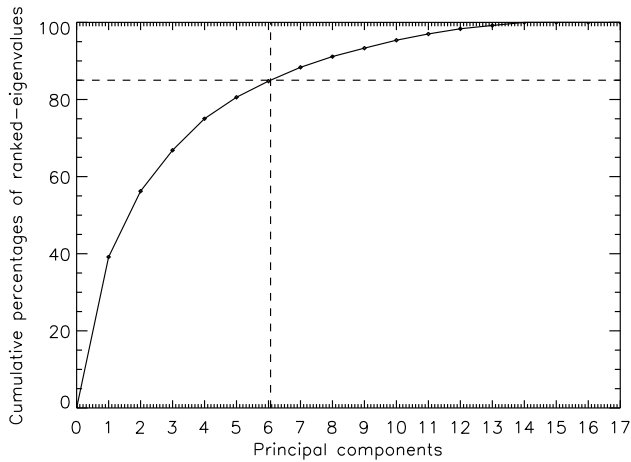}
\vspace{-0.6cm}
\caption{The ranked-eigenvalues cumulative percentages of 17 elements for the high metallicity sample.}
\label{fig:cumul-Reddy-all}
\end{figure}

Adopting a $85 \%$ cutoff for the ranked-eigenvalues cumulative percentages, as shown in Fig.~\ref{fig:cumul-Reddy-all}, we again have $6$ independent dimensions in the 17 dimensional $\mathcal{C}$-space, as for the low-metallicity sample. This is not totally unexpected. Although dimensions are lost by the homogenization of ISM at higher metallicity \citep{karl01,bla10}, there are also processes such as the AGB stars and SNe Ia that will contribute only at higher metallicity. This extra `birth-imprint' appears to compensate for the loss of dimensions due to homogenization.

\subsection[]{Open clusters}
Now we turn to the open clusters, which cover a larger volume of the Galaxy than the stellar samples discussed so far.  We do not make a separate study of their $n$-capture element subspace because there are no Eu data in the open cluster compilation.  Eu is potentially important to separate the $r$-process and $s$-process contributions. Instead, we analyze the subspace of $\alpha$-elements: Mg, Si, Ca, Ti. We then analyze the all-elements space, including the 13 elements: Al, Sc, Mg, Si, Ca, Ti, V, Cr, Co, Ni, Y, Ba, and Nd. We compare the results for the open cluster compilation with the Reddy solar neighbourhood sample by restricting the metallicity range of the solar neighbourhood sample to [Fe/H] $> -0.5$ to match the range for the open clusters.

\begin{figure}
\hspace{-0.65cm}
\includegraphics[scale=0.52, angle=0]{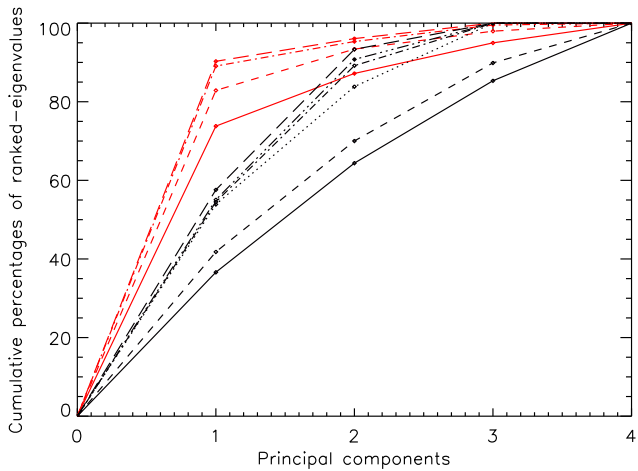}
\vspace{-0.6cm}
\caption{The ranked-eigenvalues cumulative percentages of the $\alpha$-elements, comparing open clusters ({\it black lines}) with solar neighbourhood stars ({\it red lines}). The solid lines show the observed correlation. The dashed/broken lines show the cumulative percentages derived from the estimated intrinsic correlation matrix, assuming measurement uncertainty of 0.05, 0.07, 0.10, 0.15, and 0.20 dex respectively for open clusters, and 0.02, 0.03, and 0.10 dex respectively for solar neighbourhood stars.}\label{fig:OC-MW-compare-alpha}
\end{figure}

\begin{figure}
\hspace{-0.65cm}
\includegraphics[scale=0.52, angle=0]{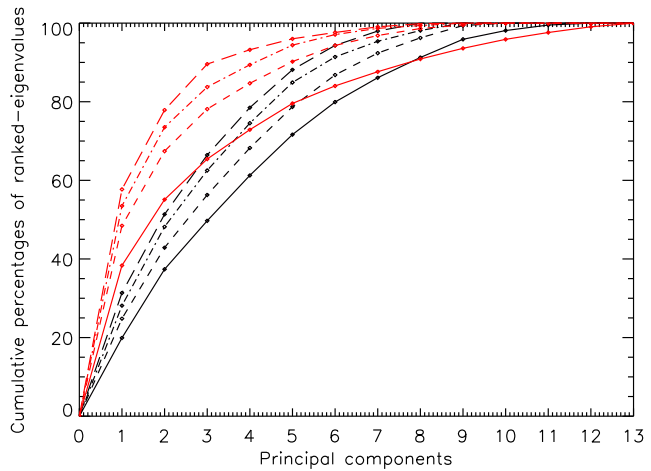}
\vspace{-0.6cm}
\caption{The same as Fig.~\ref{fig:OC-MW-compare-alpha} but for 13 elements, and assuming measurement uncertainty of 0.05, 0.07, and 0.20 dex respectively for open clusters, and 0.03, 0.05, and 0.10 dex respectively for solar neighbourhood stars.}\label{fig:OC-MW-compare-all}
\end{figure}

The ranked eigenvalues shown as the {\it solid lines} in Fig.~\ref{fig:OC-MW-compare-alpha} and Fig.~\ref{fig:OC-MW-compare-all} suggests that the $\mathcal{C}$-space for the open clusters has a higher dimensionality (about $1$ more dimension) than for the nearby metal-rich field stars, both for the restricted $\alpha$-element subspace and for the 13-element space. However, systematic differences between authors may contribute extra scatter to the open clusters abundances and the extra dimension could be spurious. To show that this is {\it not} the case, we have performed the intrinsic correlation estimates as described in Section~\ref{sec:estimate-intrinsic-scatter}. The results are over-plotted in Fig.~\ref{fig:OC-MW-compare-alpha} and Fig.~\ref{fig:OC-MW-compare-all}. 

For example in Fig.~\ref{fig:OC-MW-compare-alpha}, the {\it black dashed/broken lines} show the ranked-eigenvalues cumulative percentages derived from the estimated intrinsic correlation for the open clusters sample, assuming that the measurement uncertainty is 0.05 (short-dashed), 0.07, 0.1, 0.15, and 0.2 (long-dashed) dex, respectively. The {\it red dashed/broken lines} show the cumulative percentages derived from the estimated intrinsic correlation for the solar neighbourhood sample, assuming a measurement uncertainty of 0.02, 0.03 and 0.1 dex, respectively. Recall that for the estimated intrinsic correlation matrix, by definition, there is no residual contribution from measurement uncertainty. Therefore the cumulative percentages cutoff should be close to $100\%$ instead of $85\%$. 

The results show that our previous interpretation holds even if we make extreme assumptions about the measurement uncertainty of the two samples, such as assuming measurement uncertainty and systematic differences of $0.2$ dex for open clusters and a very small measuring uncertainty of $0.03$ dex for the solar neighbourhood stellar sample. The result is robust for the 4-dimensional $\alpha$-element subspace. This is consistent with results in the literature that the $\alpha$-element distribution may not be the same within and outside the solar neighborhood: for example the enhanced ratios of [Mg/Fe] and [Ti/Fe], but not for [Si/Fe] and [Ca/Fe] reported by \citet{yon05} at larger galactocentric distances.  For the larger 13-dimensional $\mathcal{C}$-space,  the result seems secure but is a little less robust because of the concerns described in Section~\ref{sec:estimate-intrinsic-scatter} about deriving the intrinsic correlations for weakly correlated element pairs.

Our results are consistent with \citet*{siv09} and \citet*{roe09} that the larger regions of the Galaxy covered by the open cluster sample may show more independent variations of element abundances than the immediate solar neighbourhood, even though we cannot readily identify the nature of the extra dimension of their $\mathcal{C}$-space. Although the solar neighbourhood appears chemically well mixed,  as shown by the relative tightness of say the [$\alpha$/Fe]--[Fe/H] relation, this may not prevail over larger volumes of the Galaxy.   

\subsection[]{Satellite galaxies}
\label{subsec:dwarf-galaxies-results}
Now we investigate two samples of stars, one for the Fornax dSph galaxy and one for the LMC. These are of interest because they represent different evolutionary environments from the Milky Way, and also because older versions of these galaxies may be similar to the systems which merged with the Milky Way \citep[see][]{sea78}. 

In the previous section, we showed that the $\mathcal{C}$-space dimensionality might vary as a function of the survey volume. Therefore, robust comparisons can only be made if the survey volumes are about the same. Since our homogeneous satellite galaxy samples are restricted to a small survey volume ($1$ -- $2$ kpc in diameter\footnote{Letarte's study covered 25\arcmin of the centre of Fornax \citep[about $1$ kpc, e.g.][]{riz07}.  The survey of \citet{pom08} covered $2$ kpc of the LMC inner disk}), it seems appropriate to compare the $\mathcal{C}$-space results for the satellite galaxies and the solar neighbourhood, for which the surveys cover a similar volume. For each comparison, we restricted the solar neighbourhood stars to the same metallicity range as Fornax  ($-1.2<$ [Fe/H] $<-0.6$) and  the  LMC ($-1.0<$ [Fe/H] $<-0.5$). 
\begin{figure}
\hspace{-0.5cm}
\includegraphics[scale=0.51, angle=0]{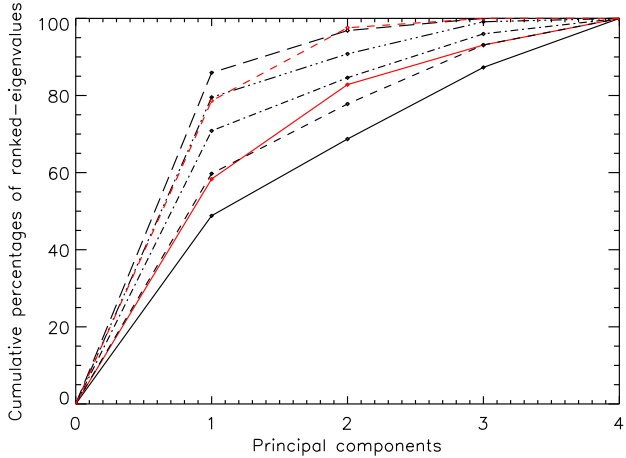}
\vspace{-0.6cm}
\caption{The $n$-capture element ranked-eigenvalues cumulative percentages comparing the Fornax dSph galaxy ({\it black lines}) with its solar neighbourhood counterparts ({\it red lines}) in the same metallicity range. The solid lines show the observed correlation. The dashed/broken lines show the cumulative percentages derived from the estimated intrinsic correlation matrix, assuming measurement uncertainty of 0.1, 0.13, 0.15, and 0.2 dex respectively for the dSph, and 0.07 dex for the solar neighbourhood counterparts.}\label{fig:dsph-vs-mw-neutron}
\end{figure}
\begin{figure}
\hspace{-0.75cm}
\includegraphics[scale=0.53, angle=0]{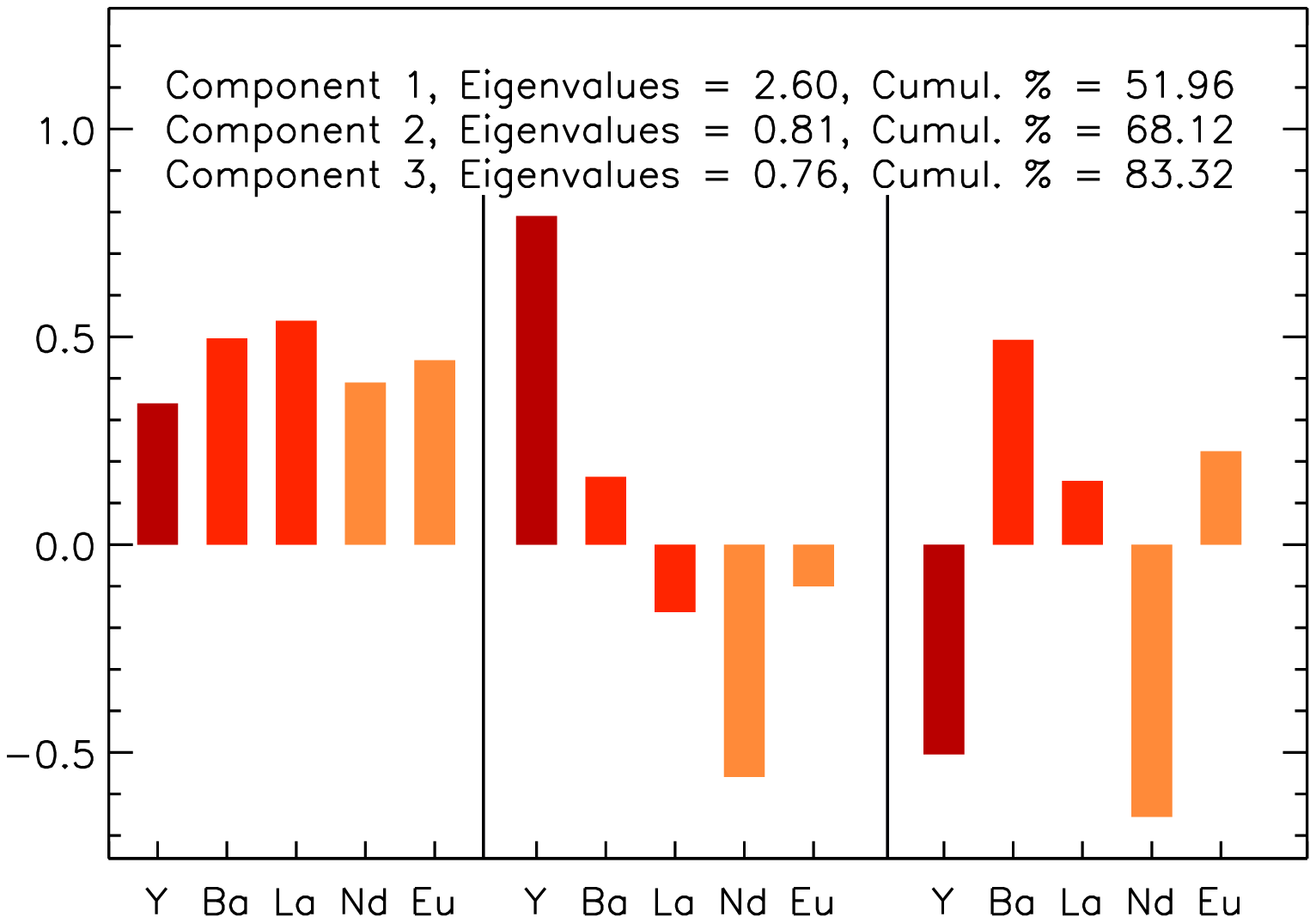}
\vspace{-0.9cm}
\caption{The composition of the first three normalized principal components for the $n$-capture elements of the Fornax dSph galaxy.}\label{fig:dsph-neutron}
\end{figure}
\begin{figure}
$\begin{array}{c}
\vspace{-0.3cm} \hspace{-0.53cm}\includegraphics[width=3.6in]{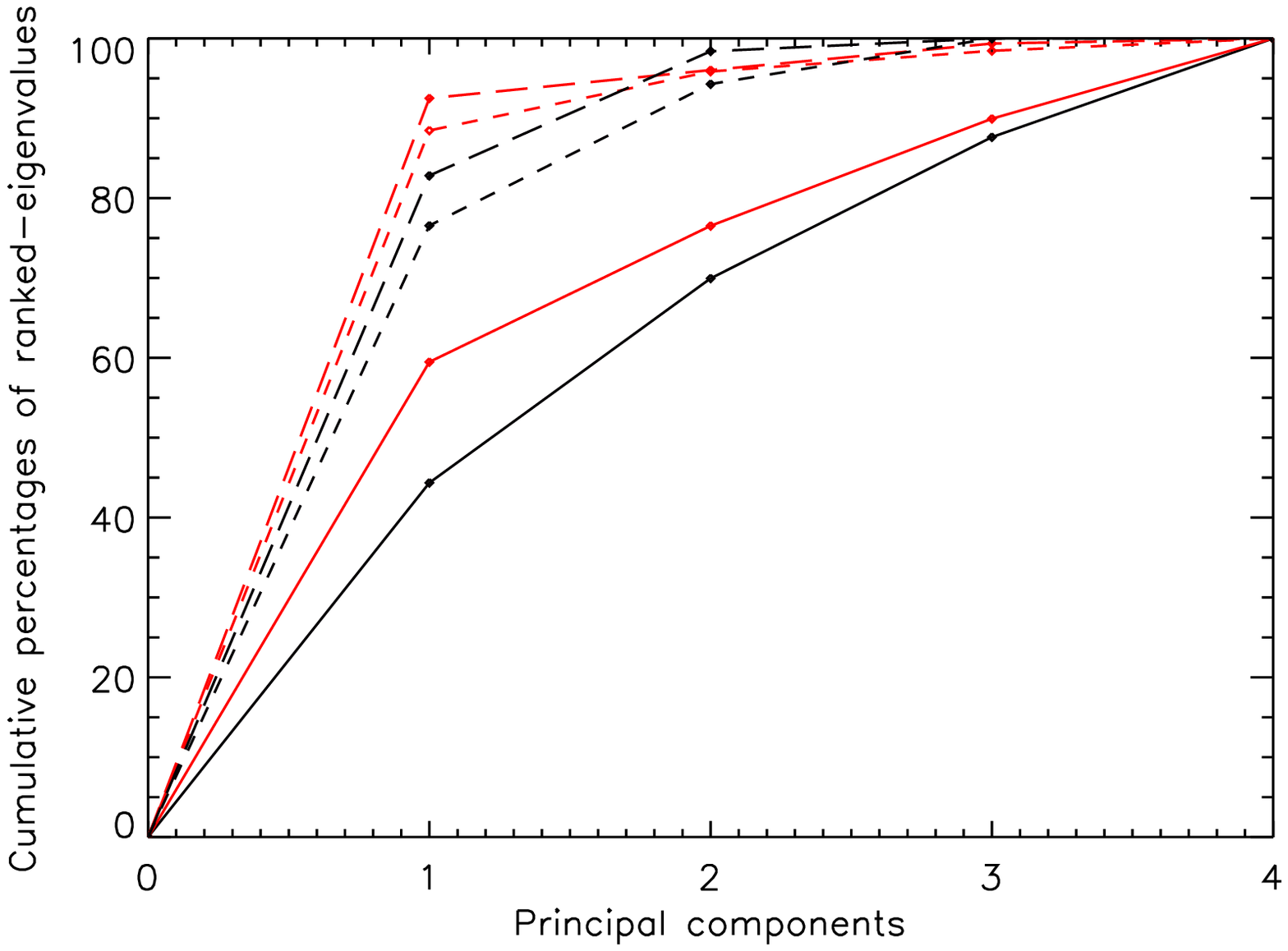}\\
\hspace{-0.53cm}\includegraphics[width=3.6in]{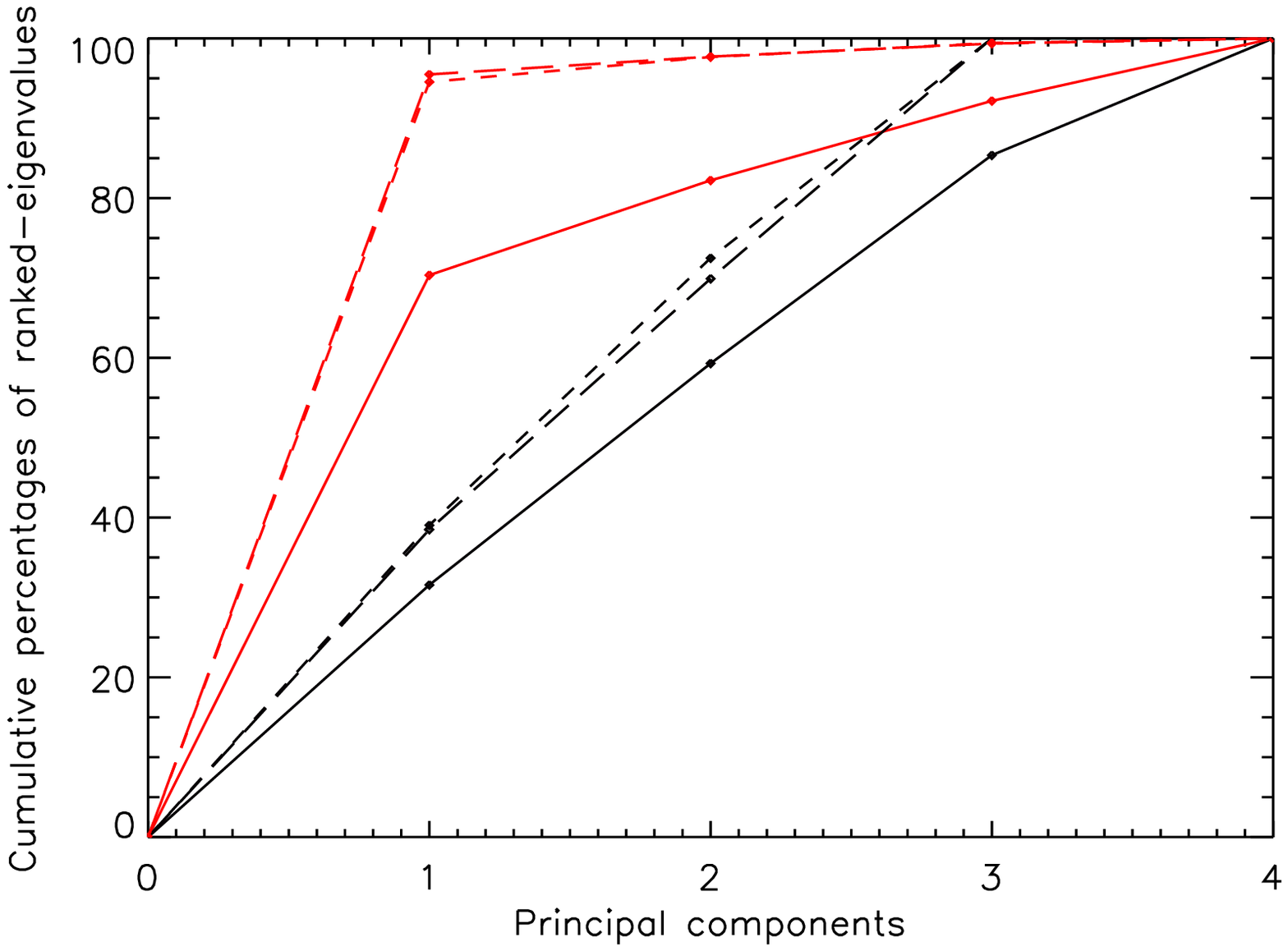}
\end{array}$
\vspace{-0.5cm}
\caption{The ranked-eigenvalues cumulative percentages for the $\alpha$-elements for the dwarf galaxies ({\it black lines}) and the solar neighbourhood ({\it red lines}) in the same metallicity range. The results for Fornax and the LMC are shown in the {\it upper plot} and {\it lower plot}, respectively. The solid lines show the observed correlation. The dashed/broken lines show the cumulative percentages derived from the estimated intrinsic correlation matrix, assuming measurement uncertainty of 0.1, 0.15 dex respectively for the dwarf galaxies and 0.05, 0.1 dex respectively for the solar neighbourhood counterparts.}\label{fig:dsph-vs-mw-alpha}
\end{figure}

First we did a PCA analysis on the $n$-capture elements (Y, Ba, Nd, and Eu). The results is shown in Fig.~\ref{fig:dsph-vs-mw-neutron}. Fornax appears to show more $\mathcal{C}$-space dimensions than the Milky Way's solar neighbourhood from the observed correlation (solid lines).  However, the extra dimension could be spurious if the measurement uncertainties for the dSph galaxy are larger. To examine this possibility, we estimated the intrinsic correlation as before and the results is over-plotted in Fig.~\ref{fig:dsph-vs-mw-neutron}.

Note that \citet{red03,red06} reported measurement uncertainty of $0.07$ -- $0.1$ dex for $n$-capture element abundances and \citet{let10} reported uncertainty of $0.15$ -- $0.25$ dex.  In the comparison with Fornax, we took the measurement uncertainty $\sigma$ for the solar neighbourhood sample to be 0.07 dex;   if we were to choose $\sigma=0.1$ dex, the PCA results for the solar neighbourhood sample in Fig.~\ref{fig:dsph-vs-mw-neutron} would further deviate from the Fornax sample. 

The simulation shows that, if we assume that the measurement uncertainty for Fornax to be $0.1$ -- $0.15$ dex,  then the Fornax $\mathcal{C}$-space has more independent dimensions than the solar neighbourhood in the same metallicity range.  However, we cannot exclude the possibility that the extra dimension is due to larger uncertainty for the Fornax sample. If we assume that the measurement uncertainty of the Fornax sample is larger than $0.2$ dex, then Fornax's dimensionality is the about the same as the solar neighbourhood. 

On the other hand, there is evidence that \citet{let10} may have overestimated their measurement uncertainty. For example, in their fig.~15, the  [Eu/Fe]--[Fe/H] and [Nd/Fe]--[Fe/H] relations are quite tight,  and the scatter of the points appears at odds with the error bar shown.   Thus, we contend that the extra apparent dimension of the $\mathcal{C}$-space for Fornax is probably not due to a larger measurement uncertainty.
\begin{figure}
$\begin{array}{c}
\vspace{-0.3cm} \hspace{-0.53cm}\includegraphics[width=3.6in]{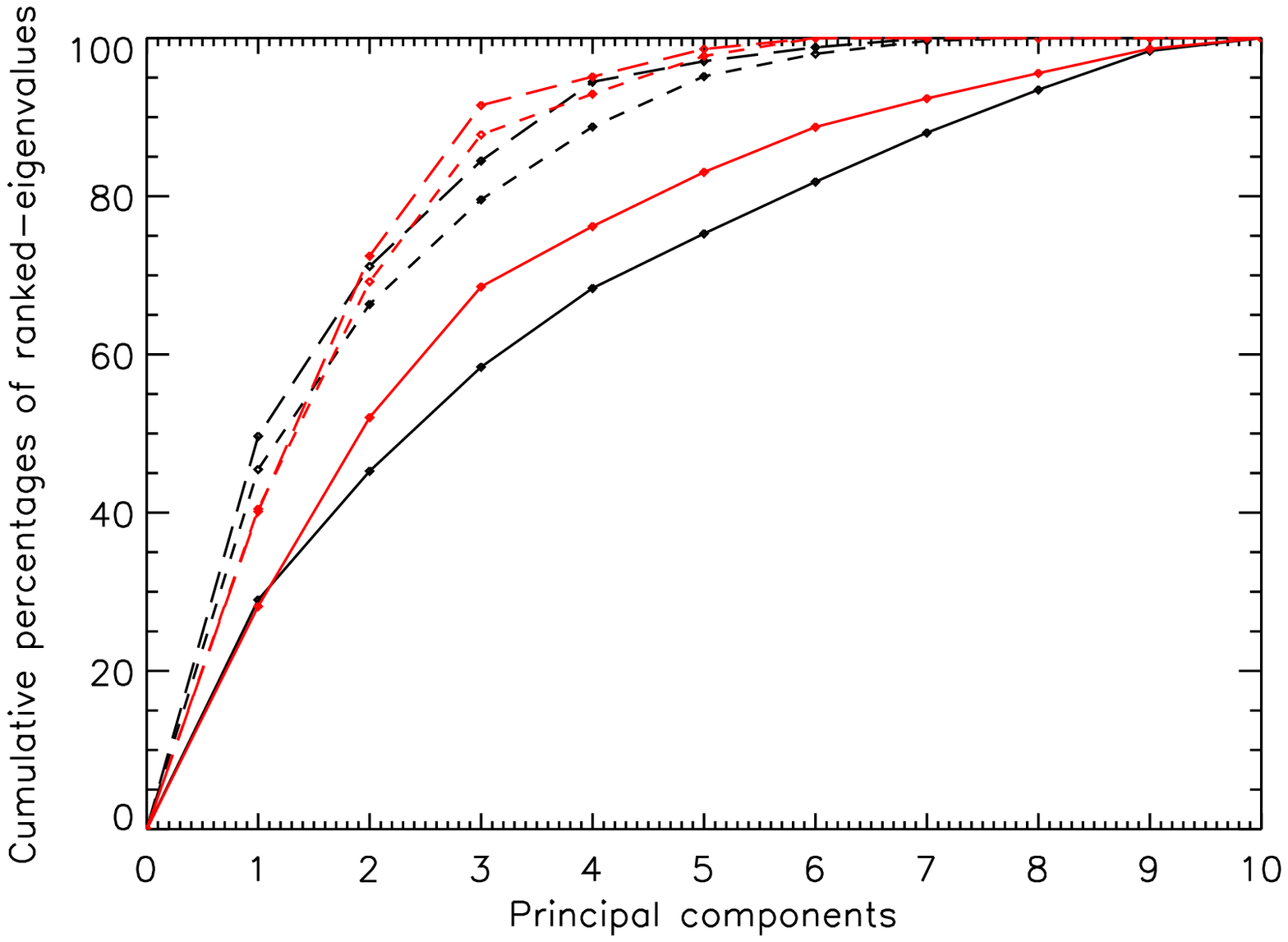}\\
\hspace{-0.53cm}\includegraphics[width=3.6in]{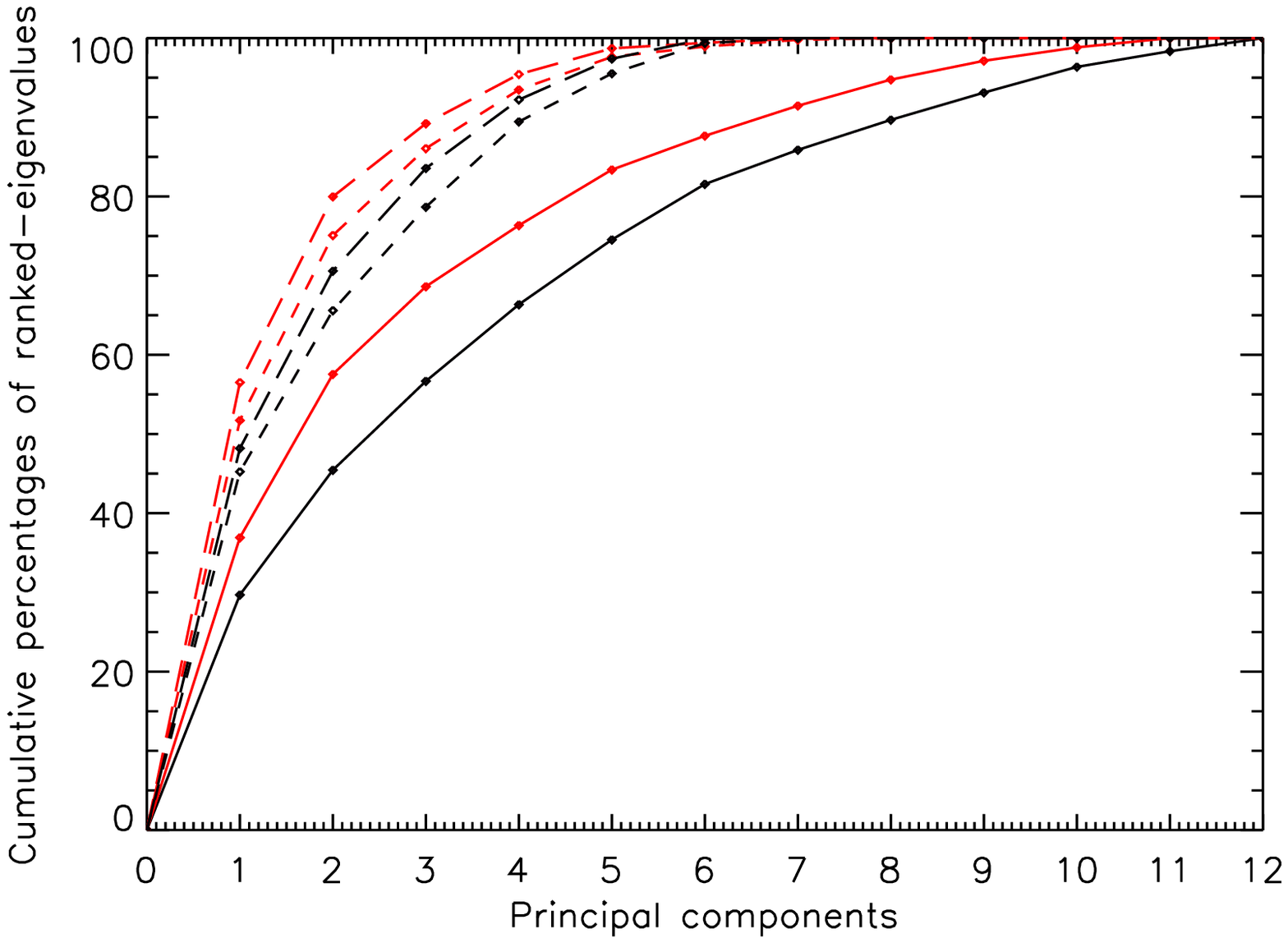}
\end{array}$
\vspace{-0.5cm}
\caption{The same as Fig.~\ref{fig:dsph-vs-mw-alpha} but for all elements in common.}\label{fig:dsph-vs-mw-all}
\end{figure}

The principal components of the PCA analysis for the Fornax $n$-capture elements (Y, Ba, La, Nd, and Eu) are shown in Fig.~\ref{fig:dsph-neutron}. The first component consists of all elements in the same sign as before. This component represents a mechanism that produces all $n$-capture elements simultaneously which probably involves the $r$-process. The second component shows an enhancement in light $s$-process elements (Y in this case) which could be the remnant of LEPP. And the third component shows enhancement in heavy $s$-process elements (Ba and La), which may be the contribution of the $s$-process in AGB stars operating at metal-poor environment. We postpone further discussion to Section~\ref{sec:discussion}.

For the LMC sample, \citet{pom08} did not measure Eu. As Eu is a crucial element to distinguish between pure $r$-process and $s$-process,  we did not study the $\mathcal{C}$-subspace of the LMC $n$-capture elements. Instead, together with the Fornax dSph galaxy, we studied the ranked-eigenvalues cumulative percentages of $\alpha$-elements (Mg, Si, Ca, Ti) and all elements measured in common: (1) for the LMC and the solar neighbourhood--  Sc, Mg, Si, Ca, Ti, V, Cr, Co, Ni, Y, Zr, and Ba; (2) for Fornax and the solar neighbourhood--  Mg, Si, Ca, Ti, Cr, Ni, Y, Ba, Nd, and Eu. The results are plotted in Fig.~\ref{fig:dsph-vs-mw-alpha} and Fig.~\ref{fig:dsph-vs-mw-all}. Both plots show that dwarf galaxies have slightly more independent $\mathcal{C}$-space dimensions than the solar neighbourhood, even though we adopt a larger measurement uncertainty for the dwarf galaxies. This is consistent with our previous discussion and results. However, to draw a firmer conclusion, dwarf galaxy samples with measurement uncertainty smaller than in the current samples are urgently needed.

\subsection[]{Globular clusters}
\begin{figure}
$\begin{array}{c}
\vspace{-0.4cm} \hspace{-0.53cm}\includegraphics[width=3.6in]{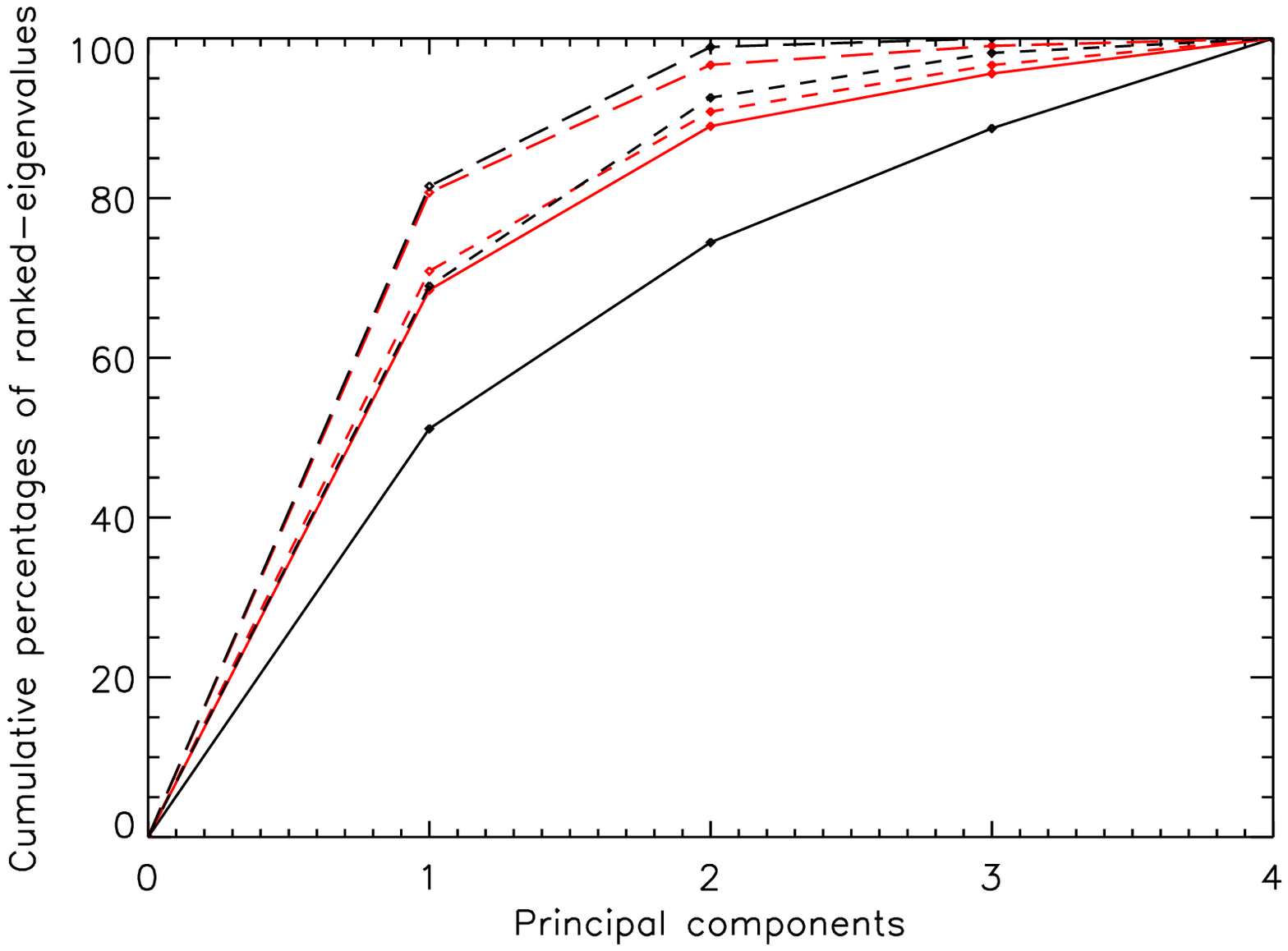} \vspace{0.35cm}\\ 
\vspace{-0.4cm} \hspace{-0.53cm}\includegraphics[width=3.6in]{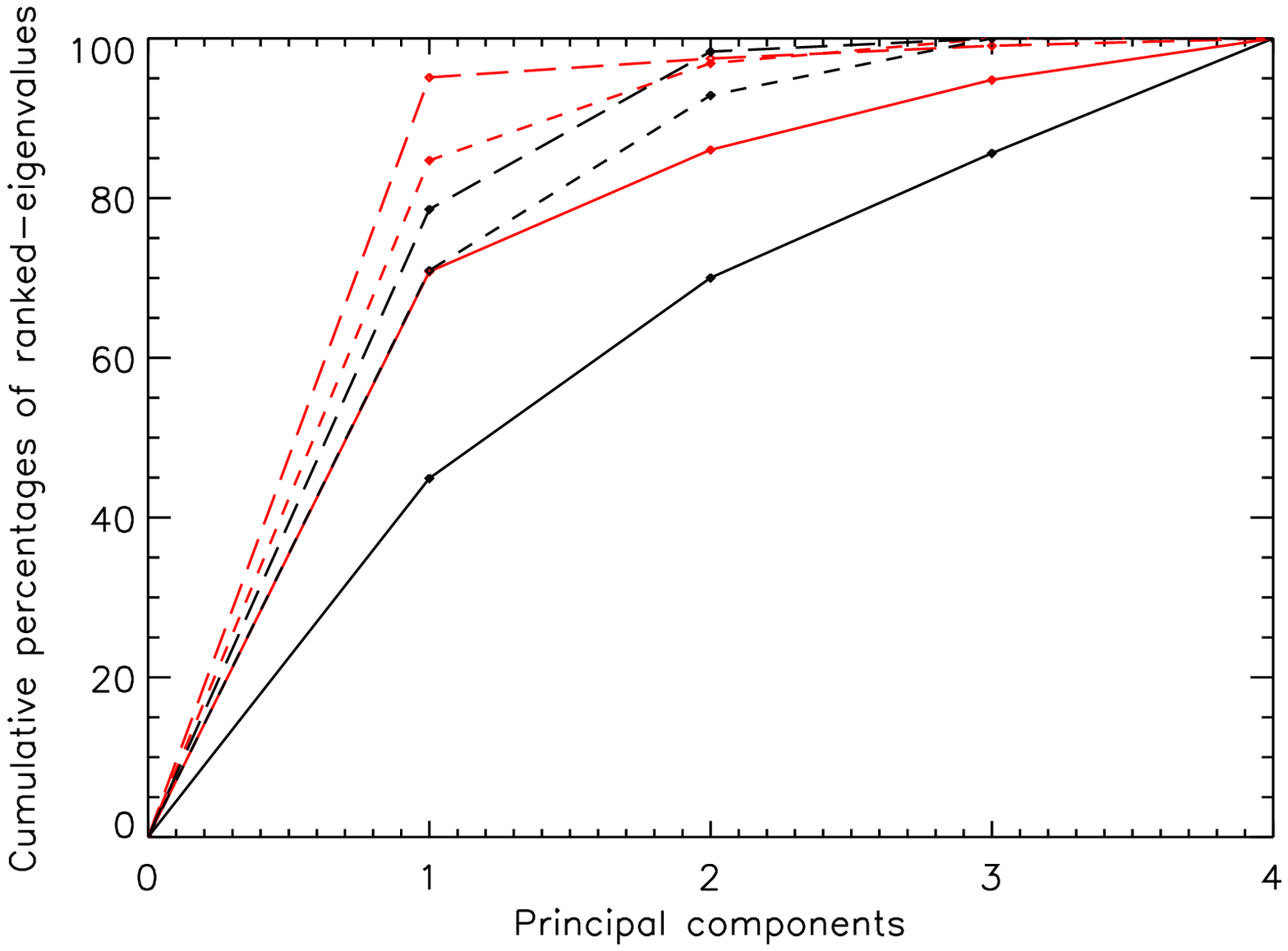} \vspace{0.35cm} \\
\hspace{-0.53cm}\includegraphics[width=3.6in]{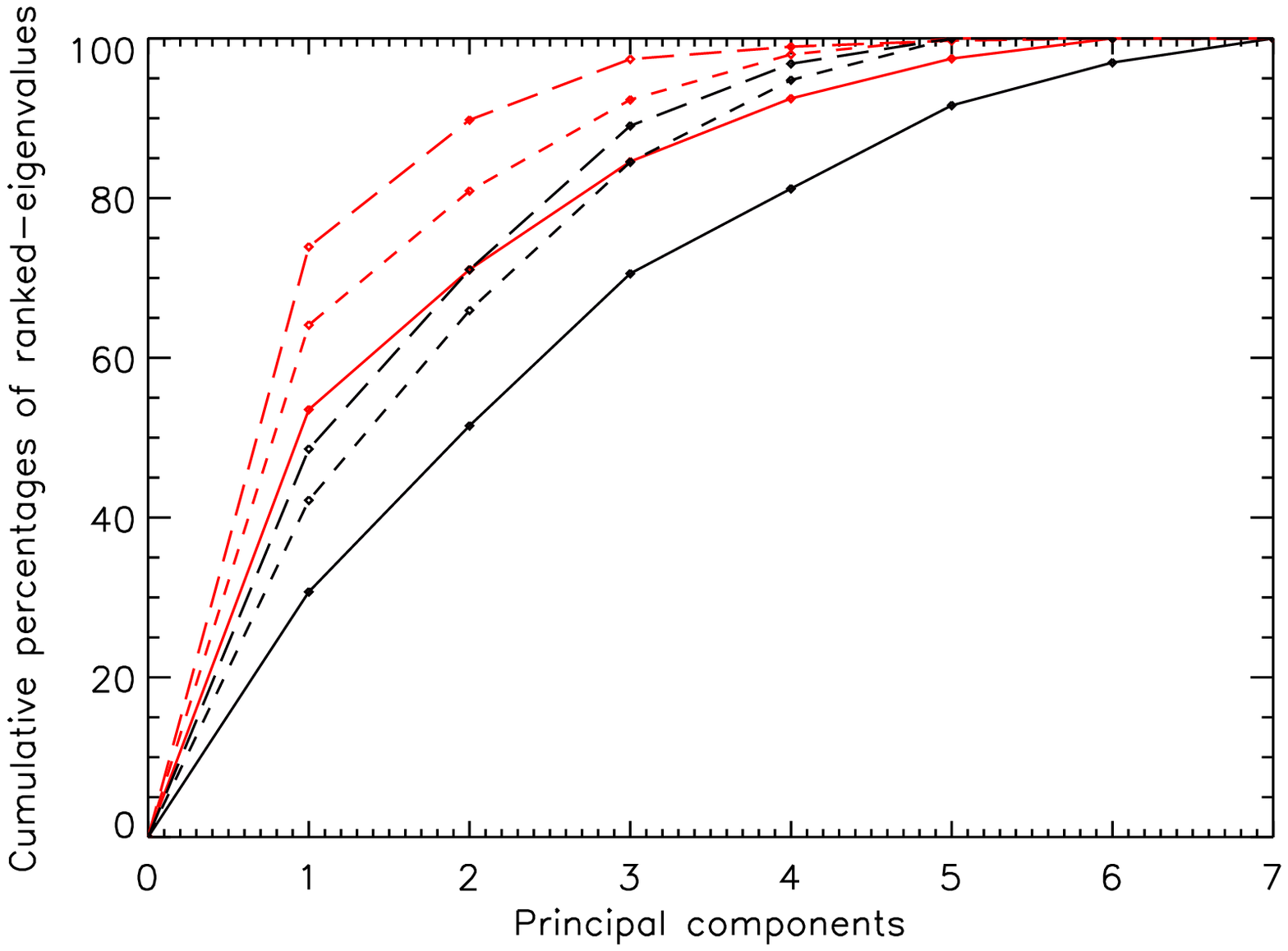}
\end{array}$
\vspace{0.1cm}
\caption{The ranked-eigenvalues cumulative percentages for the globular clusters ({\it black lines}) and the solar neighbourhood stars ({\it red lines}) in the same metallicity range: $n$-capture elements ({\it upper panel}), $\alpha$-elements ({\it middle panel}) and all elements in common ({\it lower panel}). The solid lines show the observed correlation. The dashed/broken lines show the cumulative percentages derived from the estimated intrinsic correlation matrix, assuming measurement uncertainty of 0.1, 0.2 dex respectively for globular clusters, and 0.05, 0.1 dex respectively for the solar neighbourhood stars.}\label{fig:gc-mw}
\end{figure}
Now we study the Galactic globular clusters, which are all old objects and cover a large range in Galactocentric radius. The formation process of the globular clusters remains poorly understood \citep*[for a review, see][]{gra04}, and we are interested to know what can be learned from their $\mathcal{C}$-space properties.  We use the  \citet{pri05} compilation which gives the mean abundances for each globular cluster. 

We compared $n$-capture elements (Y, Ba, La, and Eu) for globular clusters with solar neighbourhood stars from \citet{bur00}, and compared $\alpha$-elements (Mg, Si, Ca, and Ti) and then all elements in common (Mg, Si, Ca, Ti, Y, Ba, and Eu) for globular clusters with solar neighbourhood stars from \citet{ful00,ful02}, restricting all samples to the same metallicity range: $-2.5<$ [Fe/H] $<-1.0$.

The ranked-eigenvalues cumulative percentages for the $n$-capture elements are plotted in the upper panel of Fig.~\ref{fig:gc-mw}. At first sight, globular clusters ({\it black solid line}) seem to have more $\mathcal{C}$-space dimensions than the solar neighbourhood ({\it red solid line}).  As we have cautioned above, however, the sample of globular clusters is not homogeneous but rather a compilation from different authors. Therefore, we would expect more systematic uncertainty for the globular clusters sample.  We estimated the intrinsic correlation with the same method used in the open clusters and dwarf galaxies study, and the results is over-plotted in the lower panel of Fig.~\ref{fig:gc-mw}. The results show that if we assume a $0.1$ dex measurement uncertainty for solar neighbourhood stars and 0.2 dex measurement uncertainty for globular clusters, the independent dimensions of the $n$-capture element space are about the same.  On the other hand, when we examined the dimensionality for the $\alpha$-element subspace and the whole set of available elements in common to globular cluster and solar neighbourhood samples,  the dimensionality of the globular cluster space seems to be larger than for the solar neighbourhood, as shown in the middle and lower panel of Fig.~\ref{fig:gc-mw}. 

The sample of globular clusters  is small (33 clusters in the metallicity range $-2.5 <$ [Fe/H] $< -1$) and the impact of measurement uncertainty is significant. This can be seen in Fig.~\ref{fig:gc-mw} in which the estimated intrinsic correlation for globular clusters shows a significant decrease of their independent $\mathcal{C}$-space dimensions. This illustrates that the sample size is crucial to determine a robust $\mathcal{C}$-space dimensionality. We also note that our study is somewhat limited in its scope: for example, we recall the Na/O anticorrelation which is seen in globular cluster stars \citep{car09,gra10} but not among the field stars, and note that Na and O are not part of this study.

In summary,  within the limitations of the comparison,  the globular clusters appear to have slightly larger chemical space dimensionality than the solar neighbourhood stars in the same metallicity range. It is difficult to know whether this result should be expected, because we do not yet have a good perspective on what globular clusters are and how (and where) their chemical evolution took place.   A larger and homogeneous abundances survey covering more elements of globular clusters would be very desirable.
   
\section[]{Discussion}\label{sec:discussion}
\subsection[]{The $\bmath{n}$-capture elements subspace}
\subsubsection[]{The $r$-process contribution}
Several authors including \citet{hon04a,iva06} and \citet{hay09} noted that the $r$-process production ratio for the $n$-capture elements in [X/Fe] is almost constant despite the large scatter in the abundance of each element. Therefore, in the $\mathcal{C}$-space, if the major contribution is from the $r$-process, one would expect the data cloud to form a tight straight line in the $n$-capture element space.  It is pleasing to see that, at low metallicity where the $r$-process should be dominant \citep{tru81}, the PCA shows a strong first component made up of all elements positively correlated. This $r$-process component contributed  $\ga 80\%$ of the data cloud variance for the metal-poor stars. This agrees with other conclusions from the literature.

\subsubsection[]{The overabundance of light $s$-process elements}
Observations have also reported more scatter in the ratio of light $s$-process elements to heavy $s$-process elements such as [Y/Ba] \citep{bur00,nor01,iva03,aok05,hon07} at [Fe/H] $<-3$. As shown in Fig.~\ref{fig:light-s-over-abundances}, the light $s$-process elements seem to be overproduced at lower metallicity relative to the heavier elements. Various mechanisms have been proposed to explain this observation including the weak $r$-process \citep[e.g.][]{izu09} and the CPR \citep{qia07}. Our PCA method clearly show the existence of this unknown source. For the low metallicity sample (Fig.~\ref{fig:Barklem-neutron}), the second principal component is made up of light $s$-process elements anticorrelated with heavier $n$-capture elements. This suggests a second mechanism which either overproduces light $s$-process elements or overproduces heavier $n$-capture elements. From Fig.~\ref{fig:light-s-over-abundances}, the former interpretation is preferred. 

Also, as shown in Table~\ref{tab:LEPP}, the second primary source for light $s$-process elements contributes more as the metallicity decreases. In Fig.~\ref{fig:ubiquitous-r}, the First Stars Survey sample shows slightly more dimensions than the Barklem's sample. This could be interpreted as a larger contribution from the LEPP since the First Stars Survey sample has a lower mean metallicity. This decreasing trend of LEPP with increasing metallicity suggests that the culprit for LEPP is restricted to low metallicity and/or massive progenitors. If this is due to the mass dependence, this is consistent with the proposed LEPP candidates-- the weak $r$-process, CPR, collapsars or the truncated $r$-process. The relative contribution of the LEPP becomes smaller at higher metallicity and the production of $n$-capture elements is dominated by the main $r$-process and the $s$-process. The results in Table~\ref{tab:LEPP} also illustrates one of the strengths of the PCA method compared with the traditional [X/Fe]--[Fe/H] plots-- one can quantify the contributions of different mechanisms.

\subsubsection[]{Low-mass AGB contribution}
At high metallicity, one would expect a contribution from low-mass AGB stars with initial mass range of $1.5$ -- $3~\mbox{M}_\odot$. Unlike LEPP, the $s$-process in low-mass AGB stars produces both light $s$-process elements and heavy $s$-process elements such as Ba \citep[see][]{her05,kap11}, but not so much of the mostly $r$-process elements such as Nd and Eu relative to the $r$-process contribution \citep{arl99}.

In our PCA analysis, we find that the relative contribution from the first component (i.e. the $r$-process) decreases at higher metallicity (Fig.~\ref{fig:ubiquitous-r} and Fig.~\ref{fig:cumul-Reddy-neutron}). We also find that the second component is made up of both light and heavy $s$-process elements (Fig.~\ref{fig:Reddy-neutron}). Both are consistent with the $s$-process in low-mass AGB stars, which is dominant at high metallicity.

To show that this transition between high metallicity and low metallicity is real, we have also analyzed the intermediate metallicity sample taken from \citet{bur00}. We separate the the sample into two extreme metallicity bin: $-2.7<$ [Fe/H] $<-2$ and $-1.5<$ [Fe/H] $<-1$. The former shows results that resembles Fig.~\ref{fig:Barklem-neutron} and the latter shows results that is similar to Fig.~\ref{fig:Reddy-neutron}. This further confirms that the transition is indeed happening and occurs around $-2 \la$ [Fe/H] $\la -1.5$. 

\subsection[]{Satellite galaxies}
In Section~\ref{subsec:dwarf-galaxies-results}, we presented results for the dwarf galaxies samples in comparison with the solar neighbourhood stars in the same metallicity range. Recall that in this metallicity range, many stars in the solar neighbourhood are likely to be thick disk stars, and the belief is that the thick disk had a brief and intense star formation history (SFH) \citep*[see also][]{chi01,kob06}. On the other hand, the SFH is believed to be slower for dwarf galaxies \citep[e.g.][]{wei11}.

In our study, our PCA method results are consistent with the speculation that dwarf galaxies have slower SFH. We showed that dwarf galaxies seem to have more independent $\mathcal{C}$-space dimensions than the solar neighbourhood. This implies that the ISM in dwarf galaxies is impacted by a higher level of stochastic enrichment: the less vigorous history of enrichment events in the dwarf galaxies may allow more chemical traces to persist.

Now we will discuss the principal components for $n$-capture elements. Our PCA analysis shows two peculiarities of the $n$-capture elements in the dwarf galaxies. Firstly, the second principal component gives a significant contribution to the data cloud variance. We interpret the second principal component as the LEPP contribution (because it has a large contribution from light $s$-process elements).  In contrast, for the solar neighbourhood stars in the same metallicity range, the LEPP contribution is hardly visible and only contributions from the $r$-process and the $s$-process in low-mass AGB stars are seen. This is consistent with the view that the SFH in dwarf galaxies is slower and therefore retains more chemical substructure from the previous generation of ejecta.

\begin{figure}
\begin{minipage}{80mm}
$\begin{array}{c}
\hspace{-0.7cm}\includegraphics[width=3.4in]{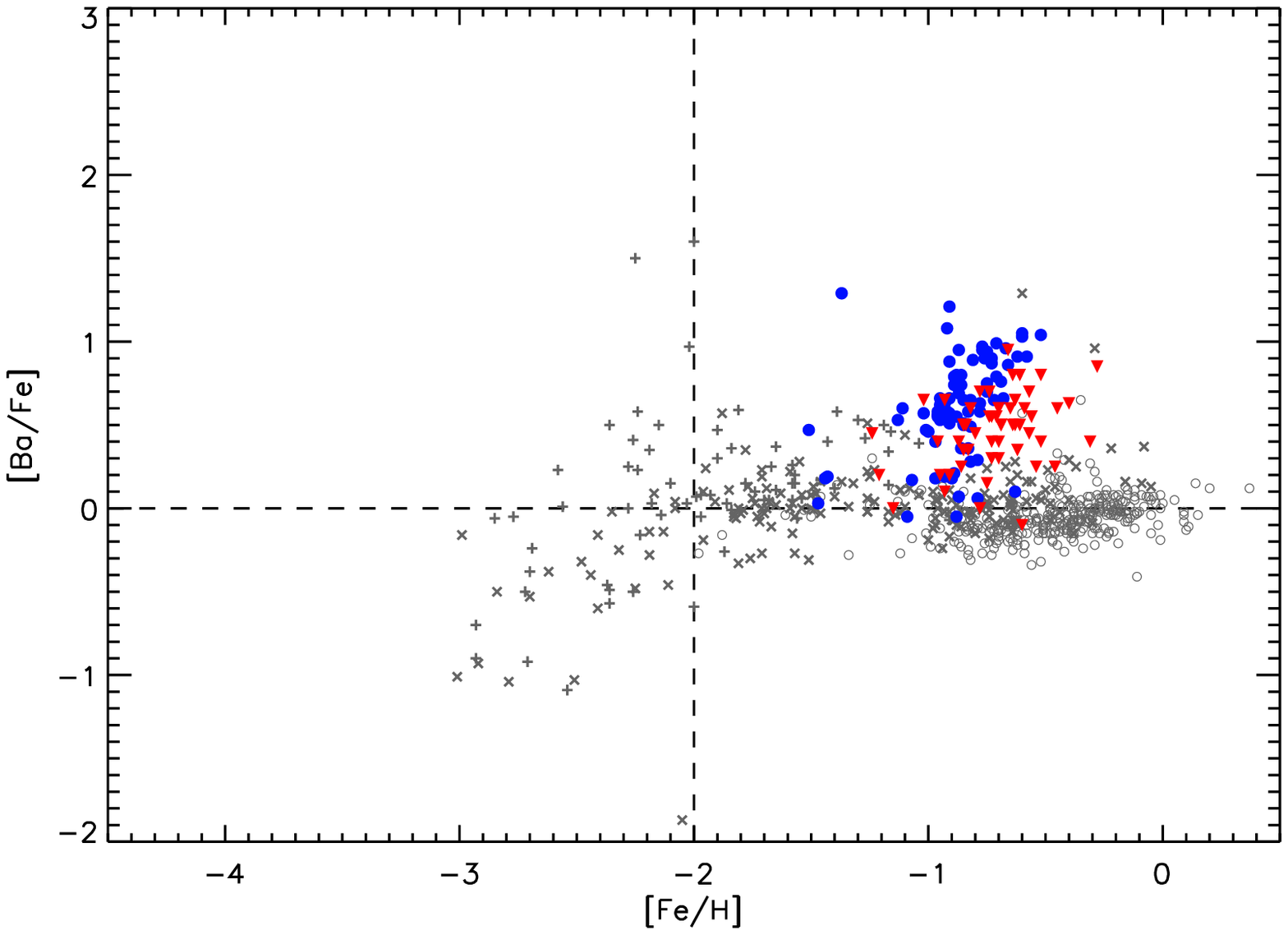}\\
\hspace{-0.7cm}\includegraphics[width=3.4in]{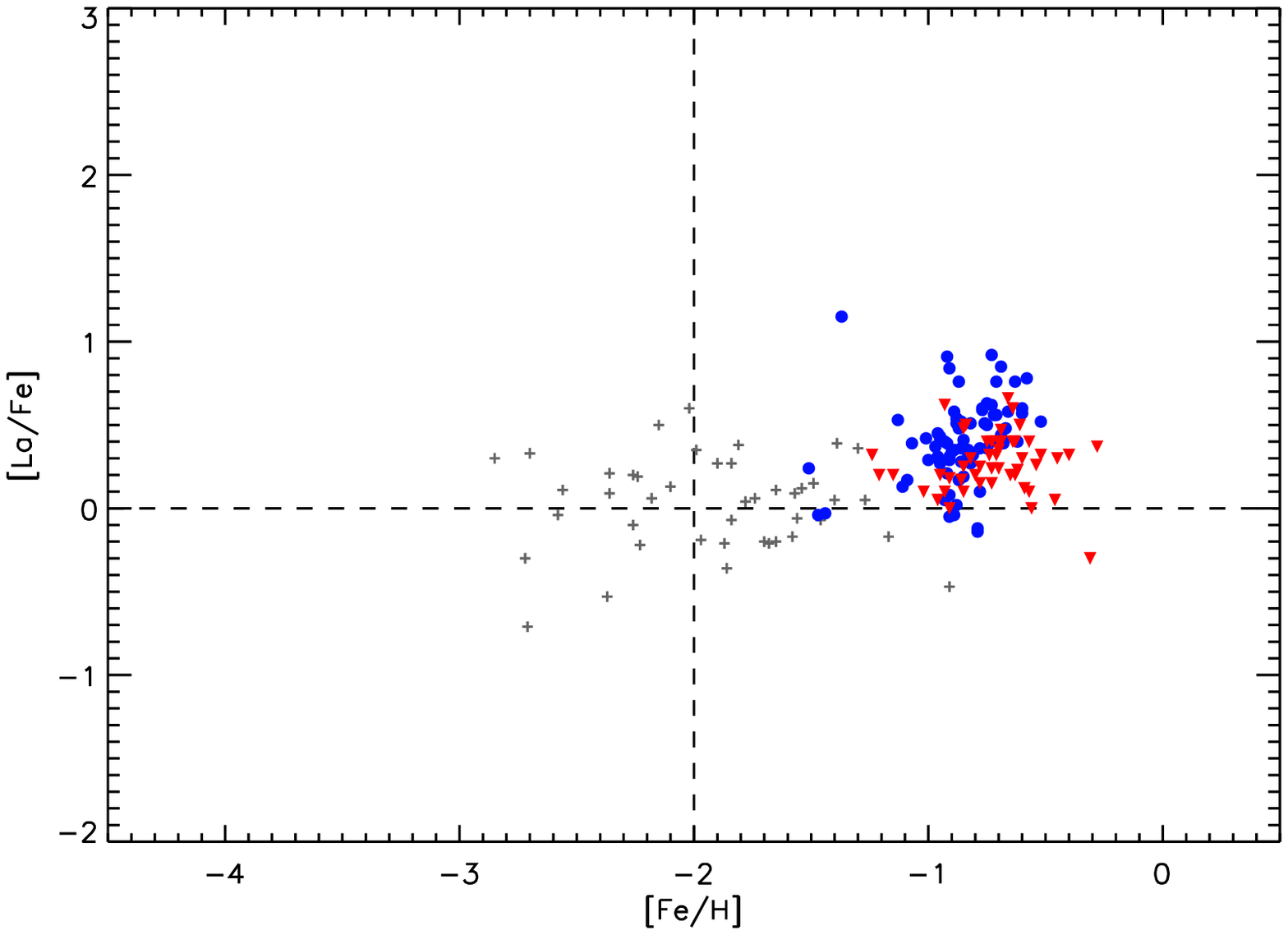}
\end{array}$
\caption{The [{\it hs}/Fe]--[Fe/H] of the solar neighbourhood, Fornax and the LMC stars. The symbols in {\it grey} represent the solar neighbourhood compilation from \citet{bur00} ({\it plus}), \citet{ful00,ful02} ({\it crosses}) and \citet{red03,red06} ({\it open circles}). The {\it blue filled circles} are Fornax stars from \citet{let10}. The {\it red filled triangles} are the LMC stars from \citet{pom08}. The two dwarf galaxies are overabundant in heavy $s$-process elements, compared to solar neighbourhood stars.}\label{fig:dwarf-light-s}
\end{minipage}
\end{figure}

\begin{figure}
\begin{minipage}{80mm}
$\begin{array}{c}
\hspace{-0.7cm}\includegraphics[width=3.4in]{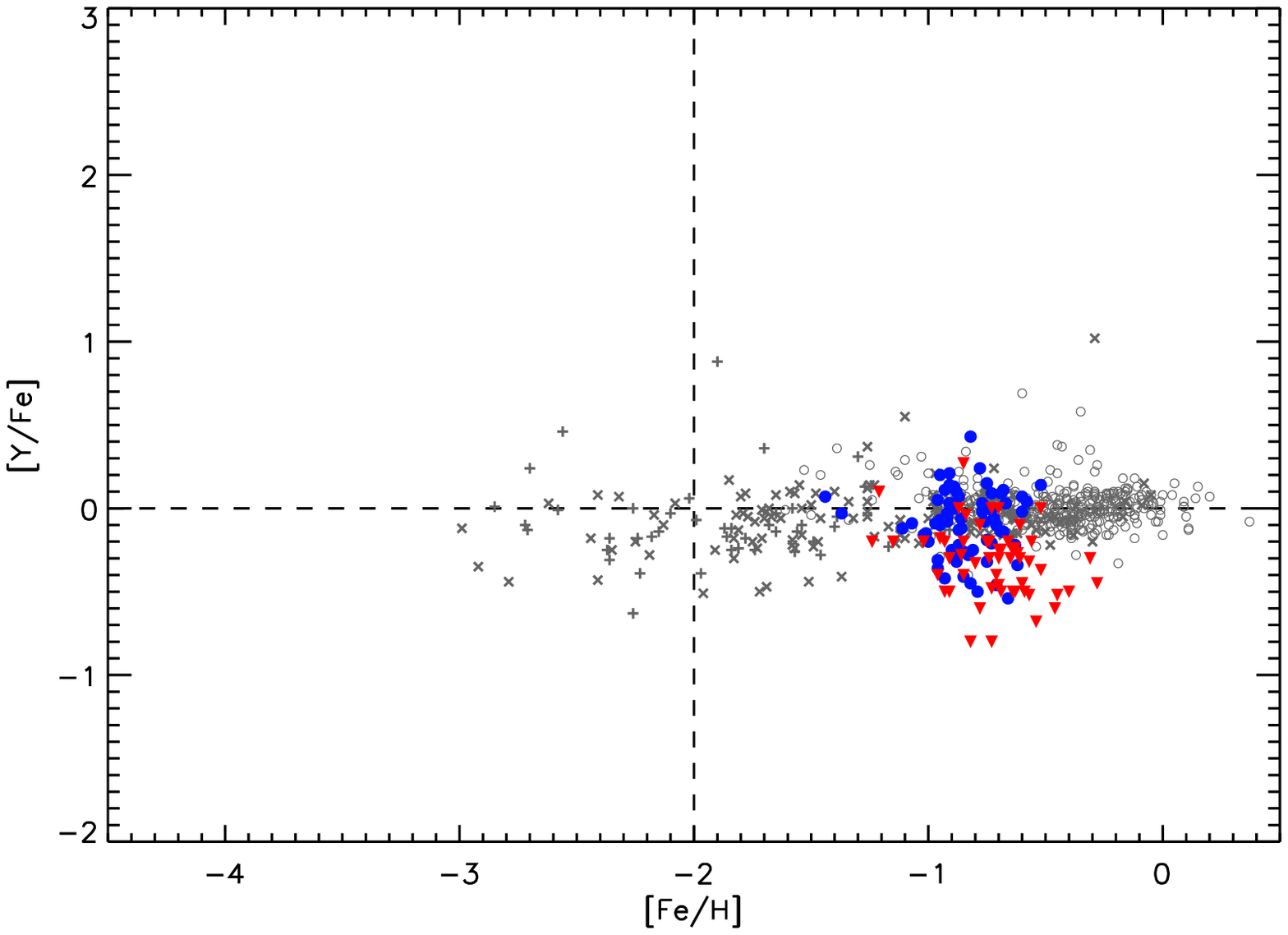}\\
\hspace{-0.7cm}\includegraphics[width=3.4in]{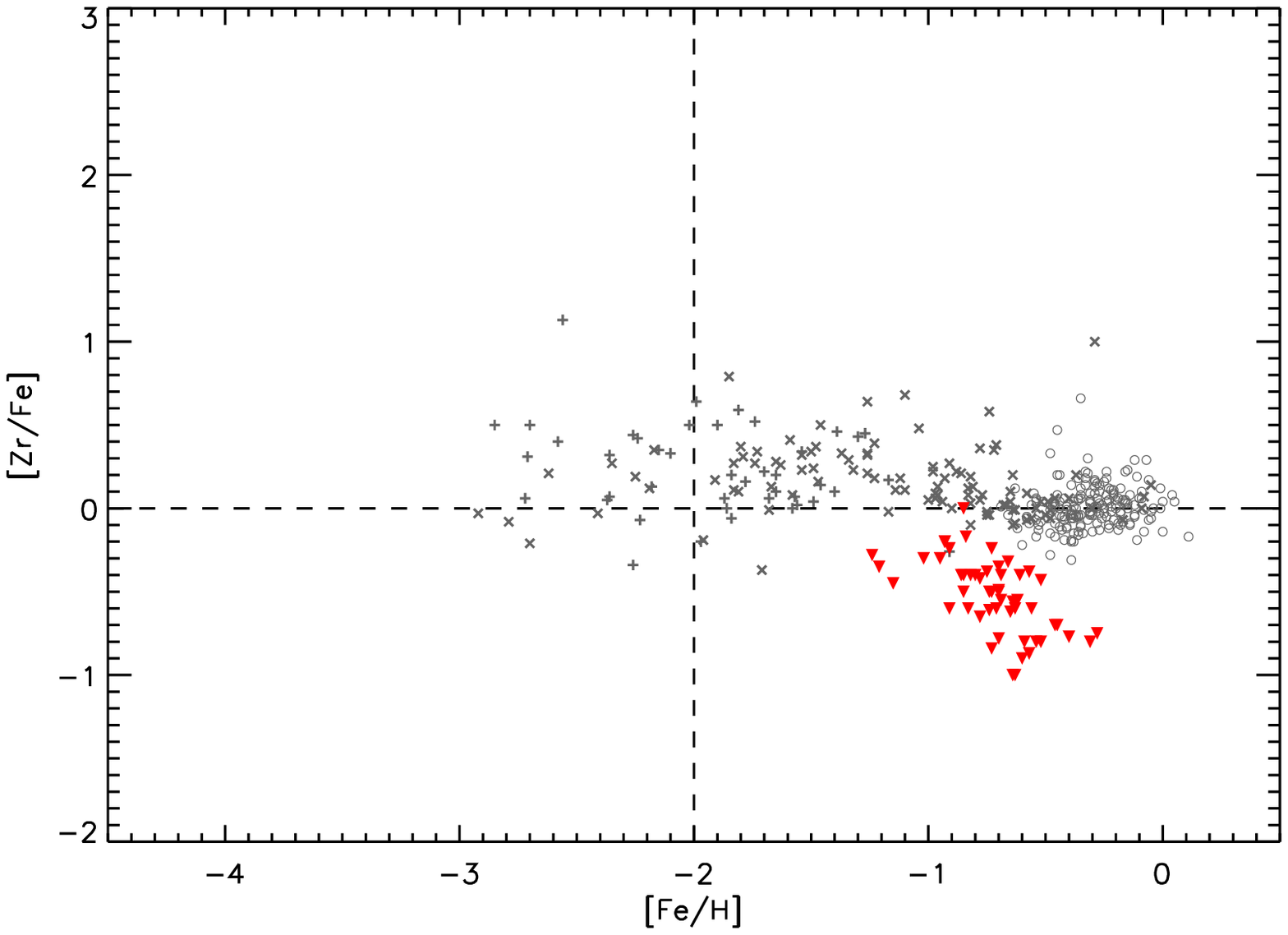}
\end{array}$
\vspace{-0.3cm}
\caption{The [{\it ls}/Fe]--[Fe/H] of the solar neighbourhood, Fornax and the LMC stars. The symbols are the same as in Fig.~\ref{fig:dwarf-light-s}. The two dwarf galaxies have similar or lower abundances of light $s$-process elements compared to solar neighbourhood stars.}\label{fig:dwarf-heavy-s}
\end{minipage}
\end{figure}

Secondly, the third principal component in this analysis was not seen before. It has heavy $s$-process elements (Ba, La) anticorrelated with the other $n$-capture elements, including the light $s$-process elements. Recall that previously at high metallicity, for $n$-capture elements in the Milky Way's solar neighbourhood, the $s$-process in low-mass AGB stars tends to produce both light and heavy $s$-process elements. If the third component represents the AGB contribution, one would expect it to have the same sign for the light and heavy $s$-process elements. However, this is not the case for the dwarf galaxies. This result is not totally unexpected. As shown in Fig.~\ref{fig:dwarf-light-s} and Fig.~\ref{fig:dwarf-heavy-s}, observations for dwarf galaxies have shown over-abundances of [{\it hs}/\mbox{Fe}] at high metallicity but similar abundances or even under-abundances for the  [{\it ls}/\mbox{Fe}].

This could be explained by the argument that the slower SFH in these galaxies allows low-mass AGB stars to contribute at lower metallicity:  AGB stars working in a metal-poor environment have higher ratio of neutrons to Fe-peak seeds and will therefore preferentially produce heavier $s$-process elements (cf. Section~\ref{sec:chemev}).

\begin{figure}
\hspace{-0.3cm}\includegraphics[scale=0.5, angle=0]{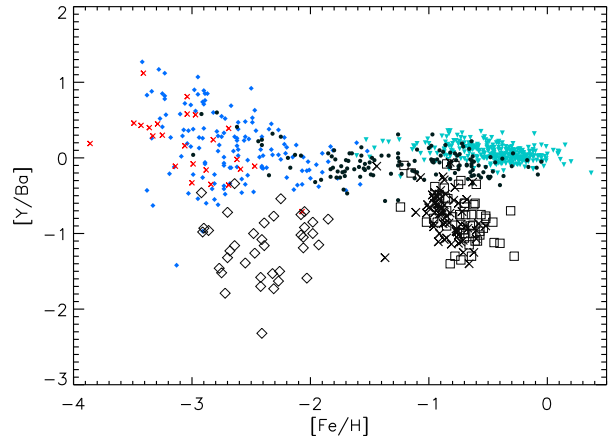}
\vspace{-0.5cm}
\caption{The Y ({\it ls}) to Ba ({\it hs}) ratios as a function of [Fe/H] for solar neighbourhood not-CEMP stars, CEMP-s stars and dwarf galaxies stars. The solar neighbourhood not-CEMP stars symbols are the same as in Fig.~\ref{fig:light-s-over-abundances}. CEMP-s stars are plotted with {\it black open diamonds}. Fornax stars are plotted with {\it black crosses} and the LMC stars are plotted with {\it black open squares}.}\label{fig:metal-poor-AGB}
\end{figure}

If this is the case, we should see some similarities between solar neighbourhood CEMP-s stars and dwarf galaxies. It has been proposed that CEMP-s stars have suffered from AGB binary mass transfer at low metallicity. AGB stars which operate at low metallicity should, by the same argument, preferentially produce {\it hs} than {\it ls}. As CEMP-s stars are rare objects, we use compilation from the SAGA database \citep{sud08}, and select stars with [C/Fe] $>1$ and [Ba/Fe] $>1$ \citep[criteria from][]{bee05}. For those stars with multiple measurements, we take the measurement from the study with the highest resolution and signal-to-noise ratio \citep[for more details, see][]{sud11}. In Fig.~\ref{fig:metal-poor-AGB}, we show the [Y/Ba] for the solar neighbourhood not-CEMP stars, CEMP-s stars and dwarf galaxies stars. This figure is consistent with our speculation. Both solar neighbourhood CEMP-s stars and dwarf galaxies stars show lower [Y/Ba] than the solar neighbourhood not-CEMP stars. This suggests that $n$-capture elements for both CEMP-s stars and dwarf galaxies could have been produced by the $s$-process in AGB stars operating in a metal-poor environment.

Alternatively, the `normal' abundances of light $s$-process element for dwarf galaxies can be explained if dwarf galaxies do not form many massive stars during their formation. This leads to a smaller contribution from LEPP and smaller production of light $s$-process elements in the formation phase of these galaxies. The under-production of light $s$-process elements from LEPP is then compensated by the onset of low-mass AGB star at low metallicities. The smaller LEPP contribution at low metallicity is not necessarily inconsistent with the PCA analysis results showing that LEPP contribution is still visible at the Fornax metallicity range ($-1.2<$ [Fe/H] $<-0.8$), as shown in Fig.~\ref{fig:dsph-neutron}, because there are also less $r$-process events to homogenize the ejecta. 

\subsection[]{All elements}
\subsubsection[]{Low metallicity}
The interpretation of the PCA analysis of all elements at low metallicity as shown in Fig.~\ref{fig:all-Barklem} can be summarized as follows:

The first principal component could be the contribution of the $r$-process since it involves substantial production of all of our $n$-capture elements. There is also substantial $\alpha$-element production together with the $r$-process production. This supports the view that the $r$-process occurs in core-collapse supernovae rather than neutron star mergers. It is generally agreed that the $\alpha$-elements at low metallicity are produced in core-collapse supernovae. If this interpretation is correct, among $\alpha$-elements, the composition of the first principal component as shown in Fig.~\ref{fig:all-Barklem} suggests that core-collapse supernovae that involve the $r$-process produce more heavier $\alpha$-elements (Ca and Ti) than light $\alpha$-elements (Mg). In the yields from \citet{kob06}, heavier $\alpha$-elements are produced more in massive progenitors and/or with higher explosion energy. Our PCA results may suggest that the $r$-process is associated with more massive progenitors and/or high explosion energy.

The second and third principal components should be discussed together. The second component shows anticorrelation between $\alpha$-elements and Fe-peak elements whereas the third component shows the production of {\it both} $\alpha$-elements and Fe-peak elements. We suggest that this could be tentatively explained by the hypernovae-supernovae model \citep{ume02,ume05,kob06}. The `normal' supernovae produce more $\alpha$-elements than iron \citep[see also][]{woo95,chi02}. However, the amounts are constrained from Galactic chemical evolution models \citep[e.g.][]{tim95, kob06}. Theoretically, since the formation of neutron stars or blackholes is uncertain, the remnant mass is determined by a mass-cut. For low-mass supernovae ($\la 20~\mbox{M}_\odot$), the ejected Fe mass is constrained to be about $0.07 M_\odot$ from the light-curves and spectra of individual supernovae. For massive supernovae ($\ga 20~\mbox{M}_\odot$), the observational data is limited, but the significant production of Fe (together with $\alpha$-elements) is supported by the light-curve and spectra. With higher explosion energy ($\ga 10 \times 10^{51}$ erg), it is possible to produce enough amounts of heavy Fe-peak elements including Zn and Co to fit the observed abundances ratios \citep{kob06}. Our PCA analysis seems to well-match this scenario: the third component could be interpreted as the contribution of hypernovae and the second component could be interpreted as the contribution from `normal' supernovae. 

We examine this idea by comparing the First Stars Survey sample which has a lower mean metallicity with the Barklem's sample. Since hypernovae are preferentially produced by more massive progenitors and therefore slightly shorter life span, \citet{kob11a} introduced metallicity dependence of the hypernovae fraction and argued that the contribution of hypernovae is larger at lower metallicity. However, the fraction of `normal' supernovae and hypernovae is uncertain. To examine these ideas, we analyze the $\mathcal{C}$-subspace of Mg, Ca, Ti, Co, Ni, Zn. In this subspace, we interpret the component that has positive contribution from all elements to represent the hypernovae contribution and the component that shows anticorrelation between the $\alpha$-elements (Mg, Ca, Ti) and heavy Fe-peak elements (Co, Ni, Zn) to be the usual supernovae contribution.  With these assumptions,  we found that for the First Stars Survey sample, the hypernovae component makes up $58.5 \%$ of the two contributions, whereas for the higher-metallicity Barklem's sample, the ratio decreases to $52.8\%$.

The $\alpha$-elements contribution appears in all of the first three components. However the second and third components do not show a visible contribution from all $n$-capture elements. This might imply that, although $r$-process production is accompanied by $\alpha$-elements yield, the reverse is not always true. If the $r$-process occurs in core collapse supernovae, then perhaps it can be triggered only if certain conditions or progenitor mass range are satisfied, as suggested by previous work \citep{wan06}. 

\begin{figure}
\begin{minipage}{80mm}
$\begin{array}{c}
\hspace{-0.65cm}\includegraphics[width=3.4in]{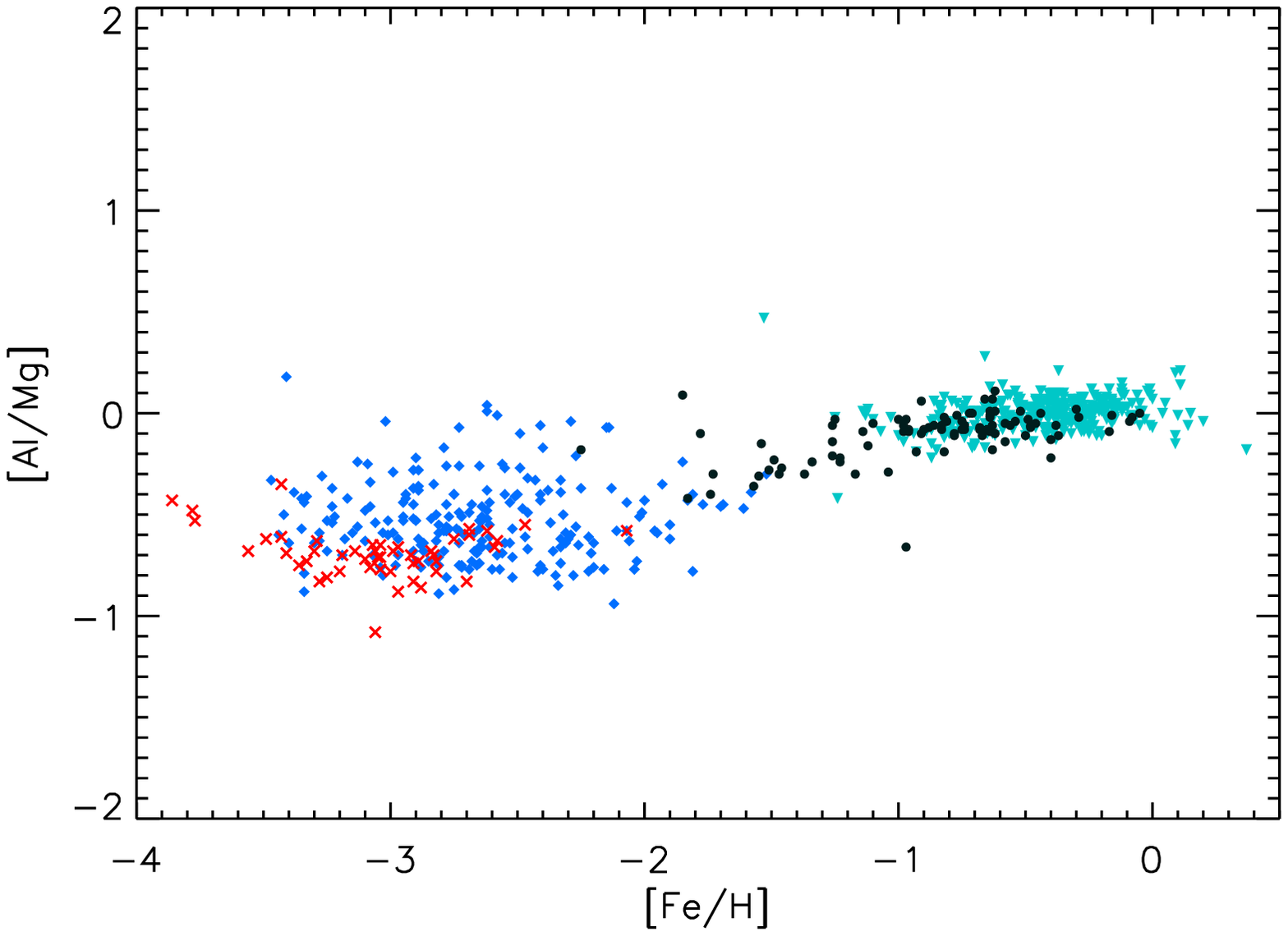} \vspace{-0.4cm}\\ 
\hspace{-0.65cm}\includegraphics[width=3.4in]{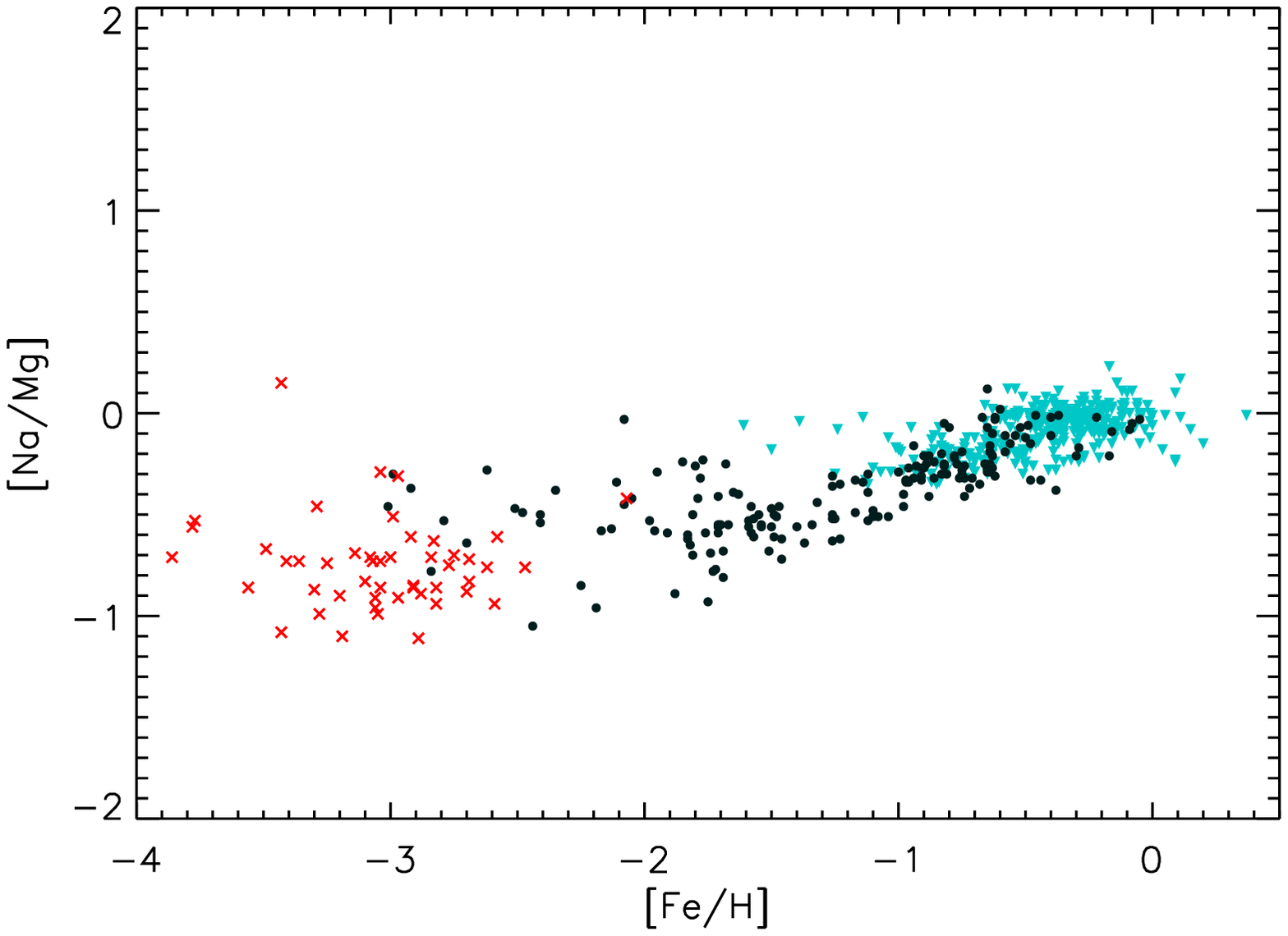}
\end{array}$
\vspace{-0.4cm}
\caption{These plot show the ratio of light odd-Z elements (Al and Na) to light even-Z elements/$\alpha$-elements (Mg) as a function of [Fe/H] of solar neighbourhood halo and disk stars. The symbols are the same as in Fig.~\ref{fig:light-s-over-abundances}. Al, Na and Mg abundances for First Stars Survey are NLTE-corrected according to \citet{and07,and08,and10} in these two plots. [Al/Fe] for Barklem's sample are NLTE-corrected by $+0.6$ dex. Both plots show an odd--even effect, i.e. light odd-Z elements are suppressed at low metallicity.}\label{fig:odd-even-effect}
\end{minipage}
\end{figure}

The Al abundance was corrected by $+0.6$ dex for NLTE effects. If this correction is not made, the results remains largely unaltered, except that Al is contributing to the second component.  Al is an light odd-Z element which comes from $^{22}$Ne, and thus the yields depends on the metallicity of progenitor stars \citep[e.g][]{kob06}; see also \citet{arn71}. Fig.~\ref{fig:odd-even-effect} shows that, relative to light even-Z elements such as Mg, light odd-Z elements such as Na and Al are indeed metallicity-dependent and are suppressed at low metallicity. This shows that the production of light odd-Z elements is indeed metallicity-dependent and is suppressed at low metallcity. Therefore, although our NLTE-correction for Al is crude and does not depend on the stellar parameters, the correction is in the correct direction and is required to give a sensible PCA result-- i.e. Al should not be contributing to the core-collapse supernovae component at low metallicity.

The fourth Cr--Mn component could be interpreted as the contribution from incomplete Si-burning, favoring the idea that the production of Cr and Mn relies on a different mass cut region as proposed by \citet{nak99}. If this interpretation is correct, this might suggest that the production ratio of elements synthesized in the incomplete Si-burning region are strongly correlated among themselves, but their production ratios are decoupled from the production of other elements such as Zn, Co, and Ni that are synthesized in the complete Si-burning region. This component should however be viewed with caution. Only \mbox{Cr\,{\sc i}} is detected in Barklem's sample, and both \mbox{Cr\,{\sc i}} and Mn are suspected to require significant NLTE correction \citep{ber08,ber10b}. Therefore this dimension could be spurious. Further discussion of this component requires the availability of a large homogeneous NLTE-corrected data set. 

\subsubsection[]{High metallicity}
Stars in this metallicity range have ages up to about $10$ Gyrs, corresponding to more than a thousand generations of core-collapse supernovae \citep[refer to fig.~8 in][]{kob09}. The PCA results for the high metallicity sample shown in Fig.~\ref{fig:all-Reddy} can be summarized as follows:

The first component is a strong component ($40 \%$ of the total variance) which consists of all light odd-Z elements, $\alpha$-elements and Fe-peak elements. This could be interpreted as the loss of chemical dimensions after many generations of short-time-scale core-collapse supernovae as shown in \citet{karl01} and \citet{bla10}. After many generations of enrichment, elements that can be formed by short-time-scale core-collapse supernovae would have been well mixed in the ISM. In short, the first component is just a measure of the metal-richness of the gas reservoirs from which the stars formed. Much information was lost after the first few generations of supernovae.

It is also interesting that Al contributes to the first component. This is different from the low metallicity sample for which the Al does not contribute. This is not unexpected. As discussed,  Al is expected to be suppressed at low metallicity, but starts to contribute as metallicity increases.

The second component can be interpreted as an $r$-process contribution, since all $n$-capture elements contribute to this component in the same sign. Unlike the low-metallicity case, $\alpha$-elements do not seem to contribute to this component. As discussed above, this may be due to the small parameter space of core-collapse supernovae that involve the $r$-process. As massive stars become fewer at high metallicity, most of the core-collapse supernovae will preferentially only produce $\alpha$-elements but not $r$-process elements. These $\alpha$-elements, as they continue to be produced and homogenized within the ISM, will only contribute to the hyperplane orthogonal to the $n$-capture elements. If this is the case, the PCA results will show a strong first component that includes supernovae ejecta and leaves the occasionally-produced $n$-captured elements to form a second component. This is also consistent with the results that the $s$-process in low-mass AGB stars starts to dominate $n$-capture elements production at high metallicity. Alternatively, the PCA results at high metallicity may imply that there is another $r$-process site that does not produce $\alpha$-elements, such as neutron star mergers \citep[e.g.][]{fre99b,ros99}

The third component includes the contribution from light and heavy $s$-process elements, and as discussed in the $n$-capture elements section,  could be interpreted as the low-mass AGB contribution. The appearance of $\alpha$-elements constribution is unclear, but could be due to the envelope mass loss releasing the natal gas with heavier elements during the AGB phase.

\subsection[]{Wider region of survey}
As shown in Fig.~\ref{fig:OC-MW-compare-all}, among the 12 elements that we investigated, compared to solar neighbourhood disk stars, we find about $1$ more dimension for open clusters which were sampled in a larger Galactic volume. The extra dimension for a larger Galactic volume is not surprising. Although stars from the solar neighbourhood could come from different regions of the galaxies via radial mixing, stars from open clusters far from the solar radius are unlikely to have undergone the same mixing processes as the solar neighbourhood stars. The larger Galactic volume ($6 \la r_G \la 20$ kpc for open clusters, compared to $7.5 \la r_G \la 8.5$ kpc for the solar neighbourhood) is likely to retain more variation in $\mathcal{C}$-space.

\subsection[]{K and Cu; \textit{APOGEE};  the Ca-triplet region}
In this section, we briefly discuss the effects of adding K and Cu to the list of elements, the dimensionality of the {\it APOGEE} $\mathcal{C}$-space, and the dimensionality of the $\mathcal{C}$-space accessible from observations of the Ca-triplet region of the spectrum.

We did not study the contribution from K and Cu, because these elements are absent from most of our samples except for Reddy's high metallicity sample. The {\it HERMES} bands are designed to include these two elements.  We found that adding K and Cu to the high metallicity sample only slightly increases the effective dimensionality-- from 6 (for 17 elements) to about 6.5 (for 19 elements)-- with K and Cu contributing equally. However, we suspect that K and Cu may add more dimensions at low metallicity because we can still distinguish contributions from different types of core-collapse supernovae. Cu is an interesting element; it is an odd Z elements preferably produced with hypernovae \citep{kob06} and may have significant contribution from the weak $s$-process \citep[like Ga \& Ge,][]{pig10}. K may be affected by aspherical explosions \citep{mae03} and also by the neutrino luminosity \citep{kob11d}. The neutrino luminosity and weak $s$-process contribution could add promising extra dimensions at low metallicity.

In this study, we focused on the {\it HERMES} $\mathcal{C}$-space.  The {\it APOGEE} high resolution near-infrared survey will measure the elements C, N, O, Na, Al, K, S, Mg, Si, Ca, Ti, V, Cr, Mn, Co, Ni but not $n$-capture elements, and we now estimate the dimensionality of this $\mathcal{C}$-space.  We exclude C, N, O, Na as before because they may be subject to internal mixing. First, we considered the 9-dimensional $\mathcal{C}$-space (Al, Mg, Ca, Ti, V, Cr, Mn, Co, Ni) measured by both Barklem's low metallicity sample and Reddy's high metallicity sample. We found that the low metallicity sample has about $4.5$ independent dimensions, whereas the high metallicity sample has $4$ independent dimensions. This is consistent with our previous observation that $n$-capture elements contribute 2 independent dimensions at high metallicity-- from $r$-process and $s$-process in low-mass AGB stars, but at low metallicity, $n$-capture elements only contribute $1$ -- $1.5$ dimensions. 
\begin{table*}
\begin{center}
\caption{Summary of effective dimensions in various environments and $\mathcal{C}$-subspace. \label{tab:summary}}
\begin{tabular}{llll}
\hline
Environment & Elements considered & Eff. dims. & No. of elements \\
\hline
& \underline{{\it Neutron-Capture Elements}} & \\
Solar neighbourhood halo stars & Y,Zr,Ba,La,Nd,Eu	& 1 -- 2  & 6 \\
Solar neighbourhood disk stars  & Y,Zr,Ba,La,Nd,Eu	& 2 & 6 \\
Fornax dSph galaxy & Y,Ba,La,Nd,Eu	& 3 & 5 \\
\\
& \underline{{\it All Elements (HERMES)}} & \\
Solar neighbourhood halo stars & Al,Sc,Mg,Ca,Ti,V,Cr,Mn,Co,Ni,Zn,Y,Zr,Ba,La,Nd,Eu	& 6 & 17 \\
Solar neighbourhood disk stars & Al,Sc,Mg,Si,Ca,Ti,V,Cr,Mn,Co,Ni,Zn,Y,Zr,Ba,Nd,Eu	& 6 & 17 \\
Solar neighbourhood disk stars & Al,K,Sc,Mg,Si,Ca,Ti,V,Cr,Mn,Co,Ni,Cu,Zn,Y,Zr,Ba,Nd,Eu	& 6 -- 7 & 19 \\
MW open clusters & Al,Sc,Mg,Si,Ca,Ti,V,Cr,Co,Ni,Y,Ba,Nd	& 6 -- 7 & 13 \\ 
MW globular clusters & Mg,Si,Ca,Ti,Y,Ba,Eu & 4 & 7 \\ 
Fornax dSph galaxy & Mg,Si,Ca,Ti,Cr,Ni,Y,Ba,Nd,Eu	& 6 & 10 \\
The Large Magellanic Cloud & Sc,Mg,Si,Ca,Ti,V,Cr,Co,Ni,Y,Zr,Ba & 6 -- 7 & 12 \\
\\
& \underline{{\it All Elements (APOGEE)}} & \\
Solar neighbourhood halo stars &  Al,Mg,Ca,Ti,V,Cr,Mn,Co,Ni & 4 -- 5 & 9 \\
Solar neighbourhood disk stars &  Al,Mg,Ca,Ti,V,Cr,Mn,Co,Ni & 4 & 9 \\
Solar neighbourhood disk stars &  Al,K,S,Mg,Si,Ca,Ti,V,Cr,Mn,Co,Ni & 5 & 12 \\
\hline
\end{tabular}
\end{center}
\flushleft
$^*$ We assume measurement uncertainty of $0.1$ dex and adopt $85 \%$ cut-off for the ranked-eigenvalues cumulative percentages to derive the effective dimensions. This might not hold for open/globular clusters and dwarf galaxies as they have higher measurement/systematic uncertainty. Therefore the dimensions for open/globular clusters and dwarf galaxies in this table should be regarded as upper limit
\end{table*}

In addition to these 9 dimensions, Reddy's sample also included Si, S and K. We found that adding Si alone does not add to the effective dimensionality, as Si is strongly correlated with the other $\alpha$-elements. However, we found that including S and K will add around $1$ additional dimension to the $\mathcal{C}$-space. Finally we summarize the effective dimensions of the $\mathcal{C}$-space (and subspaces) of various environments in Table~\ref{tab:summary}.  

The Ca-triplet region is widely used for large-scale spectroscopic surveys, so we also estimated the dimensionality of the $\mathcal{C}$-space defined by this region of the spectrum.  For the metal-rich stars, the elements Al, Mg, Si, Ca, Ti and Ni define a space with 2 independent dimensions.  For the metal-poor stars, the elements Al, Mg, Ca, Ti and Ni define a space with 3 independent dimensions. Note that this estimate excludes possible extra dimensions from the light elements C, N and O.

\section[]{Conclusion}\label{sec:conclusion}
In this paper, we used Principal Component Analysis to study and interpret stellar element abundances. We discussed a way to deal with the non-semi-definite positivity of the correlation matrix due to the incompleteness of the data.

We illustrated the power of this method by confirming the tight yield ratio of $n$-capture elements from the $r$-process, which confirms the universality of the $r$-process. We also traced the over-abundances of light $s$-process elements at low metallicity, and found that the relative contribution of the mechanism causing this over-abundances is decreasing with increasing metallicity. The site of LEPP is likely to be associated with massive stars, which is consistent with weak $r$-process and charged particle reactions. The method is also able to trace the $s$-process in low-mass AGB stars at high metallicity. The transition can be clearly shown using data with intermediate metallicity and occurs at $-2 \la$ [Fe/H] $\la -1.5$. 

Our analysis also suggests that $r$-process production sites are accompanied by $\alpha$-element production, favouring core-collapse supernovae as the $r$-process site. This can be seen at low metallicity but is no longer visible at high metallicity, which may be due to the small parameter space of the $r$-process site, or the inclusion of another site such as neutron star mergers that does not produce $\alpha$-elements.

The analysis indicates two types of core-collapse supernovae: one produces mainly $\alpha$-elements, the other produces both $\alpha$-elements and Fe-peak elements with a large enhancement of heavy Fe-peak elements. This is consistent with hypernovae. We find that the contribution of hypernovae is larger at lower metallicity, which may be important to understand the physics of hypernovae. We also discussed the Cr and Mn production from the incomplete Si burning region, and the metallicity dependence of odd-Z element (Al in our analysis), but these may be affected by NLTE effect.

To estimate the dimensionality of the C-space available to HERMES, we chose to work in [X/Fe] space, because all of the elements are highly correlated with [Fe/H].  We based our estimate of the dimensionality of [X/Fe] space on the simulations described in Section~\ref{sec:best-cut-off-4-PCA}, indicating that the eigenvectors contributing to the first 85\% of the cumulative sum of the ranked eigenvectors in [X/Fe] space are probably real. Due to the paucity of data and the potential contribution from internal mixing, we did not include Li, C, N, O and Na in our study. The PCA analysis of 17 elements shows that the chemical [X/Fe] space has about 6 dimensions both at high metallicity and low metallicity in the solar neighbourhood. We expect to have about 1 to 2 further independent dimensions of the [X/Fe] space from these elements that we excluded.  Adding K and Cu for the HERMES sample provides about another half of a dimension for the [X/Fe] space. Our ultimate goal here is to evaluate the dimensionality of the 25-element HERMES space.  Including the further dimension from [Fe/H] itself would give about 8 to 9 independent dimensions for the HERMES chemical [X/Fe]+[Fe/H] space.

Although the number of $\mathcal{C}$-space dimensions is similar at high and low metallicity, the interpretations of the principal components are very different in these two cases. For example, at high metallicity, we have an extra `birth-imprint' from low-mass AGB stars. This extra contribution compensates the dimension loss due to homogenization of the core-collapse supernovae ejecta. 

Our analysis indicates that dwarf galaxies retain more chemical inhomogeneity than the Milky Way disk. The analysis also suggests that the $s$-process in AGB stars of dwarf galaxies preferentially produce heavier $n$-capture elements which is similar to the solar neighbourhood CEMP-s stars. Both of these effects may be  produced by AGB stars working in metal-poor environment. These findings are consistent with the view that dwarf galaxies have had a slower SFH than the Milky Way disk.

The $\mathcal{C}$-space for Galactic globular clusters and open clusters, which span a large Galactic volume, appear to define a slightly higher-dimensional $\mathcal{C}$ space than that for the solar neighbourhood.  A large homogeneous sample of abundances in globular clusters and open clusters is highly desirable to confirm this conclusion.

We showed that PCA can shed some insight on the underlying nucleosynthesis mechanisms. At low metallicity where the chemical space is less homogenized, this method could be exploited to put constraints on the number of mechanisms or, more precisely, on the number of parameters that govern Galactic chemical evolution in different environments.  

For our future observational work in chemical tagging, the PCA method could combine with group-finding algorithms \citep{agr98,sha09} to increase the efficiency of the algorithms for finding substructure in chemical space. The group-finding algorithms are sensitive to redundant dimensions which decrease the density of the clusters in chemical space. We expect that our PCA method can reduce the effective dimensionality of the chemical space without compromising any useful information.

\section*{Acknowledgments}
We thank Martin Asplund for careful reading of the manuscript, Amanda Karakas, Richard Stancliffe, David Yong, Christophe Pichon, Piercarlo Bonifacio, Anna Frebel, John  Norris and Sanjib Sharma for comments and suggestions. We are grateful to Ricardo Carrera, Elena Pancino and John Fulbright for making their samples available for this study. Y.S.T. is grateful to the College of Physical and Mathematical Sciences and the Research School of Astronomy and Astrophysics at The Australian National University for their financial support throughout this project.

\appendix

\section[]{Principal Component Analysis}\label{appendix:PCA}
Consider an $n$ dimensional space with random variables $(X_1, \ldots, X_n)$. In this section, we will re-derive and show that the normalized eigenvector corresponding to the largest eigenvalue of the correlation matrix is the direction where the standardized (i.e. normalized and mean-shifted) data cloud has the largest variance. Suppose we have $m$ data points such that the random variables value of the $q$-th data point is $(X_{1,q}, \ldots, X_{n,q})$. Let $\overline{X_q}$, $\sigma_q$ be the mean and the standard deviation of the random variable $X_q$ respectively, and $\mathbfss{B}$ be the matrix such that the $p$-th column corresponds to the standardized $p$-th data point, i.e. $(\frac{X_{1,p}- \overline{X_1}}{\sigma_1}, \ldots, \frac{X_{n,p}-\overline{X_n}}{\sigma_n})$. One can show that $\mathbfss{C} = \mathbfss{B} \, \mathbfss{B}^T$ is the correlation matrix. 

Our goal is to find the normalized vector $\bmath{u}$ centered at $(\overline{X_1}, \ldots, \overline{X_n})$ such that the projection of the standardized data cloud on $\bmath{u}$ has the maximum variance. The $p$-th standardized data point projection on $\bmath{u}$ is $\Big( \frac{X_{1,p}- \overline{X_1}}{\sigma_1}, \cdots, \frac{X_{n,p}- \overline{X_n}}{\sigma_n} \Big) \cdot \bmath{u}$, so $\mathbfss{B}^T \, \bmath{u}$ is the ensemble of $\bmath{u}$-projection of each standardized data point. One can check that the mean of $\mathbfss{B}^T \bmath{u} =0$, and therefore the variance of the standardized data cloud projected on $\bmath{u}$ is $ (\bmath{u}^T \mathbfss{B})\cdot (\mathbfss{B}^T \bmath{u})/m $. To summarize, our goal is to maximize this value under the constraint $\bmath{u}^T \bmath{u} = 1$. Without lost of generality, we will maximize $(\bmath{u}^T \mathbfss{B})\cdot (\mathbfss{B}^T \bmath{u})$ instead of $(\bmath{u}^T \mathbfss{B})\cdot (\mathbfss{B}^T \bmath{u})/m$. 

We use Lagrange multiplier formalism to deal with the constraint by introducing Langrange multiplier $\alpha_1$. In this formalism, the goal is to maximize the expression as shown in Equation~\ref{eq:lagrangian-1}
\begin{eqnarray}
\label{eq:lagrangian-1}
L_1(\bmath{u}, \alpha_1) = (\mathbfss{B}^T \bmath{u})^2 - \alpha_1 (\bmath{u}^2 -1)
\end{eqnarray}

\noindent
By using the necessary condition of the maximality, i.e. $\frac{\partial L_1}{\partial \bmath{u}}= \frac{\partial L_1}{\partial \alpha_1}=0$, one can show that $\bmath{u}$ has to be a eigenvector of $\mathbfss{C}$ and $\bmath{u}^2=1$. Recall that $\mathbfss{C}$ is symmetric and therefore orthogonally diagonalizable. Let the diagonal matrix to be $\mathbfss{D}= \mbox{diag}(\lambda_i)$, i.e. $\bmath{u}^T \mathbfss{C} \bmath{u} = \bmath{u}^T \mathbfss{P}^T \mathbfss{D} \mathbfss{P} \bmath{u}$ for some orthonormal matrix $\mathbfss{P}$. Since $\mathbfss{P}$ is orthonormal and $\bmath{u}^2 =1$, it is easy to check that $(\mathbfss{P} \bmath{u})^2 = 1$. Let $\lambda_1$ be the largest eigenvalue. One has
\begin{eqnarray}
\label{eq:variance}
\max_{||\bmath{u}|| =1} \bmath{u}^T \mathbfss{P}^T \mathbfss{D} \mathbfss{P} \bmath{u} = \max_{||\bmath{v}|| =1} \bmath{v}^T \mathbfss{D} \bmath{v} = \lambda_1
\end{eqnarray}

This can be attained by choosing $\bmath{u}$ to be the eigenvector of $\mathbfss{C}$ corresponding to the eigenvalue $\lambda_1$. And thus it is the sufficient condition. Similarly, to find the second principal component $\bmath{w}$ such that it is orthogonal to the first principal component $\bmath{u}$ (i.e with constraint $\bmath{w}^T \cdot \bmath{u} = 0$) and accounts for the largest part of the rest of the variance, it suffices to maximize the expression as shown in Equation~\ref{eq:lagrangian-2}:
\begin{eqnarray}
\label{eq:lagrangian-2}
L_2(\bmath{w},\beta_1,\beta_2) = (\mathbfss{B}^T \bmath{w})^2 - \beta_1(\bmath{w}^2 -1 ) - \beta_2 (\bmath{w}^T \cdot \bmath{u})
\end{eqnarray}

By similar calculation, one can show that $\bmath{w}$ being the eigenvector corresponding to the second largest eigenvalue is necessary and sufficient to maximize this expression, and so forth for the subsequent eigenvectors. Graphically, we are looking for a orthogonal transformation of the random variables space such that after the transformation, the first axis will account for the largest part of the total variance, and second axis is orthogonal to the first axis and account for the largest part of the rest of the variance. It is important to note that the variances that they account, are given by the eigenvalues of the correlation matrix as shown in Equation~\ref{eq:variance}.

\section[]{Incomplete data set}\label{appendix:incomplete-data-set}
If a data set is incomplete, in principle we can still calculate the Pearson's correlation for any two random variables by using only the data points that have value for both random variables, and therefore we can construct the correlation matrix entry by entry. However the problem of this approach is obvious: since the correlation matrix $\mathbfss{C}$ is not $\mathbfss{B} \mathbfss{B}^T$ as before, although it is still symmetric, it might not be semi-positive-definite, i.e. it might have undesirable negative eigenvalues. Our goal is to find a semi-positive-definite matrix that is close to the correlation matrix $\mathbfss{C}$. \citet{reb99} suggested the following: 

Let $\mathbfss{S}$ to be the ensemble of eigenvectors of matrix $\mathbfss{C}$, i.e. $\mathbfss{C} \cdot \mathbfss{S} = \bmath{\Lambda} \cdot \mathbfss{S}$, where $\bmath{\Lambda} = \mbox{diag}(\lambda_i)$, and $\lambda_i$ the eigenvalues. If $\mathbfss{C}$ is not semi-positive-definite, it has at least one negative eigenvalue. We define the positive diagonal matrix $\bmath{\Lambda'}\equiv\mbox{diag}(\lambda'_i)$:
\begin{eqnarray}
 \bmath{\Lambda'} \quad : \quad \lambda'_i = \left\{ 
  \begin{array}{l l}
    \lambda_i & \quad \mbox{if $\lambda_i \geq 0$}\\
    0 & \quad \mbox{if $\lambda_i < 0$}\\
  \end{array} \right.   
\end{eqnarray}

\noindent
and the diagonal `scaling' matrix $\mathbfss{T}\equiv \mbox{diag}(t_i)$:
\begin{eqnarray}
\mathbfss{T} \quad : \quad t_i = \Big[ \sum_m s_{im}^2 \lambda'_m \Big]^{-1}
\end{eqnarray}

Let $\mathbfss{B'} \equiv \sqrt{\mathbfss{T}} \mathbfss{S} \sqrt{\bmath{\Lambda'}}$, where the square root of a diagonal matrix is defined as the square root of each of its diagonal entry. Finally we define: $\mathbfss{C'} \equiv \mathbfss{B'} \mathbfss{B'}^T$. One would expect $\mathbfss{C'}$ to be quite close to $\mathbfss{C}$ since $\mathbfss{C} = \mathbfss{S}^T \, \bmath{\Lambda} \mathbfss{S}^T$ and $\mathbfss{C'} = \sqrt{\mathbfss{T}} \mathbfss{S} \bmath{\Lambda'} \mathbfss{S} \sqrt{\mathbfss{T}}$. The lost from $\bmath{\Lambda} \rightarrow \bmath{\Lambda'}$ is compensated by the rescaling matrix $\mathbfss{T}$. There are better ways to optimize the search of $\mathbfss{C'}$ but they are mostly computational much more demanding than this method. In our case, this estimation is good enough since it gives reasonable small errors both in term of  $\epsilon_1 \equiv \sum_{ij} (C_{ij} - C'_{ij})^2$ and $\epsilon_2 \equiv \sum^n_{i=1} ( \lambda_i - \lambda'_i )^2$, where $\lambda'_i$ are eigenvalues of $\mathbfss{C'}$.

\section[]{Weighted total least square}\label{appendix:weight-TLS}
This method is adopted from \citet{kry07}. As discussed in Section~\ref{sec:estimate-intrinsic-scatter}, our goal is to minimize Equation~\ref{eq:minimize-WTLS}. Instead of considering the best fit line $y= ax+b$ using variables $a$ and $b$, \citet{kry07} suggested a change of variable $\underset{(a,b) \longmapsto (\alpha,p)}{\mathbb{R}^2 \rightarrow (-\frac{\pi}{2}, \frac{\pi}{2}) \times \mathbb{R}_+}$, where $a = \tan(\alpha)$ and $b = p/\cos(\alpha)$. They showed that in this case, Equation~\ref{eq:minimize-WTLS} becomes
\begin{eqnarray}
\chi^2 (\alpha , p) = \frac{1}{n-2} \sum^n_{k=1} \frac{(y_k \cos \alpha - x_k \sin \alpha - p)^2}{u_{x,k}^2 \sin^2 \alpha + u_{y,k}^2 \cos^2 \alpha } 
\end{eqnarray}

\noindent
For our case, we assume $u_{x,k} = u_{y,k} = \sigma$, for all $k$. Therefore we have a very neat expression:
\begin{eqnarray}
\chi^2 (\alpha , p) = \frac{1}{\sigma^2 (n-2)} \sum^n_{k=1} (y_k \cos \alpha - x_k \sin \alpha - p)^2 
\end{eqnarray}

We use Truncated-Newton Method ({\tt TNMIN.pro} in IDL, written by Craig B. Markwardt) to search for the minimal point $(\alpha_0, p_0)$. Furthermore, we can approximate the uncertainty of the parameters estimation using the inverse Hessian of $\chi^2$. More explicitly:
\begin{eqnarray}
\left( \begin{array}{cc} \sigma^2 (p) & \text{cov}(p,\alpha) \\ \text{cov}(\alpha,p) & \sigma^2 (\alpha) \end{array} \right)  = 2 \left(
\begin{array}{cc} \chi_{pp}^2 & \chi_{p\alpha}^2 \\ \chi_{\alpha p}^2 & \chi^2_{\alpha \alpha} \end{array} \right)^{-1} \Bigg|_{\alpha = \alpha_0, p =p_0}
\end{eqnarray}

\noindent
where $\chi^2_{p p}  \equiv \frac{\partial^2 \chi^2}{\partial p^2}$, $\chi^2_{\alpha \alpha} \equiv \frac{\partial^2 \chi^2}{\partial \alpha^2}$, $\chi^2_{\alpha p} = \chi^2_{ p \alpha} \equiv \frac{\partial^2 \chi^2}{\partial \alpha \partial p}$


\label{lastpage}

\end{document}